\documentclass[useAMS,usenatbib,twocolumn]{mnras}
\usepackage{graphics}
\usepackage{graphicx}
\usepackage{epstopdf}
\usepackage{amsmath}
\usepackage{amssymb}
\usepackage{xspace}
\usepackage{commath}
\usepackage{times}
\usepackage{soul}
\usepackage{bm}
\usepackage{mathtools}

\usepackage{hyperref}
\hypersetup{dvips, colorlinks=true, linkcolor=black, citecolor=black, filecolor=blue, urlcolor=black}

\usepackage{graphicx}
\usepackage{epsfig}
\usepackage{epstopdf}

\usepackage{acronym}
\newacro{DF}{distribution function}
\newacroplural{DFs}{distribution functions}
\newcommand{\DF}{\ac{DF}}
\newcommand{\DFs}{\acp{DF}}
\newacro{RR}{resonant relaxation}
\newcommand{\RR}{\ac{RR}}
\newacro{NR}{non-resonant relaxation}
\newcommand{\NR}{\ac{NR}}
\newacro{BL}{Balescu--Lenard}
\newcommand{\BL}{\ac{BL}}
\newacro{FP}{Fokker--Planck}
\newcommand{\FP}{\ac{FP}}
\newacro{QL}{quasilinear}
\newcommand{\QL}{\ac{QL}}

\newcommand{\rd}{\mathrm{d}}
\newcommand{\re}{\mathrm{e}}
\newcommand{\ri}{\mathrm{i}}
\newcommand{\bJ}{\mathbf{J}}
\newcommand{\bT}{\bm{\theta}}
\newcommand{\bO}{\mathbf{\Omega}}
\newcommand{\obJ}{\overline{\mathbf{J}}}
\newcommand{\obJp}{\overline{\mathbf{J}}^{\prime}}
\newcommand{\obn}{\overline{\mathbf{n}}}
\newcommand{\obnp}{\overline{\mathbf{n}}^{\prime}}
\newcommand{\obT}{\overline{\bm{\theta}}}
\newcommand{\obO}{\overline{\mathbf{\Omega}}}
\newcommand{\Ftot}{F_{\mathrm{tot}}}
\newcommand{\p}{\partial}
\newcommand{\bF}{\mathbf{F}}
\newcommand{\bFRR}{\mathbf{F}_{\mathrm{RR}}}
\newcommand{\bFNR}{\mathbf{F}_{\mathrm{NR}}}

\newcommand{\bn}{\mathbf{n}}
\newcommand{\bnp}{\mathbf{n}^{\prime}}
\newcommand{\mF}{\mathcal{F}}
\newcommand{\bJp}{\mathbf{J}^{\prime}}
\newcommand{\Jp}{J^{\prime}}
\newcommand{\deltaD}{\delta_{\mathrm{D}}}
\newcommand{\np}{n^{\prime}}
\newcommand{\rp}{r^{\prime}}
\newcommand{\hbr}{\widehat{\mathbf{r}}}
\newcommand{\hbrp}{\widehat{\mathbf{r}}^{\prime}}
\newcommand{\mC}{\mathcal{C}}
\newcommand{\rperi}{r_{\mathrm{p}}}
\newcommand{\rapo}{r_{\mathrm{a}}}
\newcommand{\Emin}{E_{\mathrm{min}}}
\newcommand{\Lz}{L_{z}}
\newcommand{\br}{\mathbf{r}}
\newcommand{\bv}{\mathbf{v}}
\newcommand{\brp}{\mathbf{r}^{\prime}}
\newcommand{\Rb}{R_{\mathrm{b}}}
\newcommand{\vtheta}{\vartheta}
\newcommand{\bN}{\mathbf{N}}
\newcommand{\bM}{\mathbf{M}}
\newcommand{\bI}{\mathbf{I}}
\newcommand{\thetap}{\theta^{\prime}}
\newcommand{\vphi}{\varphi}
\newcommand{\vphip}{\varphi^{\prime}}
\newcommand{\Min}{\mathrm{Min}}
\newcommand{\Max}{\mathrm{Max}}
\newcommand{\veps}{\varepsilon}
\newcommand{\rmin}{r_{\mathrm{min}}}
\newcommand{\rmax}{r_{\mathrm{max}}}
\newcommand{\up}{u^{\prime}}
\newcommand{\Ra}{R_{\mathrm{a}}}
\newcommand{\bD}{\mathbf{D}}
\newcommand{\bvp}{\mathbf{v}^{\prime}}
\newcommand{\vp}{v^{\prime}}
\newcommand{\Iinf}{I^{\mathrm{inf}}}
\newcommand{\Isup}{I^{\mathrm{sup}}}
\newcommand{\Ep}{E^{\prime}}
\newcommand{\Lp}{L^{\prime}}
\newcommand{\bOp}{\mathbf{\Omega}^{\prime}}
\newcommand{\bx}{\mathbf{x}}
\newcommand{\bxp}{\mathbf{x}^{\prime}}
\newcommand{\xp}{x^{\prime}}
\newcommand{\ep}{e^{\prime}}
\newcommand{\gp}{g^{\prime}}
\newcommand{\Card}{\mathrm{Card}}
\newcommand{\mO}{\mathcal{O}}
\newcommand{\xperi}{x_{\mathrm{p}}}
\newcommand{\xapo}{x_{\mathrm{a}}}
\newcommand{\mE}{\mathcal{E}}
\newcommand{\oxp}{\overline{x}^{\prime}}
\newcommand{\oep}{\overline{e}^{\prime}}
\newcommand{\sigmap}{\sigma^{\prime}}
\newcommand{\omegares}{\omega_{\mathrm{res}}}
\newcommand{\omegacirc}{\omega_{\mathrm{circ}}}
\newcommand{\etacirc}{\eta_{\mathrm{circ}}}
\newcommand{\omegarad}{\omega_{\mathrm{rad}}}
\newcommand{\etarad}{\eta_{\mathrm{rad}}}
\newcommand{\xprad}{x^{\prime}_{\mathrm{rad}}}
\newcommand{\xpcirc}{x^{\prime}_{\mathrm{circ}}}
\newcommand{\qb}{q_{\mathrm{b}}}
\newcommand{\Kres}{K_{\mathrm{res}}}
\newcommand{\xmax}{x_{\mathrm{max}}}
\newcommand{\qmin}{q_{\mathrm{min}}}
\newcommand{\qmax}{q_{\mathrm{max}}}
\newcommand{\xpmin}{x^{\prime}_{\mathrm{min}}}
\newcommand{\xpmax}{x^{\prime}_{\mathrm{max}}}
\newcommand{\ralpha}{r_{\alpha}}
\newcommand{\rbeta}{r_{\beta}}
\newcommand{\speri}{s_{\mathrm{p}}}
\newcommand{\sapo}{s_{\mathrm{a}}}
\newcommand{\nmax}{n_{\mathrm{max}}}
\newcommand{\ellmax}{\ell_{\mathrm{max}}}
\newcommand{\ellcrit}{\ell_{\mathrm{crit}}}
\newcommand{\ellcut}{\ell_{\mathrm{cut}}}
\newcommand{\ellcalc}{\ell_{\mathrm{calc}}}
\newcommand{\bmin}{b_{\mathrm{min}}}
\newcommand{\ncut}{n_{\mathrm{cut}}}

\newcommand{\bmax}{b_{\mathrm{max}}}
\newcommand{\tmax}{t_{\mathrm{max}}}
\newcommand{\HU}{\mathrm{HU}}
\newcommand{\Nreal}{N_{\mathrm{real}}}
\newcommand{\bFRRlowell}{\mathbf{F}_{\mathrm{RR}}^{\mathrm{low} \, \ell}}
\newcommand{\bFRRhighell}{\mathbf{F}_{\mathrm{RR}}^{\mathrm{high} \, \ell}}
\newcommand{\hbL}{\widehat{\mathbf{L}}}
\newcommand{\rH}{r_{\mathrm{H}}}
\newcommand{\omegaR}{\omega_{\mathrm{R}}}
\newcommand{\omegaM}{\omega_{\mathrm{M}}}
\newcommand{\tcross}{t_{\mathrm{cross}}}
\newcommand{\ellBL}{\ell_{\mathrm{BL}}}

\newcommand{\ImPart}{\mathrm{Im}}
\newcommand{\Lambdacut}{\Lambda_{\mathrm{cut}}}
\newcommand{\bcut}{b_{\mathrm{cut}}}
\newcommand{\bISO}{b_{\mathrm{c}}}
\newcommand{\Fd}{F_{\mathrm{d}}}
\newcommand{\Hd}{H_{\mathrm{d}}}
\newcommand{\Phid}{\Phi_{\mathrm{d}}}
\newcommand{\tF}{\widetilde{F}}
\newcommand{\tPhi}{\widetilde{\Phi}}
\newcommand{\psid}{\psi^{\mathrm{d}}}
\newcommand{\oN}{\overline{N}}
\newcommand{\oM}{\overline{M}}
\newcommand{\obN}{\overline{\mathbf{N}}}
\newcommand{\obM}{\overline{\mathbf{M}}}
\newcommand{\Htot}{H_{\mathrm{tot}}}
\newcommand{\obF}{\overline{\mathbf{F}}}
\newcommand{\half}{\tfrac{1}{2}}
\newcommand{\betap}{\beta^{\prime}}

\usepackage{soul}

\begin{document}

\title[Relaxation of globular clusters]{
Resonant and non-resonant relaxation of globular clusters
}

\author[J.-B. Fouvry, C. Hamilton, S. Rozier and C. Pichon]{
Jean-Baptiste Fouvry$^{1}$,
Chris Hamilton$^{2}$,
Simon Rozier$^{3}$,
Christophe Pichon$^{1,4,5}$
\vspace*{6pt}\\
\noindent
$^{1}$ CNRS and Sorbonne Universit\'e, UMR 7095, Institut d'Astrophysique de Paris, 98 bis Boulevard Arago, F-75014 Paris, France \\
\noindent$^{2}$ Department of Applied Mathematics and Theoretical Physics, University of Cambridge, Wilberforce Road, Cambridge CB3 0WA, UK \\
\noindent$^{3}$ Universit\'e de Strasbourg, CNRS UMR 7550, Observatoire astronomique de Strasbourg, 11 rue de l'Universit\'e, 67000 Strasbourg, France\\
\noindent$^{4}$ Korea Institute of Advanced Studies (KIAS) 85 Hoegiro, Dongdaemun-gu, Seoul, 02455, Republic of Korea\\
\noindent$^{5}$ IPhT, DRF-INP, UMR 3680, CEA, Orme des Merisiers Bat 774, 91191 Gif-sur-Yvette, France\\
}

\maketitle

\begin{abstract}
Globular clusters contain a finite number of stars.  As a result, they
inevitably undergo secular evolution (`relaxation') causing their mean
distribution function (DF) to evolve on long timescales. On one hand, this long-term
evolution may be interpreted as driven by the accumulation of local
deflections along each star's mean field trajectory --- so-called `non-resonant
relaxation'. On the other hand, it can be thought of as driven by non-local,
collectively dressed and resonant couplings between stellar orbits, 
a process termed `resonant relaxation'.
In this paper we consider a model globular cluster
represented by a spherical, isotropic isochrone DF, and compare in detail the
predictions of both resonant and non-resonant relaxation theories against
tailored direct $N$-body simulations. In the space of orbital actions (namely
the radial action and total angular momentum), we find that both resonant and
non-resonant theories predict the correct morphology for the secular evolution
of the cluster's DF, although non-resonant theory over-estimates the amplitude
of the relaxation rate by a factor ${\sim 2}$. We conclude that the secular
relaxation of hot isotropic spherical clusters is not dominated by collectively
amplified large-scale potential fluctuations, despite the existence
of a strong ${\ell = 1}$ damped mode. Instead, collective amplification
affects relaxation only marginally even on the largest scales. The
predicted contributions to relaxation from smaller scale fluctuations are essentially the
same from resonant and non-resonant theories.
\end{abstract}


\section{Introduction}
\label{sec:intro}

Predicting accurately the long-term evolution
of self-gravitating systems is a cornerstone of galactic dynamics.
Because they can (often) be modelled as isolated and fully self-gravitating systems,
globular clusters appear as ideal testbeds to challenge
our understanding of the long-term relaxation of long-range interacting systems,
 and as such have been the topic of recurrent interest.
To characterise their dynamics, one must
account for these systems' key properties.
(i) Globular clusters are inhomogeneous,
i.e.\ stars follow intricate mean field orbits.
(ii) Owing to their short dynamical time,
globular clusters are dynamically relaxed, so that their mean field distribution
can be taken as quasi-stationary.
(iii) Given that all stars contribute to the system's self-consistent
gravitational potential, globular clusters amplify perturbations,
so that potential fluctuations are dressed by collective effects.
(iv) To each orbit is associated a set of orbital frequencies,
making globular clusters resonant systems,
creating a natural time dichotomy
between the fast mean field orbital timescale and the slow
timescale of orbital distortion.
(v) Finally, globular clusters are discrete,
i.e.\ composed of a finite number of constituents.
As such, they are perturbed by Poisson
shot noise fluctuations.
It is the goal of kinetic theory to describe
the long-term fate of globular clusters
accounting for all of these features self-consistently.

The textbook approach to describing globular cluster relaxation is the one first
pioneered by~\cite{Chandrasekhar1943}
(see, e.g.\@,~\citealt{Chavanis2013a} for a detailed historical account).
In Chandrasekhar's picture, the velocity of
a given test star is weakly perturbed as it flies on a straight line through a
stationary and homogeneous background of field stars. The test star undergoes a
series of weak, local, and uncorrelated kicks from each field star it encounters
with some impact parameter $b$. Integrating over all impact parameters, one can
estimate the velocity diffusion coefficients, provided that one introduces some
appropriate cutoffs: $\bmax$ to avoid a large-scale divergence associated with
the system's finite extent, and $\bmin$ to avoid a small-scale divergence
associated with hard encounters. The resulting diffusion coefficients are
proportional to the Coulomb logarithm, ${\ln \Lambda \!=\! \ln (\bmax /
\bmin)}$. The cluster's overall long-term relaxation is interpreted as the
result of the superposition of a large number of accumulated deflections felt by
each star while following its underlying unperturbed mean field orbit. This
(orbit-averaged) \NR\ theory provides the canonical picture of cluster
relaxation~\citep{HeggieHut2003,BinneyTremaine2008}.

Of course, some of the intrinsic limitations of Chandrasekhar's \NR\ theory
should not be so easily dismissed. (i) It ignores the system's inhomogeneity when
describing the star's unperturbed trajectories, which would typically require
the introduction of angle-action coordinates\footnote{The \NR\ theory still
partially accounts for inhomogeneity through its orbit-average of the local
homogeneous diffusion coefficients.}. (ii) Owing to the quasi-periodic nature of
the orbits, stellar encounters can be resonant and correlated. (iii) The \NR\
theory neglects collective effects, i.e.\ it neglects the ability of the cluster
to amplify its own intrinsic self-generated fluctuations. This dressing of
potential fluctuations is of prime importance on the cluster's largest scale,
given the attractive nature of the gravitational force, and may be described
using linear response theory (see \S{5.3} in~\citet{BinneyTremaine2008}).
Fortunately, recent theoretical efforts have provided us with a more generic
kinetic theory that can account for all these additional physical ingredients:
the (inhomogeneous) \BL\ equation~\citep{Heyvaerts2010,Chavanis2012}. As such,
we now have at our disposal a self-consistent approach that accounts
simultaneously for a system's inhomogeneity, resonances, and self-gravity. We
generically call such a framework the \RR\ theory\footnote{
This terminology was first introduced in
the context of galactic nuclei~\citep{RauchTremaine1996},
where all orbits satisfy the same global resonance condition
of the form ${ \bn \!\cdot\! \bO (\bJ) \!=\! 0 }$.}, to emphasise its ability to
capture the contributions from long-range, amplified, resonant and correlated
fluctuations.

Despite its shortcomings, the \NR\ theory is relatively easy to implement in
practice and is used routinely to describe the long-term evolution of globular
clusters~\cite[see, e.g.\@,][and references therein]{Vasiliev2015}. Implementing
the \RR\ theory is much more difficult, as it requires one to characterise the
system's orbits, linear response, and resonance structure. As such, it has not
been applied widely to self-gravitating systems, except in the cases of
razor-thin stellar discs~\citep{Fouvry+2015}, galactic
nuclei~\citep{BarOr+2018}, and globular clusters~\citep{Hamilton+2018}. Here, we
focus on the case of globular clusters, which are the archetypes of isolated
self-gravitating spherical stellar systems. Benefiting from recent improvements
to the effective implementation of \RR\ theories, we revisit the calculations
from~\cite{Hamilton+2018} to place them
on much firmer numerical ground. In
addition to these analytical developments, it is now possible to perform ever
larger numerical simulations by integrating directly the dynamics of globular
clusters with realistic numbers of stars~\citep[e.g.\@,][]{Wang+2015}. Relying
on such tailored simulations,
we are able to carefully test the \NR\ and \RR\
theories, and to examine the influence of resonances
and collective effects, by computing the
evolution of the \DF\ in action space.

We have two main goals in this paper: (i) to determine the importance of
collective effects in accelerating the cluster's large-scale resonant
relaxation, and (ii) to compare the two main theories of relaxation
in spherical clusters (\NR\ and \RR\@) against detailed direct $N$-body
simulations. Our work is organised as follows. In \S\ref{sec:Relaxation}, we
present the key concepts of both \NR\ and \RR\ theories. We apply these theories
to isotropic isochrone clusters in \S\ref{sec:Application}, and discuss them in
\S\ref{sec:Discussion}. Finally, we conclude in \S\ref{sec:Conclusion}, and
discuss the relative merits and flaws of \NR\ and \RR\ in the more general
context of  galactic dynamics. Throughout these sections we keep technical
exposition to a minimum, and refer the reader
to the relevant appendices for the details.

\section{Relaxation of spherical stellar systems}
\label{sec:Relaxation}

We consider a set of $N$ stars of individual mass ${ \mu \!=\! M / N }$, with
$M$ the system's total mass. We assume that the system's mean potential $\psi$
is spherically symmetric, i.e.\ ${ \psi \!=\! \psi (r) }$.
Owing to spherical symmetry, unperturbed stellar orbits
in the mean potential $\psi$
are each confined to a two-dimensional plane.
They can therefore be characterised by their orientation
(i.e.\ the direction of their orbital angular momentum vector)
as well as two action variables
\begin{equation}
\bJ = (J_{r} , L) ,
\label{bJ}
\end{equation}
with $J_{r}$ the radial action,
and $L$ the norm of the angular momentum.
We spell out explicitly all our conventions
for the angle-action coordinates in \S\ref{sec:MF}.

The fact that there is a finite number $N$ of stars in the system means that the
exact potential is not equal to ${ \psi (r) }$, but instead fluctuates around
${ \psi (r) }$.  As a result, stars are gradually nudged to new mean field
orbits, i.e.\ they slowly drift to new values of $\bJ$. To describe this
evolution statistically we introduce the total \DF\@, ${ \Ftot \!=\! \Ftot (\bJ)
}$ --- with $\br$ the position and $\bv$ the velocity --- defined so that $\rd
\br \rd \bv \Ftot(\bJ(\br,\bv))$ is the mass enclosed in the ${6D}$ phase space
volume element $\rd \br \rd \bv$. Integrating over all phase space, we then have
${ \!\int\! \rd \br \rd \bv \Ftot \!=\! M }$. Moreover, as shown in
\citet{Hamilton+2018} one can integrate out the variables corresponding to the
orbital orientations and focus exclusively on the evolution of the system in the
${2D}$ $\bJ$-space (see also~\S\ref{sec:3Dto2D}). To this end we define the reduced \DF\
\begin{equation}
F (\bJ) = 2 L \, \Ftot (\bJ) .
\label{def_DFred}
\end{equation}
The average number
of stars within the phase space area element ${ \rd \bJ }$ is then equal to
$[(2\pi)^3/\mu]F(\bJ) \rd \bJ$, and the secular evolution of the phase space
density is determined via a diffusion equation of the form (see \S4 of \citealt{Hamilton+2018}):
\begin{equation}
\frac{\p F (\bJ)}{\p t} = - \frac{\p }{\p \bJ} \!\cdot\! \bF (\bJ) .
\label{def_Flux}
\end{equation}
The flux ${ \bF (\bJ) }$ describes the speed and direction
at which stars drift, on average, through action space.
The primary job of kinetic theory is to provide an expression for $\bF$.

Recent works (see the Introduction)
have highlighted the existence of two (connected)
theoretical frameworks to describe
the self-consistent long-term relaxation
of a self-gravitating system such as a globular cluster.
As a result the flux ${ \bF (\bJ) }$ can be computed 
via  two distinct methods, so that 
\begin{equation}
\bF (\bJ) =\left\{ \begin{aligned}
 {} & \bFNR (\bJ) 
\\
 {} & \quad \text{or}
\\
{} & \bFRR (\bJ) .
\end{aligned}\right.
\label{split_bF}
\end{equation}
Here, ${ \bFNR (\bJ) }$ is the prediction of the orbit-averaged non-resonant
(NR) relaxation theory, while ${ \bFRR (\bJ) }$ is that of the (dressed)
resonant relaxation (RR) theory. A key goal of the present paper is to assess
which of these formalisms is the most apt at describing the relaxation of
stellar clusters, and to clarify the connections between them. Let us now
briefly review each of them in turn.

\subsection{Non-Resonant Relaxation}
\label{sec:NR}

One contribution to relaxation comes from `local' two-body scattering events.
Here, the word `local' denotes interactions that can be considered local in
space and instantaneous in time, so that they can be treated using an impulse
approximation \citep{BinneyTremaine2008}. In this case, each star undergoes a
series of independent two-body encounters that result in small modifications to
its velocity by some amount ${ \delta \bv }$.  In particular, ${ \delta \bv }$
is a function of the impact parameter $b$ of the encounter in question. Summing
up all such encounters by integrating over all possible $b$, converting to
angle-action space and averaging over stellar orbits results in the
orbit-averaged \FP\ flux~\citep[][\S{7.4.2}]{BinneyTremaine2008}
\begin{equation}
\bFNR (\bJ) = \bD_{1} (\bJ) \, F (\bJ) - \frac{1}{2} \frac{\p }{\p \bJ} \!\cdot\! \bigg[ \bD_{2} (\bJ) \, F (\bJ) \bigg] ,
\label{def_bF_NR}
\end{equation}
where the first-order diffusion vector ${ \bD_{1} (\bJ) }$ and the second-order
diffusion tensor ${ \bD_{2} (\bJ) }$ are given by
\begin{align}
\bD_{1} (\bJ) \!&=\!\!
\begin{pmatrix}
\!\big\langle \Delta J_{r} \big\rangle \!
\\
\!\big\langle \Delta L \big\rangle \!
\end{pmatrix} \!\! ,\\
\;
\bD_{2} (\bJ) \!&=\!\!
\begin{pmatrix}
\!\big\langle \big( \Delta J_{r} \big)^{2} \big\rangle\! & \!\!\big\langle \Delta J_{r} \Delta L \big\rangle \!
\\
\!\big\langle \Delta J_{r} \Delta L \big\rangle\! & \!\!\big\langle \big( \Delta L \big)^{2} \big\rangle \!
\end{pmatrix} \!\! ,
\label{D1_D2}
\end{align}
and ${ \langle \cdot \rangle }$ denotes the average increment of a given
quantity per unit time, once averaged over an orbital period. 
We note that the \DF\ appearing in the r.h.s.\ of
Eq.~\eqref{def_bF_NR} is the reduced \DF\ from Eq.~\eqref{def_DFred}, 
as it is proportional to the density of stars in $\bJ$-space.

For details of how $\bD_{1,2}$ are computed in \NR\ theory
we refer to~\S\ref{sec:FP}.
Here we merely emphasise that unlike in the \RR\ theory
(\S\ref{sec:RR}), the diffusion coefficients $\bD_{1,2}$ do not involve any resonance condition nor require any basis
function expansion for their computation. As such, the \NR\ flux ${ \bFNR (\bJ) }$ is much easier
to compute than the \RR\ flux ${ \bFRR (\bJ) }$, to which we now turn.

\subsection{Resonant Relaxation}
\label{sec:RR}

The other contribution to the relaxation that we consider here
is that from long-range resonant couplings
between stars and fluctuations
as they stream along their mean field orbital motion.
More precisely, two stars with actions $\bJ$ and $\bJp$ will 
resonate if there exist ${ \bn , \bnp \!\in\! \mathbb{Z}^{2} }$ such that 
\begin{align}
   \bn \cdot \bO (\bJ) - \bnp \cdot \bO (\bJp) = 0.
\end{align}
where $\bO (\bJ)$, $\bO (\bJp)$ are the dynamical frequency vectors
(\S\ref{sec:MF}).
In addition, these resonantly interacting pairs of stars should not be treated
as an isolated 2-body system. Instead one must account for the fact that
resonant interactions are conveyed through the `dielectric medium' of the other
${ N \!-\! 2 }$ stars, so that the corresponding behaviour is
\textit{collective}. In the analogous setting of an electrostatic plasma, these
collective effects lead to the phenomenon of Debye shielding --- the Coulomb
interaction between two particles is greatly diminished (screened) on scales
longer than the Debye length because the collective motion of many other
particles reacts to keep the plasma quasineutral. On the other hand, in a stellar
system, where the pairwise interaction is attractive, collective effects tend to
amplify the strength of the interaction on large scales rather than diminish it.

We note that when collective effects are included one need not drop the
`2-body' concept completely.  Instead, Rostoker's principle tells us that
collective amplification acts to effectively `dress' the bare 2-body
interactions.  In this view the system evolves via a superposition of two-body
resonant encounters,
but with the Newtonian interaction
potential replaced by an effective `dressed'
potential~\citep{Gilbert1968,Hamilton2021}.

The theory that accounts for pairwise resonant interactions dressed by
collective effects is the inhomogeneous \BL\
theory~\citep{Heyvaerts2010,Chavanis2012}.
When applied to spherical stellar
systems, the \BL\ theory provides a flux \citep{Hamilton+2018}
\begin{align}
\bFRR (\bJ) {} & = \sum_{\ell} \bFRR^{\ell} (\bJ)\,,
\nonumber
\\
{} & = \sum_{\ell} \sum_{\bn , \bnp} \bn \, \mF^{\ell}_{\bn\bnp} (\bJ) ,
\label{sum_bF}
\end{align}
where ${ \bn , \bnp \!\in\! \mathbb{Z}^{2} }$
are the resonance numbers, and
${ \ell \!=\! 0,1,2,... }$ is 
 an index corresponding to the spherical harmonic
expansion of the interaction potential,
capturing the fact that pairs of orbits are typically non-coplanar\footnote{Roughly speaking this index
is Fourier conjugate to the relative angle  
between two given orbital planes --- see~\S\ref{sec:orbit-orbit}
and~\S\ref{sec:3Dto2D} for more details.}.
In the above expression $\mF_{\bn\bnp}^{\ell} $ is given by
\begin{align}
\mF_{\bn\bnp}^{\ell} {} & (\bJ) \equiv \frac{\pi (2 \pi)^{3}}{2 \ell + 1} \mu \!\! \int \!\! \rd \bJp \, L \, \Lp \, \big| \Lambda_{\bn\bnp}^{\ell} (\bJ , \bJp , \bn \!\cdot\! \bO (\bJ)) \big|^{2}
\label{def_mF}
\\
&\hskip -0.5cm{}  \times \deltaD (\bn \!\cdot\! \bO (\bJ) \!-\! \bnp \!\cdot\! \bO (\bJp)) \, \bigg(\! \bnp \!\cdot\! \frac{\p }{\p \bJp} \!-\! \bn \!\cdot\! \frac{\p }{\p \bJ} \!\bigg) \frac{F (\bJ)}{L} \frac{F (\bJp)}{\Lp}.
\nonumber
\end{align}
Here the coefficient ${ \Lambda_{\bn\bnp}^{\ell} (\bJ , \bJp , \omega) }$
captures the $\ell$-harmonic strength of the interaction of orbits with actions
$\bJ$ and $\bJp$ coupled via the resonance ${ (\bn , \bnp) }$ at frequency
$\omega$, and includes the effect of the collective amplification. We present
the key steps to derive Eq.~\eqref{def_mF} in \S\ref{sec:BLEqDerivation}; in
particular we highlight the appearance of the summation over $\ell$
harmonics in Eq.~\eqref{sum_bF} --- in other words, the fact that contributions
from different angular scales contribute independently to the flux. 
We provide
the explicit expression for ${ \Lambda_{\bn\bnp}^{\ell} }$ in \S\ref{sec:BLEq}. 

When collective amplification can be considered negligible,  the expression for
$\mF_{\bn\bnp}^{\ell}$ is unchanged except that one substitutes new
(frequency-\textit{independent}) `bare' coefficients ${ \Lambda_{\bn\bnp}^{\ell} (\bJ ,
\bJp) }$ in place of the dressed coefficients ${ \Lambda_{\bn\bnp}^{\ell} (\bJ ,
\bJp , \omega) }$ --- see \S\ref{sec:LandauEq}. In that limit, the inhomogeneous
\BL\ flux reduces to the inhomogeneous Landau flux~\citep{Chavanis2013}. As
we will see in \S\ref{sec:BL_divergence}, one   difficulty of  Eq.~\eqref{sum_bF} is its
appropriate convergence, or divergence, w.r.t.\ the sum over infinitely many harmonics $\ell$, as well as
w.r.t.\ the sum over infinitely many resonance vectors ${ (\bn , \bnp) }$.
\\
\\
The purpose of this paper is to compare the \NR\ and \RR\ predictions for
${ \p F/ \p t }$ for a spherical cluster, which are driven by the fluxes given in
Eqs.~\eqref{def_bF_NR} and~\eqref{sum_bF} respectively. While they appear fairly
different at first glance, the question at hand is to determine whether or not
they reflect a different physical diffusion mechanism, and if so what can one
learn from their detailed comparison. Some of the differences between both
fluxes may indeed be superficial, since an early choice of canonical
angle-action variables in the \RR\ case naturally highlights resonances, but a
detailed summation over all resonant couplings should formally equate to the
classical Newtonian interaction. Yet, the \NR\ flux assumes local deflections
(before an incoherent orbit-averaging), whereas the \RR\ flux accounts for
non-local and resonant coupling across the whole cluster. As such, \NR\
decouples collisions and phase mixing, while \RR\ treats them consistently. In
addition, the \RR\ theory, because it directly deals with orbits, exhibits no large-scale
divergence, while the \NR\ theory does formally diverge and so must
rely on an ad-hoc large-scale truncation $\bmax$. Finally, self-gravity is
accounted for in the \RR\ flux, whereas it is ignored in the \NR\ flux. As such,
the \RR\ theory is expected to be more realistic than its \NR\ counterpart, but
may prove needlessly complicated for following cluster relaxation in practice.

In what follows, we find a remarkable agreement between both theories,
up to an overall amplitude mismatch,
when applied to the prediction of the divergence of the diffusion flux of an
isotropic spherical isochrone cluster. This suggests that the summation over
$\ell$ in Eq.~\eqref{sum_bF} is dominated by high-order harmonics which reflect
local coupling. We also show that, in the inner regions of the cluster, the
(dressed) \BL\ flux closely resembles its (bare) Landau counterpart. This
suggests that self-gravity (i.e.\ collective amplification) has little effect on the cluster's
overall relaxation in its central regions. As such, it implies that isotropic
spheres are dynamically hot, as they involve numerous resonances with
gravitational couplings on a wide range of scales,
reflected in the need to account for many harmonics
in Eq.~\eqref{sum_bF}.

\section{Application to the spherical isochrone }
\label{sec:Application}

\begin{figure}
   \centering
  \includegraphics[width=0.45 \textwidth]{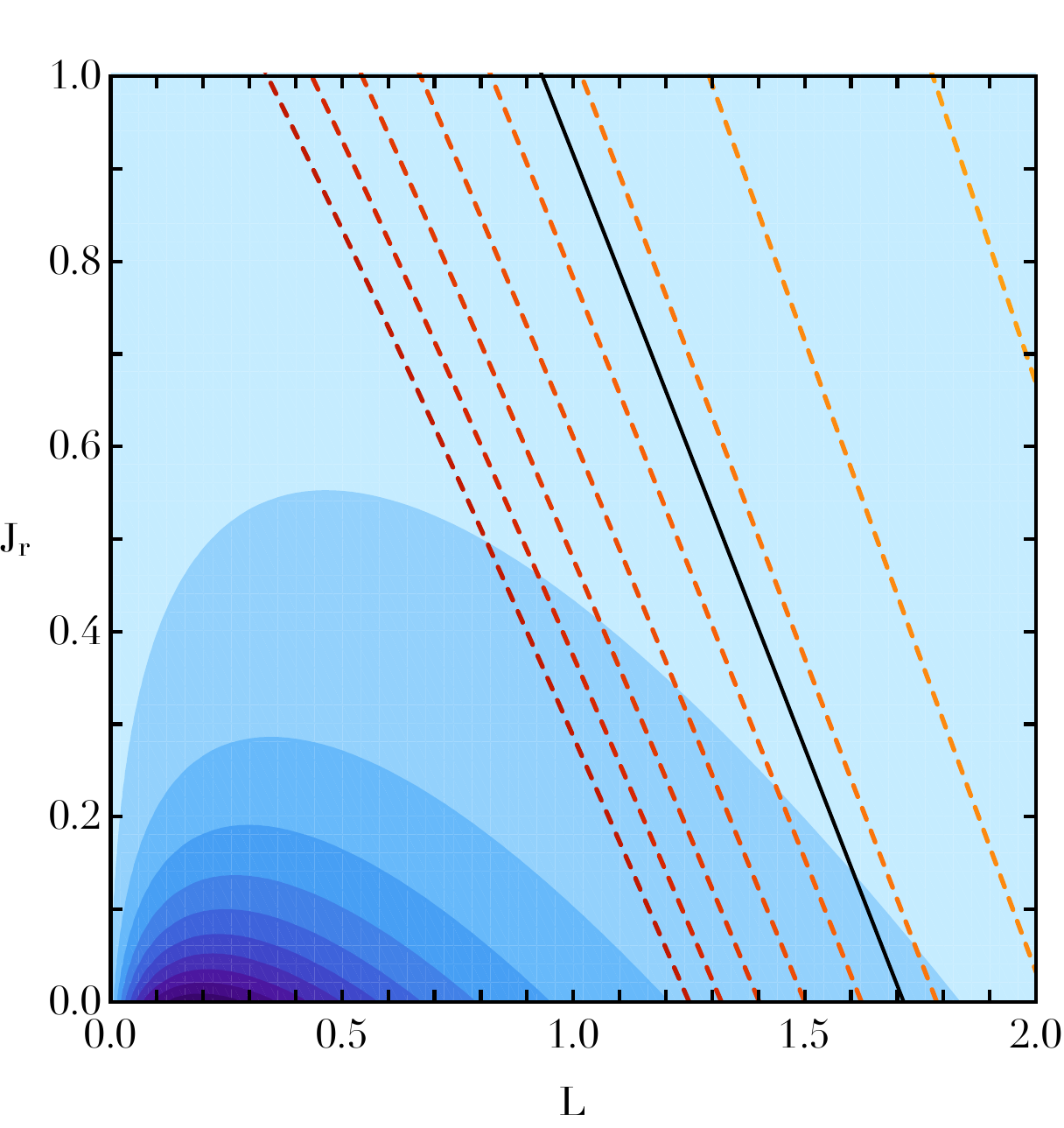}
  \caption{Illustration of the reduced \DF\@, ${ F (\bJ) }$,
  as defined in Eq.~\eqref{def_DFred},
  for an isotropic isochrone cluster, in units ${ G \!=\! M \!=\! \bISO \!=\! 1 }$.
  Blue contours are spaced linearly between 5\% and 95\%
  of the function maximum.
  Dashed contours represent the level lines
  of the resonance frequency ${ \omega \!=\! \bn \!\cdot\! \bO (\bJ) }$,
  for ${ \bn \!=\! (1,-1) }$. Contours are spaced linearly
  between ${ \omega \!=\! 0.04 \, \Omega_{0} }$ (dark color)
  and ${ \omega \!=\! 0.005 \, \Omega_{0} }$ (light color).
  As such the strong ${ \ell \!=\! 1 }$ self-gravitating amplification,
  that occurs for ${ \omega \!\simeq\! 0.017 \, \Omega_{0} }$
  (black line, see also Fig.~\ref{fig:NMat}),
  can only resonantly couple
  to a very poorly populated location of action space,
  hence the inefficiency of collective effects
  to accelerate the \RR\ relaxation (\S\ref{sec:RoleCollective effects}). 
  }
  \label{fig:DF}
\end{figure}

So far our results have been applicable to any stable stellar system with a
spherically symmetric mean field. Hereafter we will use the isochrone potential
${ \psi (r) \!=\! -GM/(\bISO \!+\! \sqrt{\bISO^2 \!+\! r^2}) }$, where $M$ is the total cluster mass and $\bISO$ the scale radius.
We use a self-consistent
\DF\@ for the isochrone model, and assume ${ N \!=\! 10^{5} }$. Moreover, to ease the computation of the
\NR\ flux, we let the \DF\ have an isotropic velocity distribution, i.e.\
${ \Ftot \!=\! \Ftot(E) }$, as illustrated in Fig.~\ref{fig:DF}.
Other details of the model are given
in \S\ref{sec:IsochronePotential}.

In order to test Eq.~\eqref{split_bF}
and the associated kinetic theories,
we will now compare the \NR\ and \RR\
predictions to direct measurements in $N$-body simulations.
In addition, in order to better highlight the importance
of long-range couplings,
we will consider two sets of clusters,
either driven by the traditional Newtonian interaction,
or by a softened Plummer interaction.
Before diving into this comparison,
a bit more work is required.

\subsection{Curing divergences}

First, as both \NR\ and \RR\ theories describe the dynamics of perturbations at
linear order, e.g.\@, through the linearised Vlasov equation, they both suffer
from divergences associated with strong interactions (i.e.\ interactions at very
short lengthscales). The \NR\ theory also diverges at large scales.  We now show
how the various divergences may be cured.

\subsubsection{\NR\ relaxation and Coulomb logarithm}
\label{sec:Truncation Coulomb}

One drawback of the orbit-averaged \FP\ theory is that it exhibits two
divergences: one arising from stellar encounters with very small impact
parameters, and one from encounters with very large impact parameters.
As a
result, the final answer, ${ \bFNR (\bJ) }$,
is necessarily proportional to the Coulomb
logarithm
\begin{equation}
   \ln \Lambda = \ln (\bmax / \bmin) \,,
   \label{exp_lnLambda_Classical}
   \end{equation}
in which the minimum/maximum impact parameters $b_{\mathrm{min}/\mathrm{max}}$
have to be prescribed by hand. The first of these is traditionally taken to be
the scale of $90^{\circ}$ deflections, ${ 2G\mu/\sigma^{2} }$, in the case 
of purely Newtonian interaction)
or by the considered softening length $\veps$ (as in
Eq.~\eqref{Usoft}) for a softened interaction:
\begin{equation}
\bmin =
\begin{cases}
\displaystyle \frac{2 G \mu}{\sigma^{2}} & \text{(Newtonian interaction)} ,
\\[1.0ex]
\displaystyle \veps & \text{(Softened interaction)} ,
\end{cases}
\label{choice_bmin}
\end{equation}
with $\sigma$ the cluster's velocity dispersion.
Meanwhile, the maximum impact parameter
is normally taken to be roughly the scale of the system itself;
a reasonable choice
is
\begin{equation}
\bmax \sim \bISO ,
\label{choice_bmax}
\end{equation}
with $\bISO$ the typical lengthscale of the considered cluster (e.g.\@, the
lengthscale entering the isochrone potential, see Eq.~\eqref{def_psi_iso}). Of
course, one should already be suspicious that interactions on these lengthscales
do not satisfy the key assumptions of Chandrasekhar's theory (see the Introduction), as they
cannot seriously be considered either local or impulsive.

In practice, for the particular case ${ N \!=\! 10^{5} }$,
and the parameters considered in our numerical simulations
(see \S\ref{sec:NumericalSimulations})
we readily find from Eqs.~\eqref{sigma_iso} and~\eqref{d_iso}
that the classical Coulomb logarithm reads
\begin{equation}
\ln \Lambda \simeq
\begin{cases}
\displaystyle 8.69 & \text{(Newtonian interaction)} ,
\\[1.0ex]
\displaystyle 3.60 & \text{(Softened interaction)} .
\end{cases}
\label{val_lnLambda}
\end{equation}
As a result, for such a large value of $N$,
strong encounters are drastically suppressed
by softening,
hence slowing down the evolution by a factor ${ \sim\! 2 }$.

\subsubsection{\RR\ and divergence at small scales}
\label{sec:BL_divergence}

In the \RR\ theory, the spatial scale of each interaction is essentially set by
the harmonic number $\ell$ (see \S\ref{sec:orbit-orbit} for further discussion).
Because of this, the resonant flux $\bFRR$ does not suffer from a large-scale
divergence: the largest scales in the problem are set by the minimum harmonic
number ${ \ell = 0 }$ from which there stems a finite contribution.  However,
$\bFRR$ still exhibits a small-scale divergence, associated with ${ \ell \!\to\!
+ \infty }$ and the improper accounting of hard interactions, that one must
heuristically cure. We now explore how this divergence arises, and offer a
prescription for dealing with it in practice.

\begin{figure}
   \centering
  \includegraphics[width=0.45 \textwidth]{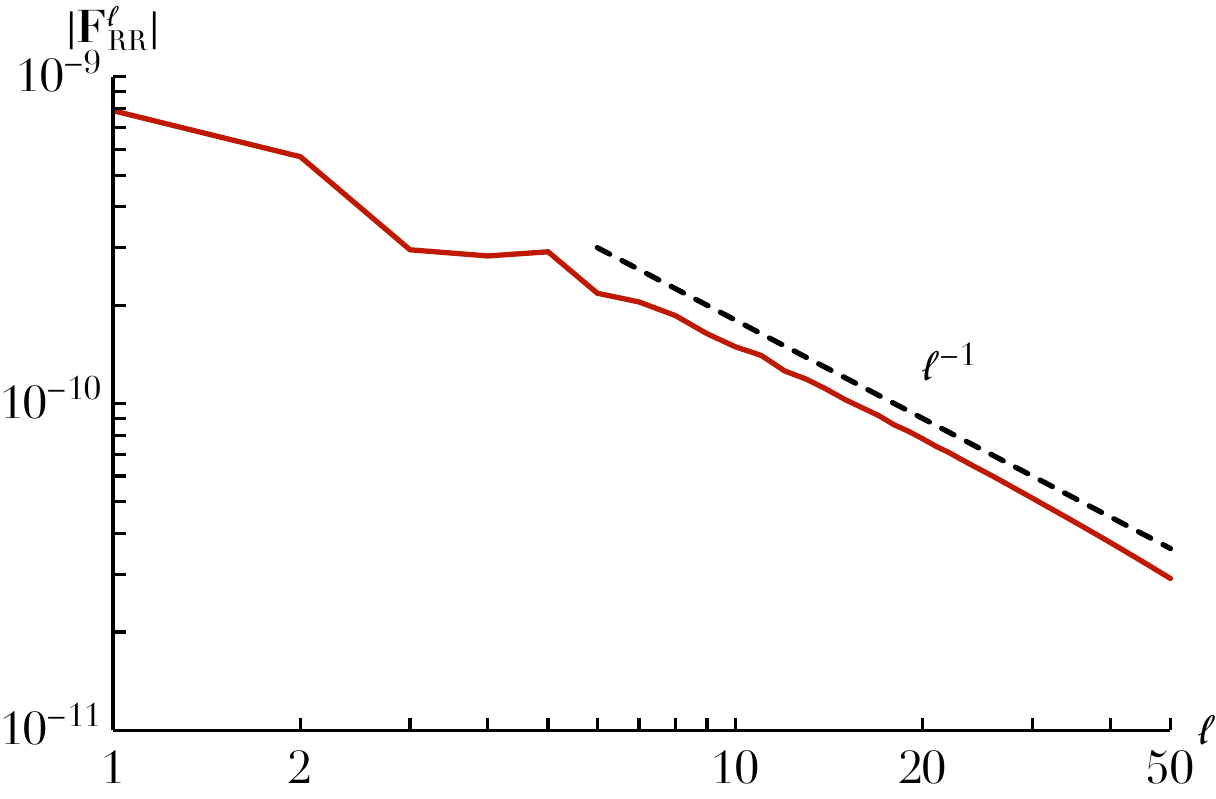}
  \caption{Illustration of the respective contributions
  from a given harmonic $\ell$ to the resonant flux from Eq.~\eqref{sum_bF},
  in the absence of collective effects,
  in units ${ G \!=\! M \!=\! \bISO \!=\! 1 }$.
  Here, the \RR\ flux has been computed at the orbital location
  ${ (J_{r} , L) \!=\! (0.1,0.5) }$, with ${ 0 \!\leq\! \ell \!\leq\! 50 }$,
  and ${ |n_{1}| , |\np_{1}| \leq 200 }$,
  using ${ K \!=\! 500 }$ points to compute the orbit-average,
  and ${ \Kres \!=\! 500 }$ points to construct the resonance lines.
  For ${ \ell \!=\! 50 }$, this amounts
  to considering more than $10^{8}$
  resonance pairs ${ (\bn , \bnp) }$
  possibly contributing to the flux.
  For ${ \ell \!\gtrsim\! 6 }$,
  we recover ${ |\bFRR^{\ell}| \!\propto\! 1/\ell }$,
  i.e.\ the logarithmic divergence on small scales
  associated with the unsoftened Newtonian interaction.
  }
  \label{fig:Coulomb}
\end{figure}

We first note that, all things being equal,
from the prefactor of Eq.~\eqref{def_mF}
we expect the flux $\bFRR^\ell$ to be
proportional to ${ 1/\ell }$ for large $\ell$,
a scaling already noticed by~\cite{Weinberg1986}
in the context of resonant dynamical friction.
At this point one might argue that the presence of the coefficient ${ \vert \Lambda^{\ell}_{\bn\bnp} \vert^2 }$
may change this simple picture; however, in practice it turns out that
${ \bFRR^\ell \!\propto\! 1/\ell }$ is a good relation.
We confirm this prediction
numerically in Fig.~\ref{fig:Coulomb} for the particular case of the spherical
isochrone potential.
In this figure, we plot the value of the resonant flux\footnote{Strictly speaking, for this calculation we ignored collective effects,
so $\bFRR^\ell$ here is the Landau flux not the \BL\ flux.  However identifying these fluxes is a good approximation since
we are only interested in the large $\ell$ behaviour where collective effects
are unimportant, see Fig.~\ref{fig:dFdtBL}.} $\vert \bFRR^\ell
\vert$ at a particular phase space location $\bJ$ as a function of $\ell$.
(To understand how
we computed
$\bFRR^\ell$, itself a significant technical challenge,
see \S\ref{sec:BLEq}--\ref{sec:IsochronePotential}).
We see that for the purely Newtonian
interaction, we find the expected scaling
${ \vert \bFRR^\ell \vert \!\propto\! 1/\ell }$
for ${ \ell \!\gtrsim\! \ellcrit \!\equiv\! 6 }$.
This scaling naturally leads to a logarithmic divergence in the calculation of
the \RR\ flux, since for $\ellmax$ large enough,
one has ${ \sum_{\ell = 0}^{\ellmax} \!1/\ell \!\propto\! \ln (\ellmax) }$.
We also point out that, for a fixed $\ell$,
there is no divergence of Eq.~\eqref{sum_bF}
w.r.t.\ the infinite sum ${ \sum_{\bn , \bnp} }$,
as illustrated in Fig.~\ref{fig:Coulombn1max}.
The problem is therefore to choose the ${ \ellmax }$
at which the infinite harmonic sum
from Eq.~\eqref{sum_bF} should be truncated.

To find the appropriate $\ellmax$, we first note that the efficiency of resonant
interactions is determined by the coupling coefficients ${
\Lambda_{\bn\bnp}^{\ell} }$.  For large $\ell$ we expect collective
amplification to be unimportant as high frequency oscillations cancel out
long-range effects (see Fig.~\ref{fig:dFdtBL}), so we consider only the bare
coefficients defined in Eq.~\eqref{def_LambdaBare}. From Eq.~\eqref{def_Ul} we
know that the efficiency of the coupling between two locations $r$ and $\rp$ is
proportional to
\begin{equation}
U_{\ell} (r , \rp) \propto \frac{1}{\rmax} \, \bigg(\! \frac{\rmin}{\rmax} \!\bigg)^{\ell} ,
\label{shape_U_Leg}
\end{equation}
with ${ \rmin \!=\! \min (r , \rp) }$ and ${ \rmax \!=\! \max (r , \rp) }$. As
$\ell$ increases, this function gets sharper so that only very local
interactions get picked up by the resonant interaction.
Let us then consider one such interaction in the
core of the cluster, and let us take ${ \rmax \!=\! \bISO }$ (the typical
lengthscale of the cluster's density), and ${ \rmin \!=\! \bISO (1 \!-\! \alpha) }$,
with ${ \alpha \!>\! 0 }$. For two stars to have a close encounter necessarily
requires that $\alpha$ is very small. Therefore in the limit of interest
(${ \alpha \!\ll\! 1, \ell \!\gg\! 1 }$), Eq.~\eqref{shape_U_Leg} becomes
\begin{equation}
U_{\ell} (\alpha) \propto \frac{1}{\bISO} \, \big( 1 \!-\! \alpha \big)^{\ell} \simeq \frac{1}{\bISO} \, \re^{- \ell \alpha} .
\label{asymp_U_Leg}
\end{equation}
The typical separation associated with this interaction
is that given by its half-width,
i.e.\ the value of $\alpha$
such that ${ U_{\ell} (\alpha)/U_{\ell} (0) \!=\! \half }$.
One naturally gets ${ \alpha \!=\! \ln (2) / \ell }$.
For a given harmonic $\ell$, ${ \bISO \alpha }$
then corresponds to the smallest scale of separation that is effectively
resolved by the coupling coefficients. As a consequence, equating this interaction scale with $\bmin$,
we may then truncate the \RR\ harmonics expansion at
\begin{equation}
\ellmax = \ln (2) \, \frac{\bISO}{\bmin} ,
\label{def_lmax}
\end{equation}
hence heuristically curing the small-scale divergence
of the \RR\ theory.

For the parameters considered in our numerical simulations,
(see \S\ref{sec:NumericalSimulations}),
we therefore truncate the \RR\ flux computation at
\begin{equation}
\ellmax \simeq 
\begin{cases}
\displaystyle 4115 & \text{(Newtonian interaction)} ,
\\[1.0ex]
\displaystyle 25 & \text{(Softened interaction)} .
\end{cases}
\label{choice_lmax}
\end{equation}
We note that the introduction of softening strongly reduces the range of
harmonics that contribute to the dynamics.

\subsection{The role of collective effects in RR}
\label{sec:RoleCollective effects}

A central feature of the \BL\ formalism is that it accounts for the
collective amplification (`dressing') of potential fluctuations. Mathematically,
this amplification is captured in the \RR\ flux from Eq.~\eqref{def_mF}
through the frequency-dependent dressed coupling coefficients ${ \Lambda_{\bn\bnp}^{\ell}
(\bJ , \bJp , \omega) }$, which are defined
in Eq.~\eqref{def_Lambda}. These coefficients in
turn depend on the susceptibility matrix ${ \mathbf{N}_\ell(\omega) }$,
defined in Eq.~\eqref{def_N}.
If one ignores collective effects then
${ \mathbf{N}_\ell(\omega) \!\to\! \bI }$,
the dressed coupling coefficients become
the bare coupling coefficients (\S\ref{sec:LambdaBare}),
and the \BL\ flux reverts to the Landau flux.  

It is natural to ask what impact the collective amplification has upon secular
evolution in spherical systems --- in other words, how does the \BL\ prediction
differ from that of Landau? In this section we argue that the difference between
\BL\ and Landau predictions is marginal on the largest scales, and is otherwise
negligible, so that collective effects have only a minor role to play
in the bulk evolution of dynamically hot stellar systems.  

\begin{figure}
   \centering
  \includegraphics[width=0.45 \textwidth]{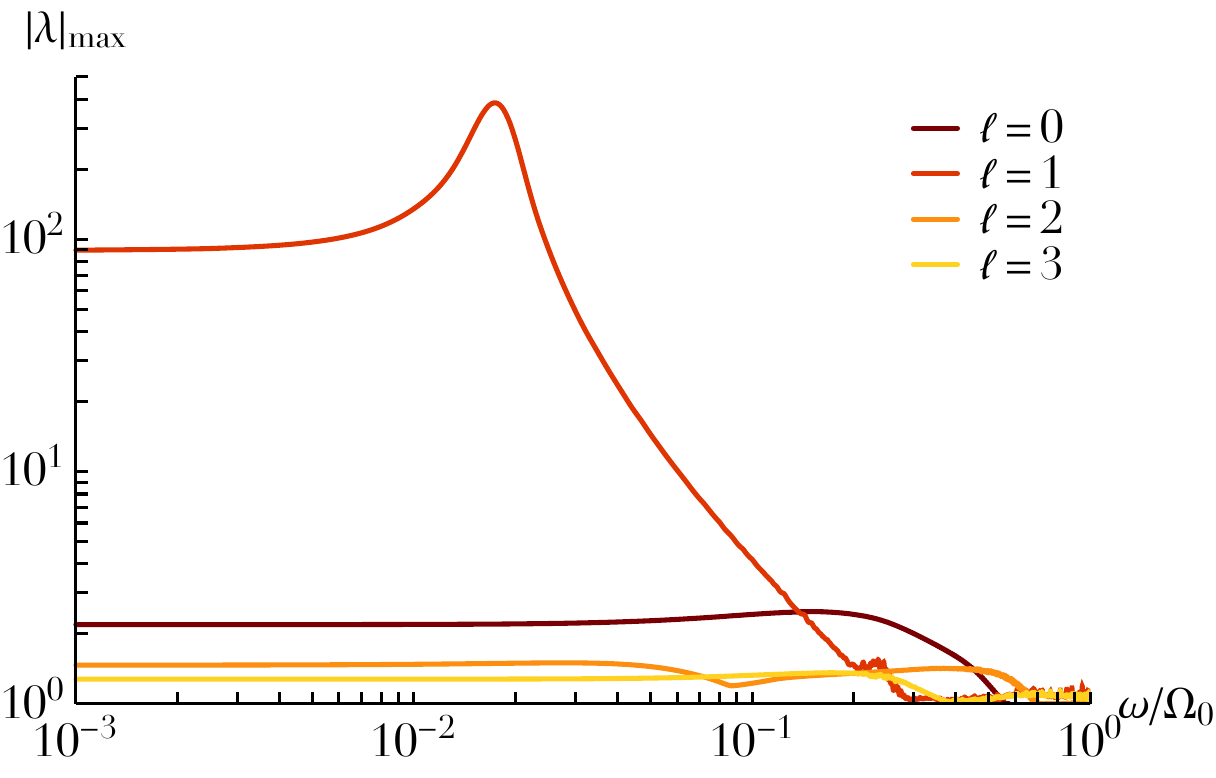}
    \includegraphics[width=0.45 \textwidth]{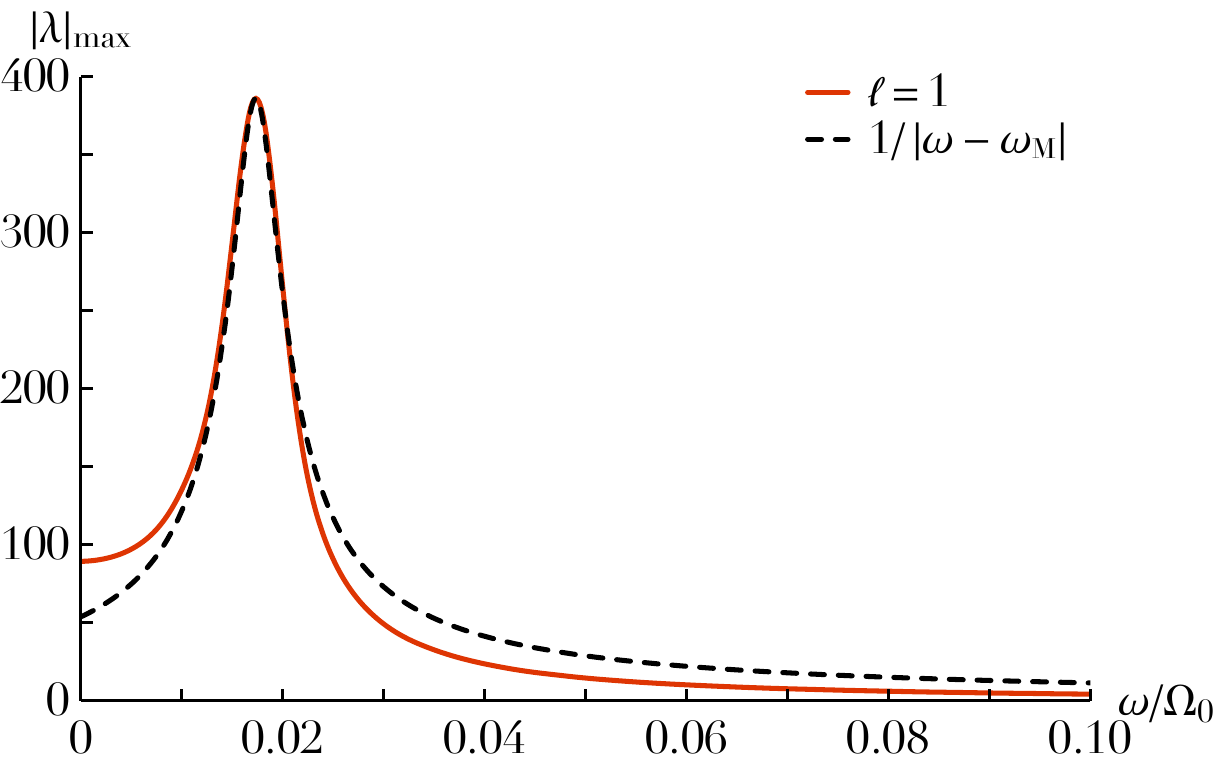}
  \caption{Top panel: Illustration of the maximum eigenvalue norm,
  ${ |\lambda (\omega)|_{\max} }$,
  of the susceptibility matrix, ${ \bN_{\ell}
  (\omega) }$, as a function of the frequency $\omega$,
  and for various harmonics $\ell$. Both axes are logarithmic.
  Bottom panel: Same as the top panel except on linear axes,
  and zoomed around the frequency of the ${ \ell \!=\! 1 }$
  weakly damped mode,
  along with a function ${ \!\propto\! 1/|\omega \!-\! \omegaM| }$.
  As already put forward
  in~\protect\cite{Weinberg1994},
  globular clusters generically support a weakly
  damped ${ \ell \!=\! 1 }$ mode,
  here estimated with the complex frequency
  ${ \omegaM / \Omega_0 \!\simeq\! 0.017 \!-\! 0.0024\,\ri }$,
  using the peak's position and full width at half maximum.
  The \BL\ equation accounts for the
  amplification of fluctuations at all real frequencies, and naturally the
  biggest effect is around that frequency,
  leading to a very efficient self-gravitating dressing
  of the perturbations.
  Yet, such a low frequency is only commensurate
  with outer orbits in the cluster,
  hence the inefficiency of collective effects
  to accelerate the \RR\ relaxation in the centre of the cluster.
  See \S\ref{sec:RepMat}
  for the details of the numerical parameters used.
  }
  \label{fig:NMat}
\end{figure}

To see this, we begin by considering the top panel of Fig.~\ref{fig:NMat}, in
which we plot the eigenvalue of ${ \mathbf{N}_{\ell} (\omega) }$ that has the greatest modulus, which we call $\vert \lambda \vert_\mathrm{max}$, as a function of $\omega$
for different $\ell$.
Clearly, if collective amplification is to be an important effect (i.e.\ if
${\mathbf{N}_\ell(\omega) }$ is to differ significantly from $\bI$) for any
given $\ell$ and $\omega$, then this eigenvalue must differ significantly from
unity. We see from the plot that in all cases, ${ \vert \lambda
\vert_\mathrm{max} \to 1 }$ for $\vert \omega/\Omega_0 \vert \gtrsim 1$, meaning
collective effects are unimportant at high frequencies. Meanwhile ${ \vert
\lambda \vert_\mathrm{max} \!\gg\! 1 }$ is achievable for ${\ell \!=\! 1 }$,
i.e.\ dipole fluctuations can be greatly enhanced by collective amplification
\citep{Weinberg1994,Lau+2019,Heggie+2020}.
However, even for ${ \ell \!=\! 1 }$ the
amplification is large (${\vert \lambda \vert_\mathrm{max} \!\gtrsim\! 10 }$)
only for very small frequencies, ${\omega/\Omega_0 \!\lesssim\! 0.1 }$.  In the
lower panel of Fig.~\ref{fig:NMat} we demonstrate that the ${ \ell \!=\! 1 }$
curve exhibits a clear and narrow amplification around the frequency ${\omega
\!\simeq\! 0.017 \, \Omega_0 }$. This peak is to be interpreted as the imprint
along the real frequency line of the cluster's ${ \ell \!=\! 1 }$ weakly damped
mode~\citep{Weinberg1994}, i.e.\ a pole of the susceptibility matrix,
${\bN_{\ell = 1} (\omega) }$, in the lower half of the complex frequency plane.
More precisely, following Eq.~{(139)} of~\cite{Nelson+1999},  it is sensible to
approximate 
\begin{equation}
   {|\lambda (\omega)|_{\max} \, \propto \, 1/|\omega \!-\! \omegaM| },
\label{eqn:pole_approximation}
\end{equation} 
 around this peak, where $\omegaM$ is the complex frequency of the mode, 
with ${ \ImPart[\omegaM] \!<\! 0 }$. By fitting the analytical form
from Eq.~\eqref{eqn:pole_approximation}
to match the amplitude and full-width half-maximum
of our numerical results, we estimate ${\omegaM / \Omega_{0}
\!\simeq\! 0.017 \!-\! 0.0024 \, \ri }$. 
To summarise, the dressed coupling coefficients
${\Lambda_{\bn\bnp}^{\ell} (\bJ , \bJp , \omega) }$
may differ markedly from their corresponding bare ones
only for ${ \ell \!=\! 1 }$ and ${ \omega/\Omega_0 \!\lesssim\! 0.1 }$, and the strongest amplification will be centred on 
$\omega/\Omega_0 \simeq 0.017$. 

We can now use this information to pinpoint the likely impact of collective
effects on the secular evolution. Considering again Eq.~\eqref{def_mF}, we see
that the coupling coefficients contribute to the flux ${ \bF_{\bn \bnp}^{\ell}
(\bJ) }$ at the resonance frequency ${ \omega \!=\! \bn \!\cdot\!
\mathbf{\Omega} }$. Moreover, we know from \S\ref{sec:Lambda} that for ${ \ell
\!=\! 1 }$ the only vectors $\mathbf{n}$ that contribute are of the form ${ \bn
\!=\! (n_{1}, 1) }$, with ${ n_{1} \!\in\! \mathbb{Z} }$ any
integer\footnote{Strictly speaking ${ (-n_{1}, -1) }$ also contributes, but this
does not change our argument.}. Putting these two facts together with the
requirement ${ \omega/\Omega_0 \!\lesssim\! 0.1 }$, we see that ${ \bF_{\bn
\bnp}^{\ell} (\bJ) }$ may undergo significant collective amplification only if
${ \ell \!=\! 1 }$, ${ \bn \!=\! (n_{1}, 1) }$, and
\begin{equation}
   n_1 \Omega_{1} + \Omega_{2}
   \lesssim 0.1 \, \Omega_0.
   \label{eqn:collective_amplification_requirement}
\end{equation}
Given that ${ \Omega_{1}, \Omega_{2} \!>\! 0 }$ and that orbits in
cored spherical systems always have ${ 1/2 \!\leq\! \Omega_{2}/\Omega_{1} \!\leq\!
1}$, the only practical value of $n_1$ for which some stars will be capable of
satisfying Eq.~\eqref{eqn:collective_amplification_requirement} is ${ n_1 \!=\!
-1 }$. This can be checked easily in the case of the isochrone potential,
for which we have explicit expressions for the frequencies
(\S\ref{sec:IsochronePotential}), but should hold for all sensible cored
spherical potentials. In Fig.~\ref{fig:DF} we plot contours of the reduced
isotropic \DF\ of the isochrone model, ${ F (\bJ) }$.  Overplotted with dashed
lines are contours of $\vert { (-1,1) \!\cdot\! \mathbf{\Omega} } \vert$ spaced linearly
from a maximum of ${ 0.04 \, \Omega_0 }$ (dark) to a minimum of ${ 0.005 \,
\Omega_0 }$ (light). Since ${ \ell \!=\! 1 }$ fluctuations are
amplified most strongly around ${\simeq\! 0.017 \, \Omega_0 }$, we
see that even for ${ n_1 \!=\! -1 }$, the condition from
Eq.~\eqref{eqn:collective_amplification_requirement} holds only in a very
sparsely occupied region of action space.
 
To complete our argument, we look once again at Eq.~\eqref{def_mF}.
The Dirac delta
function in its right hand side demands that a star with action $\bJ$ and
frequency ${ \bn \!\cdot\! \bO }$ couples to another star with action
$\bJp$ and frequency ${ \bnp \!\cdot\! \bOp }$ only if
${ \bn \!\cdot\! \bO \!=\! \bnp \!\cdot\! \bOp }$.
But for
this interaction to be strongly amplified by collective effects,
we require Eq.~\eqref{eqn:collective_amplification_requirement}
to be true,
which we have just seen means that 
${ F(\bJ) }$ and ${ F (\bJp) }$ and their gradients are very small.
This fact severely suppresses the flux from Eq.~\eqref{def_mF}
at the locations where strong self-gravitating
amplification is possible. Moreover, since there is nothing particularly special about the isochrone potential or its
isotropic \DF\@, these conclusions ought to hold for all sensible spherical systems
even with mildly anisotropic \DFs\@\footnote{It is possible that they do
not hold for strongly anisotropic clusters --- see the Discussion.}. However, in
isotropic systems ${ \Ftot \!=\! \Ftot (E) }$,
we note that
${ \bn \!\cdot\! \p F/\p \bJ \!=\! (\bn \!\cdot\! \bO) \, \rd F/\rd E }$.
Since ${ \omega \!=\! \bn \!\cdot\! \bO }$ needs to be very
small for the coupling coefficients ${ \Lambda_{\bn\bnp}^{\ell} (\bJ , \bJp , \omega) }$
to be amplified significantly, an additional small 
factor necessarily enters the flux computation at these frequencies,
further suppressing the effect.

To summarise: (i) In near-isotropic spherical clusters, the only potential
fluctuations that are greatly amplified by collective effects are ${ \ell \!=\!
1 }$ (dipole) fluctuations at very low frequencies. (ii) For ${ \ell \!=\! 1 }$
the only resonance vector that allows meaningful coupling to these very low
frequency fluctuations is ${ \bn \!=\! \pm(-1,1) }$. (iii) The only stars that
can resonantly couple to these low frequency fluctuations are on rather loosely
bound orbits, which are sparsely populated. (iv) Secular evolution occurs only
if two such stars couple to one another, which at these positions in $\bJ$ space
is exceedingly rare.  Thus we conclude that there are simply not enough pairs of
stars able to resonate with one another at sufficiently low frequency for the
collective dressing to be dominant. Put another way, collective effects will
have at most a marginal impact on the largest scales, and will be totally absent on smaller
scales.

\begin{figure*}
    \centering
\includegraphics[width=0.99 \textwidth]{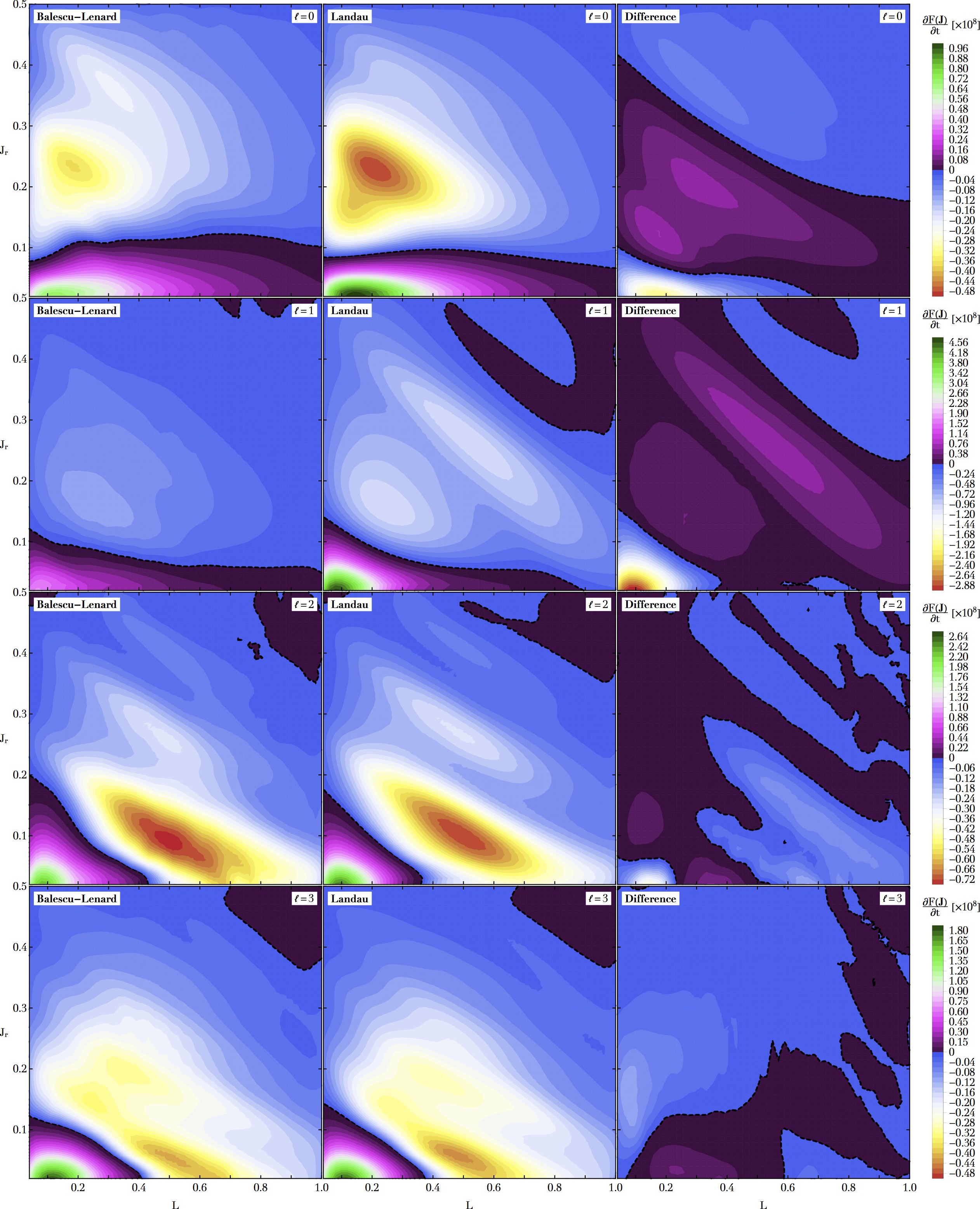}
   \caption{Illustration of the divergence of the \RR\
   diffusion flux, ${ - \p / \p \bJ \!\cdot\! \bFRR^{\ell} (\bJ) \!=\! \p F / \p t}$,
   as defined in Eq.~\eqref{sum_bF},
   in units ${ G \!=\! M \!=\! \bISO \!=\! 1 }$,
   as a function of the considered harmonics $\ell$.
   The left panel corresponds to the dressed \RR\ flux,
   as given by the \BL\ equation, the centre panel
   is its bare analog given by the Landau equation,
   while the right panel corresponds to their differences.
   The number of orbits is predicted to increase
   in the central regions of action space,
   and decrease in the outer ones.
   We refer to \S\ref{sec:LandauEq}
   for the numerical details.
   As expected, as $\ell$ increases,
   collective effects become less and less important.
      We also note that collective effects tend
   to decrease the efficiency of the \RR\ relaxation,
   a paradoxical point already raised in~\protect\cite{Weinberg1989}.
   }
   \label{fig:dFdtBL}
\end{figure*}

In order to test these claims, we compare in Fig.~\ref{fig:dFdtBL} the \RR\
predictions for ${ \p F /\p t }$ for different harmonics $\ell$, in the
presence (i.e.\ \BL\@; left column) and absence (i.e.\ Landau; centre column) of collective effects.
In the right column we plot the difference between these two predictions
(i.e.\ `BL minus Landau'). 
The numerical computations that give rise to this figure are highly non-trivial 
--- for a discussion of
their convergence, see \S\ref{sec:BasisFunctionExpansion}.
We note that the ${ (L , J_{r}) }$ range shown
in Fig.~\ref{fig:dFdtBL} barely includes
any of the resonance lines shown in Fig.~\ref{fig:DF},
However, we emphasise that the action domain
covered by Fig.~\ref{fig:dFdtBL}
still contains about ${ \sim 53\% }$ of the total mass of the cluster.

From the bottom two rows of Fig.~\ref{fig:dFdtBL} we see that for ${ \ell \!=\! 2, 3 }$
the impact of collective effects is already very small. Since self-gravity
operates less efficiently on smaller scales, we can be confident that for
${ \ell \!\geq\! \ellBL \!\equiv\! 4 }$, collective effects may be neglected
completely~\citep{Weinberg1989}.
In other words, for ${\ell \!\geq\! \ellBL }$, we may use the Landau
equation to safely compute the \RR\ prediction for ${ \p F / \p t }$. 
This greatly alleviates the numerical difficulty
of future \RR\ computations
(\S\ref{sec:computing_RR}), as it is far easier to compute the bare coupling
coefficients ${\Lambda_{\bn\bnp}^{\ell} }$ than the dressed ones --- see
\S\ref{sec:Multipole}. 

Moreover, Fig.~\ref{fig:dFdtBL} shows us that even on the smallest scales
${ \ell \!=\! 0,1 }$, the collective amplification is a marginal effect.
Somewhat paradoxically,
collective effects tend to reduce the efficiency of
\RR\@, in particular for ${ \ell \!=\! 0, 1 }$.
Such a conclusion was already reached in~\cite{Weinberg1989}
(see Fig.~{7} therein), which showed that self-gravity
tends to reduce the magnitude
of the ${ \ell \!=\! 1 }$ dynamical friction
in spherical clusters.
Such a trend
was interpreted in~\cite{Weinberg1989}
as being due to the fact that the self-gravitating
wake generated by a perturber
is symmetric and closely in phase with it,
so that this wake cannot generate itself
any significant torque back on the perturber.
Interestingly, we note that for ${ \ell \!=\! 1 }$,
collective effects also lead to the fading
of a diagonal `ridge' that is present in the bare prediction,
whereas, inversely, in self-gravitating discs, collective effects are what 
give rise to striking ridges in the action space
diffusion map~\citep{Fouvry+2015}.

\subsection{Computing the RR flux}
\label{sec:computing_RR}

\begin{figure*}
    \centering
\includegraphics[width=0.99 \textwidth]{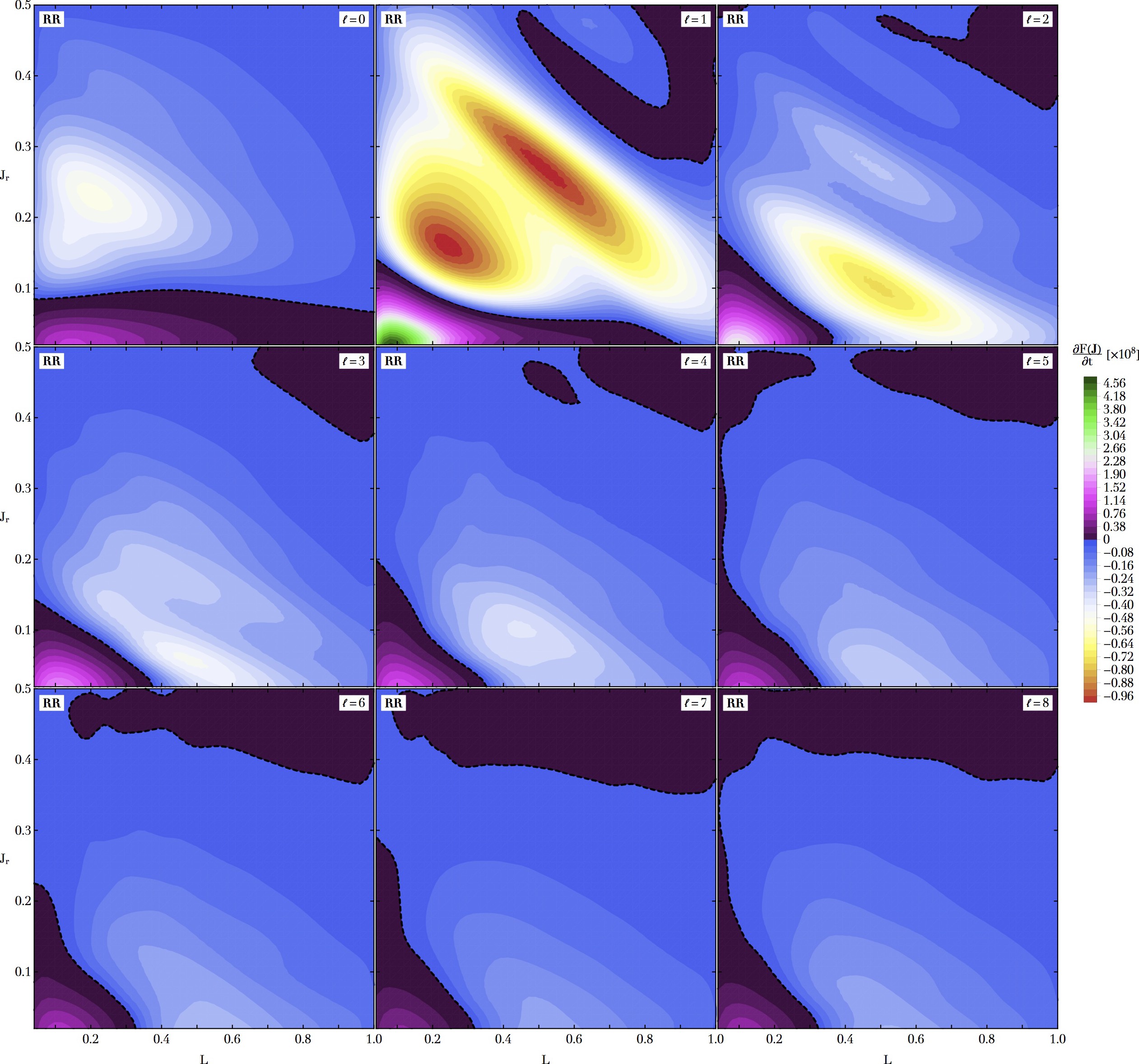}
   \caption{Illustration of the divergence of the bare (Landau) \RR\
   diffusion flux, ${ - \p / \p \bJ \!\cdot\! \bFRR^{\ell} (\bJ) \!=\! \p F / \p t}$,
   as defined in Eq.~\eqref{sum_bF},
   as a function of the considered harmonics $\ell$,
   without collective effects,
   using the same conventions as in Fig.~\ref{fig:dFdtBL}.
   For $\ell$ large enough, the similarities in shape
   with the \NR\ prediction presented in the central panel
   of Fig.~\ref{fig:dFdtNewtonian} is striking.
   }
   \label{fig:dFdtLandauMultipole}
\end{figure*}

In Fig.~\ref{fig:Coulomb},
we determined the critical harmonic number, ${ \ellcrit \!=\! 6 }$,
at which the logarithmic scaling
${ \bFRR^{\ell} \!\propto\! 1 / \ell }$ starts to appear.
In addition, in Eq.~\eqref{choice_lmax}
we determined the maximum harmonic number $\ellmax$
that must be considered in the infinite sum over harmonics.
Of course for the sake of numerical feasibility,
one can only estimate numerically the diffusion fluxes $\bFRR^{\ell}$
for small enough values of $\ell$.
In practice we are able to do this for
${ \ell \!\leq\! \ellcalc \!\equiv\! 11 }$.
Such individual fluxes
are illustrated in Fig.~\ref{fig:dFdtLandauMultipole},
in the absence of any collective effects.

Given these constraints, one may estimate
the associated total \RR\ diffusion flux as follows.
To begin we decompose the flux into low- and high-order
harmonic contributions: 
\begin{align}
\bFRR {} & = \bFRRlowell + \bFRRhighell
\label{calc_FSUM}
\end{align}
where 
\begin{equation}
\bFRRlowell \equiv \sum_{\ell = 0}^{\mathclap{\ellcrit - 1}} \bFRR^{\ell} ;
\quad
\bFRRhighell \equiv  \sum_{\mathclap{\ell = \ellcrit}}^{\ellmax} \bFRR^{\ell} ,
\label{def_bFlowhigh}
\end{equation}
We can now calculate these two contributions separately. For the low $\ell$
(i.e.\ large-scale) contribution, following Fig.~\ref{fig:dFdtBL} we have shown
that for ${ \ell \!\geq\! \ellBL \!=\! 4  }$, collective effects may be safely
neglected. Thus we may approximate
\begin{equation}
\bFRRlowell \simeq \sum_{\ell = 0}^{\mathclap{\ellBL - 1}} \bFRR^{\ell} [\mathrm{BL}] + \sum_{\mathclap{\ell = \ellBL}}^{\mathclap{\ellcrit - 1}} \bFRR^{\ell} [\mathrm{Landau}] ,
\label{calc_Flowell}
\end{equation}
i.e.\ collective effects are only accounted for for the harmonics ${ 0 \!\leq\!
\ell \!<\! \ellBL }$. Meanwhile for the high $\ell$ (i.e.\ smaller-scale)
contribution, on account of the logarithmic scaling
for ${ \ell \!\gtrsim\! \ellcrit }$,
we can approximate
\begin{equation}
   \label{eqn:high_ell_approx}
   \bFRRhighell \simeq \kappa \sum_{\mathclap{\ell = \ellcrit}}^{\ellcalc} \bFRR^{\ell} ,
\end{equation}
where 
\begin{align}
   \label{def_kappa}
\kappa &\equiv\bigg[ \sum_{\ell = \ellcrit}^{\ellmax} 1/ \ell \bigg] / \bigg[ \sum_{\ell = \ellcrit}^{\ellcalc} 1 / \ell \bigg] .
\\ &= 
\begin{cases}
\displaystyle 8.98 & \text{(Newtonian interaction)} ,
\\
\displaystyle 2.08 & \text{(Softened interaction)} .
\end{cases}
\label{val_kappa}
\end{align}
where we took ${ \ellcrit \!=\! 6 }$,
and used the particular values
of $\ellmax$ from Eq.~\eqref{choice_lmax}.
The quantities on the right hand sides
of Eqs.~\eqref{calc_Flowell} and~\eqref{eqn:high_ell_approx} are what we compute numerically. We get the total \RR\ flux by summing them according to Eq.~\eqref{calc_FSUM}.

\subsection{Comparing RR, NR and N-body evolution}

\begin{figure*}
\centering
\includegraphics[width=0.99 \textwidth]{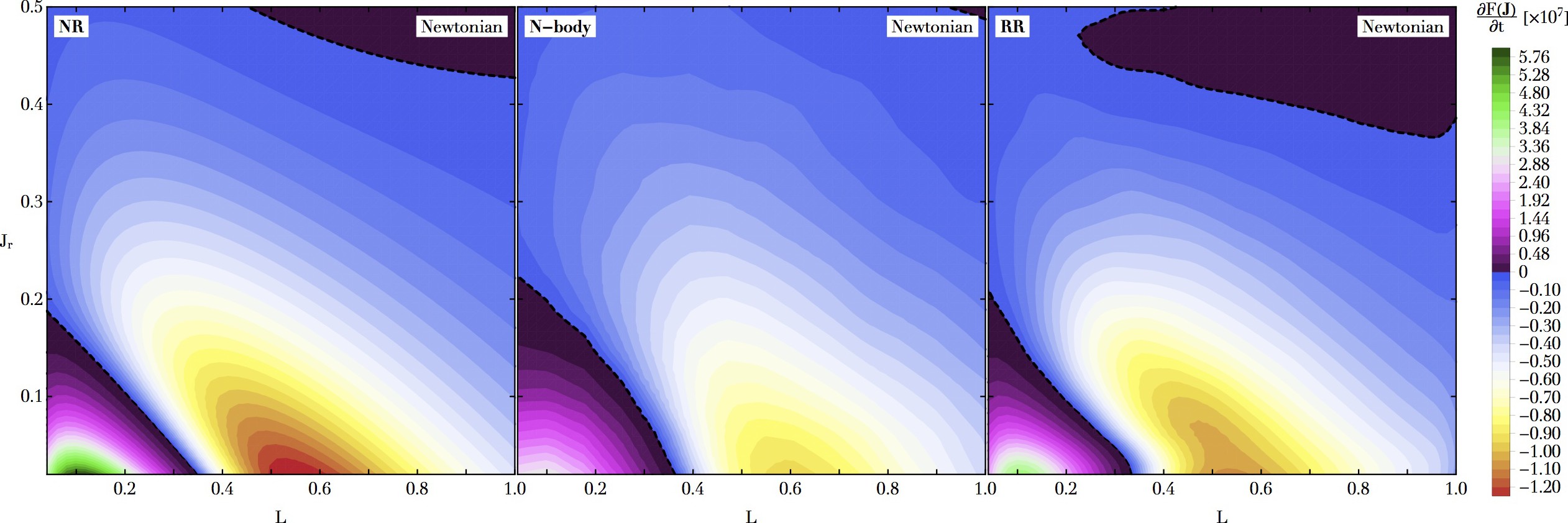}
   \caption{Illustration of the divergence of the
   diffusion flux, ${ - \p / \p \bJ \!\cdot\! \bF (\bJ) \!=\! \p F / \p t}$,
   as predicted by the \NR\ theory (left),
   measured in direct numerical simulations (centre),
   and predicted by the \RR\ theory (right),
   using the same convention as in Fig.~\ref{fig:dFdtBL}.
   The qualitative agreement between the three panels is striking. 
   Note that the overall amplitude of \NR\
   is about twice that of the other panels.
   }
   \label{fig:dFdtNewtonian}
\end{figure*}

We are now in a position
to compare the \NR\ and \RR\ predictions
with direct measurements from $N$-body simulations
(see \S\ref{sec:NumericalSimulations}
for the details of our numerical setup).
Our main result is presented in Fig.~\ref{fig:dFdtNewtonian}, in which we plot
contours of ${ \p F/\p t }$ predicted by the \NR\ (left) and \RR\ theories
(right) to those measured in $N$-body simulations (centre).
While the shape of the contours differs slightly
between each panel, it is reassuring
to note that both \NR\ and \RR\ theories are in satisfactory agreement
with $N$-body measurements.
We note however that the \NR\ theory
over-estimates the efficiency of the relaxation
by a factor ${ \sim 2 }$,
which is reminiscent of an observation made by~\cite{Theuns1996}.
This amplitude mismatch is much reduced
in the \RR\ predictions.
Of course, one should recall
that the harmonic small-scale truncation from Eq.~\eqref{def_lmax},
and the associated self-similar summation from Eq.~\eqref{calc_FSUM}
still remain somewhat heuristic,
and are likely responsible for some of the remaining mismatches
present in Fig.~\ref{fig:dFdtNewtonian}.

The agreement between the $N$-body measurement on the one hand and the \RR\
prediction on the other is a clear vindication
of the \BL\@/Landau kinetic theory.
This is a non-trivial result, because
implementing the \BL\@/Landau formalism in practice
involves significant technical challenges:
as discussed above, it requires summing fluxes over many harmonics and extremely large
numbers of resonances, which in turn requires very good accuracy for each
contribution to the sum.

The relative match of the \NR\ prediction
with the numerical simulations
comes as a pleasant surprise,
since it is far simpler to implement than the \RR\ theory.
Yet, even if taken at face value,
the \NR\ prediction requires some significant tuning
of ${ \ln \Lambda }$ (by a factor $\sim{2}$)
for which there exists no generic, effective
and systematic prescription.
In addition, it does raise some fundamental
questions since the \NR\ theory effectively ignores non-local resonances,
which are properly accounted for by the \RR\ theory.
We further discuss all these elements
in \S\ref{sec:Discussion}.

\subsection{The impact of softening}
\label{sec:soft}

Having investigated in the previous section
the relaxation of a cluster governed
by the Newtonian pairwise interaction,
we now briefly turn our interest
to the case of a softened interaction,
as defined in Eq.~\eqref{Usoft}.
We refer to \S\ref{sec:gyrfalconSIM}
for the details of our numerical setup.

\begin{figure}
   \centering
    \includegraphics[width=0.45 \textwidth]{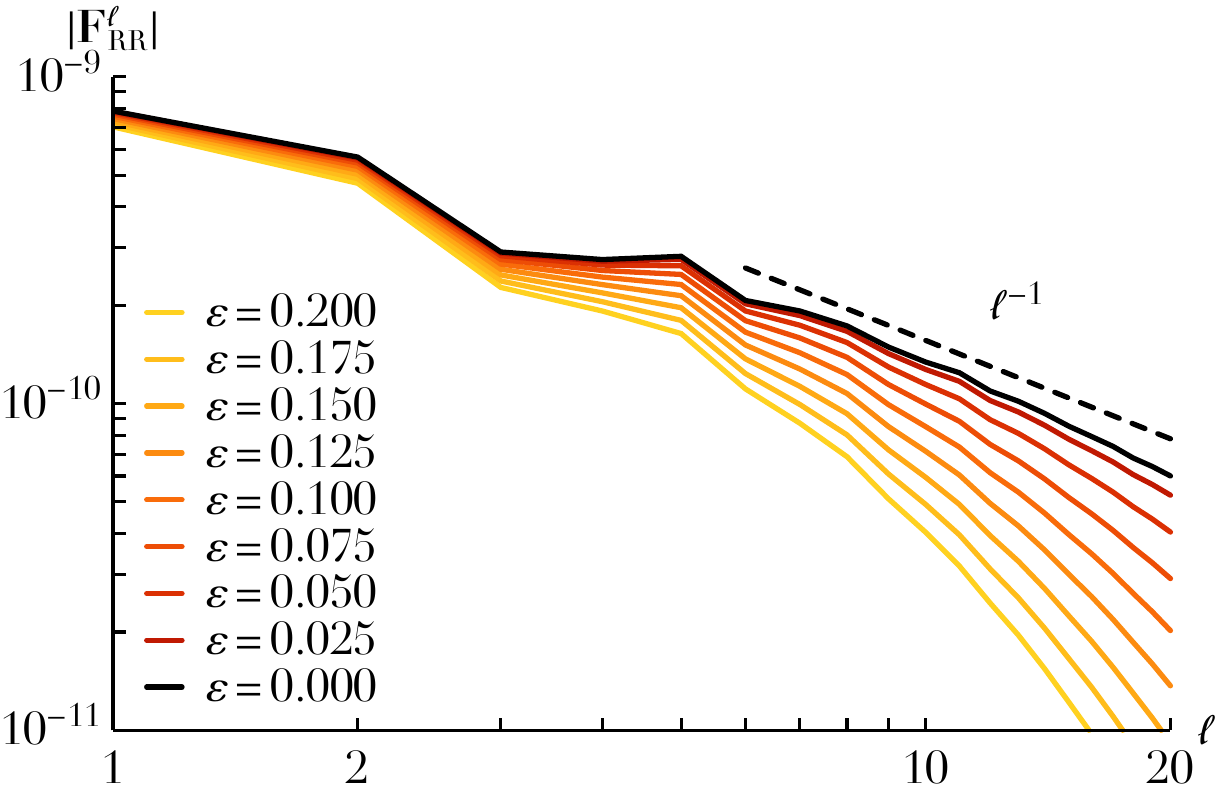}
  \caption{Same as in Fig.~\ref{fig:Coulomb}
  but for a softened pairwise interaction, as in Eq.~\eqref{Usoft}, with various
  softening lengths, $\veps$. Here, we used ${ 0 \!\leq\! \ell \!\leq\! 20 }$,
  and ${ |n_{1}| , |\np_{1}| \leq 40 }$, using ${ K \!=\! 100 }$ and ${ \Kres
  \!=\! 100 }$. The larger the softening length,
  the more rapidly the flux
  contribution decays with $\ell$ for higher-order harmonics,
  as advocated in Eq.~\eqref{asymp_softened}.}
  \label{fig:CoulombSoft}
\end{figure}

In Fig.~\ref{fig:CoulombSoft} we illustrate the impact of softening that
cures the logarithmic divergence of the \RR\ flux for ${ \ell \!\to\! + \infty
}$. (This result was already demonstrated in Fig.~{5} of~\cite{Weinberg1986}).
The larger $\veps$, the stronger the softening,
and therefore the stronger the dampening
of the \RR\ flux for large $\ell$,
i.e.\ the smaller the contributions
from small scales.

\begin{figure*}
    \centering
\includegraphics[width=0.99 \textwidth]{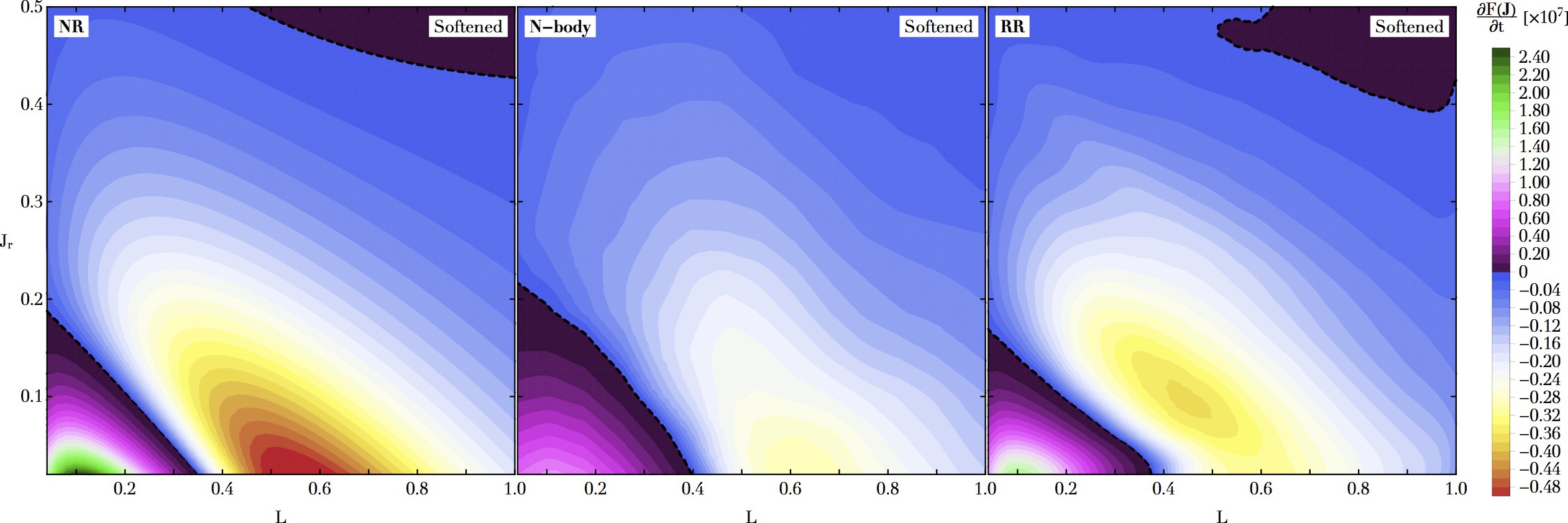}
   \caption{Same as in Fig.~\ref{fig:dFdtNewtonian}
   but for a softened pairwise interaction
   with its respective prescriptions to deal
   with the \NR\ and \RR\ divergences.
   }
   \label{fig:dFdtSoftened}
\end{figure*}

Following the prescriptions
from Eqs.~\eqref{val_lnLambda} and~\eqref{choice_lmax},
we present in Fig.~\ref{fig:dFdtSoftened}
the associated diffusion maps,
as predicted by the \NR\ and \RR\ theories
and measured in numerical (collisionless) simulations.
As expected, we recover that
the introduction of softening,
because it smears out the contributions
from small scales and hard encounters,
leads to a reduction in the overall diffusion rate.
Similarly to Fig.~\ref{fig:dFdtNewtonian},
we note that the \NR\ theory still
over-predicts the rate of relaxation
by a factor ${ \sim\! 2 }$.
The \RR\ kinetic theory presents once again
a welcome satisfactory agreement with numerical simulations.

\section{Discussion}
\label{sec:Discussion}

Let us now discuss our results in steps:
first the connection between \RR\ and \NR\@ theories (\S\ref{sec:RR2NR}), then how one may interpret our results in terms of orbit coupling (\S\ref{sec:orbit-orbit}), 
how our results connect to those in previous works (\S\S\ref{sec:H18}-\ref{sec:literature}),
and finally possible future extensions (\S\ref{sec:Future}).

\subsection{From RR to NR}
\label{sec:RR2NR}

We recall that Fig.~\ref{fig:dFdtLandauMultipole} shows the contribution to the Landau prediction for
${ \p F/\p t }$ from harmonic $\ell$, namely ${ - \p /\p \bJ
\!\cdot\! \bFRR^{\ell} }$, for ${ \ell \!=\! 0,...,8 }$.
One key result of this figure is that for ${ \ell \!\gtrsim\! \ellcrit \!=\! 6 }$,
the map of ${ - \p /\p \bJ \!\cdot\! \bFRR^{\ell} }$ begins to 
resemble the \NR\ prediction ${ - \p /\p \bJ \!\cdot\! \bFNR }$
(see the left panel of Fig.~\ref{fig:dFdtNewtonian}),
up to an overall amplitude.
In other words, the \RR\ theory and the \NR\ theory
give qualitatively equivalent results on small scales.
Heuristically, this may be
understood as follows. 

On the one hand, at large scales ${ \ell \!\lesssim\! \ellcrit }$,
the pairwise coupling between orbits is not a very sharp function of
their respective separation, as highlighted in Eq.~\eqref{shape_U_Leg}. As a
consequence, for such low-order harmonics, long-range resonant couplings are
possible, leading to the non-trivial diffusion maps presented in the first
panels of Fig.~\ref{fig:dFdtLandauMultipole}. We also recall that these maps
maps are distorted by collective effects for the smallest $\ell$, as
in Fig.~\ref{fig:dFdtBL}, which the \NR\ theory
has no hope of accounting for. 

On the other hand, for ${ \ell \!\gtrsim\! \ellcrit }$, the pairwise coupling
becomes a sharp function of the stars' separations. As a consequence, for such
high-order harmonics, relaxation is made possible only through local scatterings,
i.e.\ the form of relaxation captured by the \NR\ theory from Eq.~\eqref{def_bF_NR}. As
highlighted in the last panels of Fig.~\ref{fig:dFdtLandauMultipole}, this
allows for the maps of $\bFRR^{\ell}$ to greatly resemble the ones from $\bFNR$
(see Fig.~\ref{fig:dFdtLandauMultipole}), up to an overall change in the
amplitude, that follows the logarithmic scaling recovered in
Fig.~\ref{fig:Coulomb}.

As such, one of the key improvements
of the \RR\ theory over the \NR\ one is to offer
a better estimation of the diffusion flux
for low-order harmonics
(i.e.\ the contributions from large scales).
In addition, this inhomogeneous \RR\ prediction
also naturally cures the large-scale divergence
present in the \NR\ theory.
While this does not significantly affect the overall structure
of the maps of ${ \p F / \p t }$,
it does improve the estimation of the overall amplitude
of the diffusion flux,
as highlighted in Fig.~\ref{fig:dFdtNewtonian}.

Benefiting from this self-similarity between the \NR\ theory and the large $\ell$ contribution to the \RR\ theory, 
we may improve upon Eq.~\eqref{calc_FSUM}
and propose a simpler effective approach,
combining both \RR\ and \NR,
to estimate the total diffusion flux.
As such, we write
\begin{equation}
\bF = \bFRR^{< \ellcut} + \bFNR^{\bcut} .
\label{effective_sum}
\end{equation}
In that expression, the contribution from low-order harmonics,
${ \bFRR^{< \ellcut} }$, is computed as
\begin{equation}
\bFRR^{< \ellcut} = \sum_{\ell = 0}^{\mathclap{\ellcut - 1}} \bFRR^{\ell} ,
\label{def_FRRcut}
\end{equation}
where, following Eq.~\eqref{calc_Flowell},
collective effects are also accounted for
in low-order harmonics.
In Eq.~\eqref{effective_sum}, we also introduced
$\bFNR^{\bcut}$ as the \NR\ flux
computed with a Coulomb logarithm given by
${ \ln \Lambdacut \!=\! \ln (\bcut / \bmin) }$,
where the minimum impact parameter, $\bmin$,
is given by Eq.~\eqref{choice_bmin},
while the maximum impact parameter, $\bcut$,
is a function of $\ellcut$,
and follows from Eq.~\eqref{def_lmax} reading
\begin{equation}
\bcut = \frac{\ln (2)}{\ellcut} \, \bISO .
\label{def_bcut}
\end{equation}

\begin{figure}
   \centering
    \includegraphics[width=0.385 \textwidth]{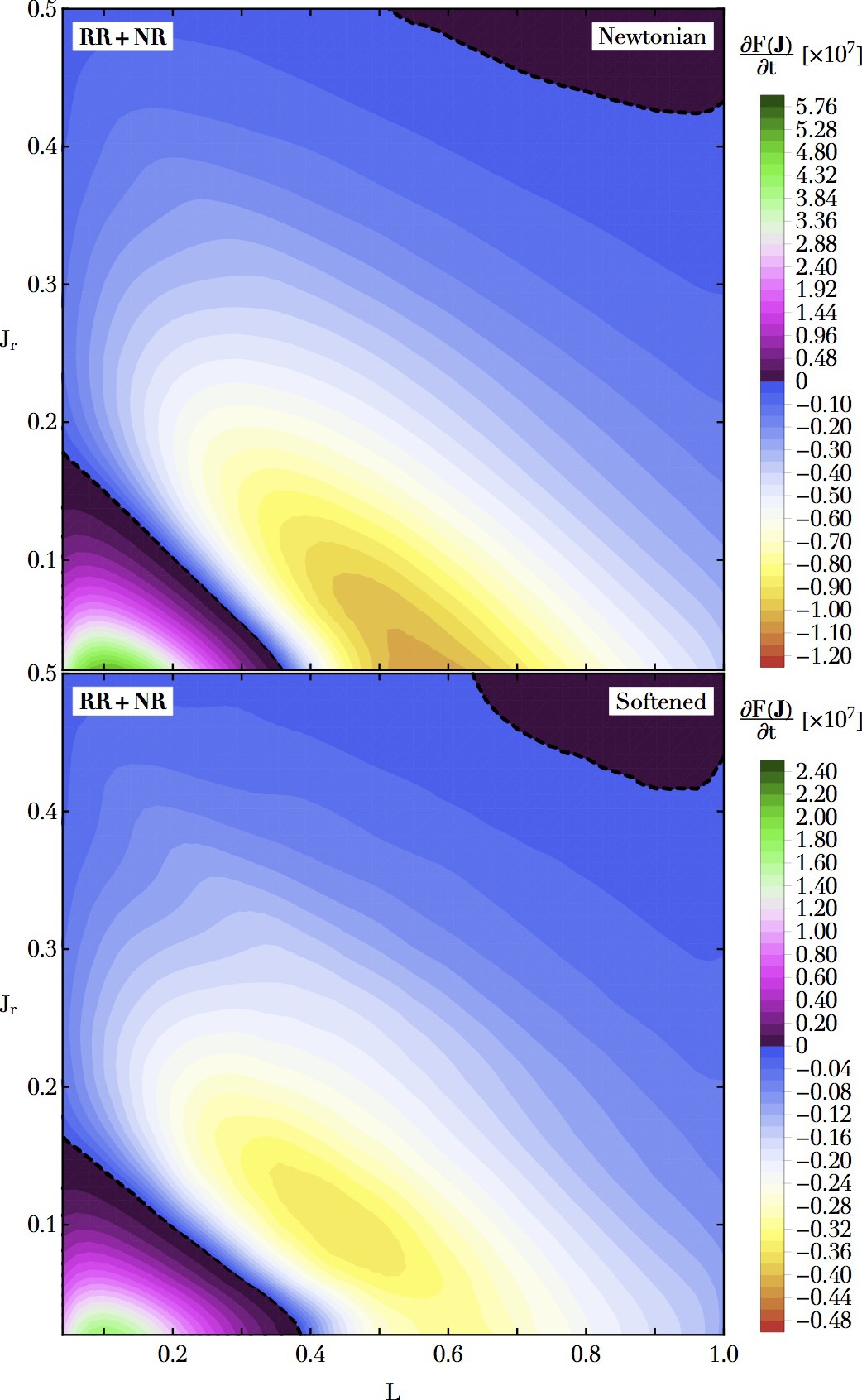}
  \caption{Same as Figs.~\ref{fig:dFdtNewtonian} and~\ref{fig:dFdtSoftened}
  but using the effective diffusion flux from Eq.~\eqref{effective_sum}
  that combines the \RR\ and \NR\ fluxes,
  for a Newtonian interaction (top) and a softened interaction (bottom),
  using a truncation at ${ \ellcut \!=\! 12 }$.
  Accounting for the large-scale contributions to the flux using \RR\
  allows for an improved agreement with the numerical simulations,
  compared to the direct \NR\ prediction.
  }
  \label{fig:dFdtEffective}
\end{figure}

Such an effective calculation is presented in Fig.~\ref{fig:dFdtEffective},
for both a Newtonian and softened interaction potential.
In that figure, we note, on the one hand, that accounting
for the large-scale contributions to the flux using \RR\ rather than \NR\
partially alleviates the amplitude mismatch that was observed
in Figs.~\ref{fig:dFdtNewtonian} and~\ref{fig:dFdtSoftened}
when comparing the \NR\ prediction
with the numerical simulations.
On the other hand, accounting for the small-scale contributions
to the flux using \NR\ rather than \RR\
significantly alleviates the computational difficulty of the prediction,
as one does not need to solve any non-local resonance condition,
nor sum over numerous high-order resonance pairs.
All in all, the effective approach from Eq.~\eqref{effective_sum}
appears as a promising way to effectively and simultaneously account
for the joint effects of large-scale, resonant, dressed, and non-local contributions
(as captured by \RR\@),
and small-scale, non-resonant, bare, and local contributions
(as captured by \NR\@).

\subsection{Qualitative interpretation via orbit-orbit torques}
\label{sec:orbit-orbit}

Let us now attempt to explain the physical origin of the observed agreement between the
\RR\ and \NR\ maps. The two basic questions we are trying to answer are as
follows.
(A) Why is it that the distinct conceptual pictures of (i) long-lived
resonant encounters of stars with small scale
potential fluctuations, and (ii) instantaneous local non-resonant two-body
encounters between individual stars, ultimately end up being equivalent here?  
(B) Why do interactions of stars with small-scale
(high $\ell$) potential fluctuations dominate the \RR\ flux, rather than interactions with
large-scale (low $\ell$) collectively dressed fluctuations?

To begin to answer these questions, let us start with (i) and  
argue why it is the same as (ii), at least for the system at hand.
Let us also
 stress that in contrast to the result of the previous section, what follows in
 this sub section is only  offered as a broad heuristic explanation that
 warrants further work.
 We shall ignore collective effects, since these are
 unimportant on small scales (see Fig.~\ref{fig:dFdtBL}). Then the mathematical
 formalism behind (i) is Landau theory, i.e.\ with the same flux as in
 Eq.~\eqref{def_mF} but with the dressed coupling coefficients replaced by bare
 coefficients. We begin to make a connection with star-star scattering when we
 realise that the bare coupling coefficients entering the Landau flux are merely
 Fourier transforms of the interaction potential between pairs of stars w.r.t.\
 both sets of angles~\citep[see, e.g.\@,][]{pichon94,Chavanis2013}. Concomitant
 with this, Rostoker's principle \citep{Gilbert1968,Hamilton2021} tells us that
 Landau theory is nothing more than a theory of bare two-body resonant
 interactions between stars on mean field orbits. Furthermore, we know that any
 star's mean field orbit can be labelled by its ${3D}$ action ${ \obJ \!=\!
 (J_r,L,L_z) }$ and its ${3D}$ angle variable at some reference time, $\obT_0$,
 and then written as a Fourier series: ${ \br(t) \!=\!
 \!\sum_{\obn}\!\br_{\obn}(\overline{\bJ}) \exp(\ri \obn \!\cdot\! \obT)}$, with
 ${ \obT \!=\! \obT_0 \!+\! \obO(\overline{\bJ})t }$. Taking this Fourier series
 literally, we could equivalently think of replacing each star on mean field
 orbit ${ (\obT_0, \overline{\bJ}) }$ by a superposition of many (less massive)
 quasi-stars each labelled by $(\obT_0, \obJ, \obn)$, and having orbits ${
 \br(t) \!=\! \br_{\obn}(\overline{\bJ}) \exp(\ri \obn \!\cdot\! \obT) }$. From
 this viewpoint the interaction between any two stars ${ (\obT_0,
 \overline{\bJ}) }$ and ${ (\obT_{0}^{\prime}, \overline{\bJ}^{\prime}) }$ can
 be thought of as a superposition of interactions between all possible pairs of
 quasi-stars labelled by $\obn$, $\obnp$. Thus, Landau theory is a theory of
 bare interactions between all possible resonant quasi-stars. (Note that since
 we are now considering angles and actions in ${3D}$, we have not yet thrown
 away any information about the relative inclination of the orbital planes of
 these quasi-stars).
 
\begin{figure}
   \centering
    \includegraphics[clip, width=0.485 \textwidth]{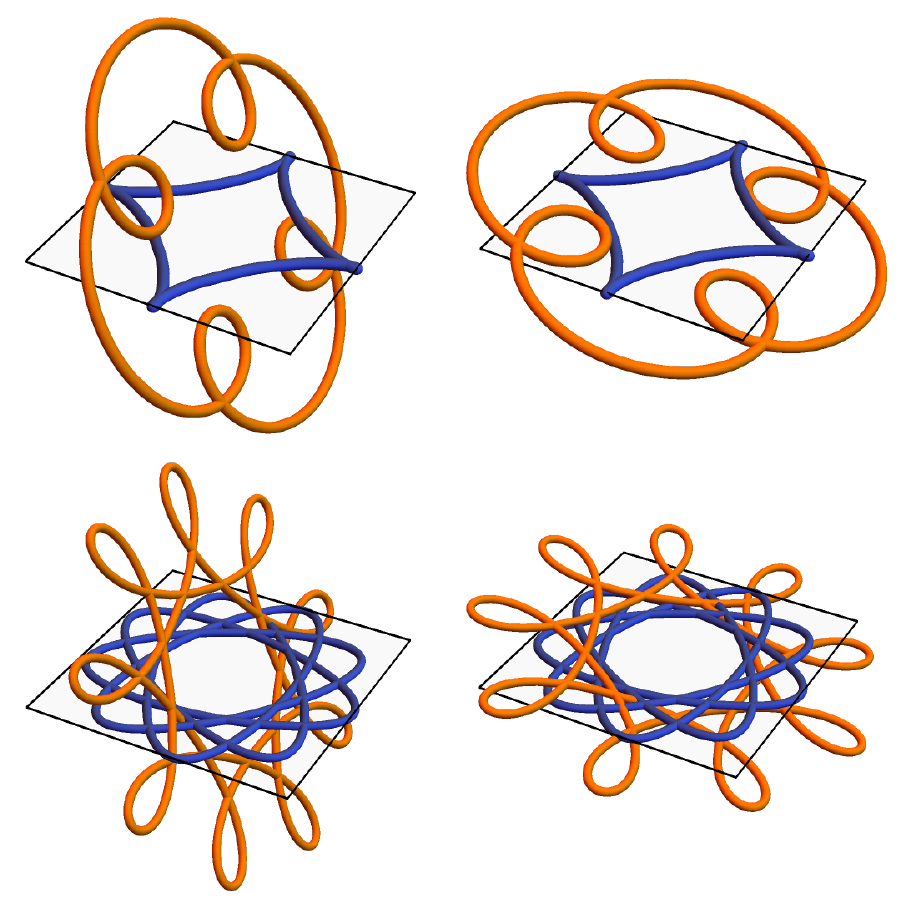}
  \caption{Illustration of the four basic types of interaction between
  quasi-star orbits with ${ (J_r,L)/\sqrt{G M \bISO} }$
  equal to resp.\ ${ (0.04,0.2) }$ (blue) and ${ (0.11,0.4) }$ (orange).
  Roughly speaking we have
  (top left) out-of-plane interactions via a low-order resonance;
  (top right) close-to-in-plane interactions via a low-order resonance;
  (bottom left) out-of-plane interactions via a high-order resonance;
  (bottom right) close-to-in-plane interactions via a high-order resonance.
  }
  \label{fig:shapeoforbits}
\end{figure}

What sort of resonant interactions can pairs of quasi-stars have?   
To start with, given the corresponding ${2D}$ actions and resonant numbers in each corresponding plane $\bJ$, $\bJp$, $\bn$ and $\bnp$ we can always find a
rotating reference frame in which both quasi-star orbits are closed. As viewed
in this reference frame there are then roughly four qualitative types of
resonant interaction, stemming from the fact that the orbital orientations can
be either `in-plane' or `out-of-plane', and that the resonances can
be either `high-order' or `low-order'. To illustrate what we mean,
in the top two panels of Fig.~\ref{fig:shapeoforbits},
we show two
typical quasi-star orbits in the isochrone potential, with ${ \bn \!=\! (-1,4)}$
in blue and ${ \bnp \!=\! (1,4) }$ and in orange.
Note that we have transformed to
the aforementioned rotating frame, so that both orbits are closed. On the top
left we take the orbital planes to be inclined w.r.t.\ each other by an angle
${\pi/3}$, so that they are significantly `out-of-plane'. On the top right we
have simply changed the relative angle to ${\pi/16 }$ so that the two orbits are
instead almost coplanar. Finally, the bottom panels show
something similar except for a `higher-order' resonance, namely taking ${ \bn
\!=\! (-4,9)}$ (blue) and ${\bnp \!=\! (-2,9) }$ (orange). 

Obviously, the `order' of the resonance here corresponds roughly to the number
of `loops' in these diagrams.
Thus, low-order resonances (top row) correspond to
relatively small $\bn$, $\bnp$ while high-order resonances correspond to
relatively large ${ \bn, \bnp }$ (bottom row).  Moreover, as explained further in
\S\ref{sec:3Dto2D}, the {minimum} relative inclination angle between quasi-star orbits is
roughly set by ${ \pi/\ell }$, so that large $\ell$
can capture
  near coplanar
configurations (right)
while small $\ell$
are limited to
 to out-of-plane
interactions (left).

We can now plausibly answer question (A).
Large $\ell$ means a typical small
angular-scale ${\pi/\ell }$ between the two planes.  Also, while there are very
many high-order $\bn$, $\bnp$ to choose from like in the bottom right panel of
Fig.~\ref{fig:shapeoforbits}, there are not so many low-order ones like in the
top right. Thus we expect the large-$\ell$ (small scale) contribution to the \RR\
flux to mostly consist of interactions like those in the bottom right panel.
These mimic local deflections: indeed, when the overlap between the two
quasi-star orbits is strongest, the local trajectories follow paths as
though they (locally) deflect one another. At this stage we already begin to
suspect a similarity between (i) and (ii). Taking it a step further, the
orbit-averaged \NR\ theory assumes that the sequence of deflections can be
considered as uncorrelated. The \RR\ theory considers  ${ |n_{2}| \!\leq\! \ell }$ sequential deflections
per azimuthal period in the rotating frame (see~\S\ref{sec:Lambda}), and must be summed over many such $\ell$s
(bottom-right panel of Fig.~\ref{fig:shapeoforbits}). Since the type of
deflection decorrelates from one pair ${ (\bn,\bnp) }$  of orbital configurations  to the next, it seems likely that the sum over such pairs  in Eq.~\eqref{sum_bF} induces the same level of decoherence as the a posteriori orbital average
implemented  in Eq.~\eqref{D1_D2}  for \NR\ theory.
Thus we end up with the heuristic expectation
that the secular evolution predicted 
by large-$\ell$ \RR\ theory behaves
qualitatively the same as that predicted by \NR\ theory.

To answer question (B), we ask more generally: which of the four configurations
 do we expect to dominate the secular evolution?
 We notice that there are simply
 not so many low values of $\ell$ to choose from, whereas the number of high
 $\ell$ is in principle infinite.
 Also, as we have seen, the value of $\ell$
 puts a limit on the order of the resonance that may contribute to the \RR\ flux, ${ \vert n_2 \vert , \vert \np_2 \vert \!\leq\! \ell }$, so that configurations like
 the bottom-left panel of Fig.~\ref{fig:shapeoforbits} are rare. In
 addition to this, the lack of orbital overlap for low $\ell$
 means that at a
 typical time in their mutual orbits the quasi-stars will typically be
 physically far apart, leading to rather weak interactions.
 Conversely, the
 large $\ell$ in-plane interactions (right panels in Fig.~\ref{fig:shapeoforbits}) allow for multiple localised
 overlaps, particularly from the numerous available high-order resonances
 (bottom right). This hand-waving argument suggests that the dominant
 contributions to the flux could come from many configurations
 like the bottom right
 panel, since they are strongest and most numerous.
 
When attempting to draw a closer connection to \S\ref{sec:Application},  recall that when $\ell$ is
 small, the interaction kernel ${ U_{\ell} (r , \rp) }$ is wide (see
 Eq.~\eqref{def_Ul} and Appendix~\ref{sec:BasisMethod}), so that orbits can
 still couple at fairly different $r$ and $\rp$,  whereas high-$\ell$ display a
 very narrow  kernel. This is the basis-sampling\footnote{Because gravity is a
 long-range interaction, one has in this context to expand configuration space
 over sets of non-local basis elements, and sample the quasi-stars on those
 elements.} counter-part to the   geometric argument of `capturing
 in-and-out-of-plane orbit overlap'.  If $\ell$ is large, a narrow kernel is
 sufficient to sample small local loops which can get close to each other
 in-plane; if $\ell$ is small the wider kernel accounts for the  less
 locally-averaged impact of the  other orbit: such terms will contribute  to the
 sum, but, it turns out, less so.
 In fact, \S\ref{sec:Application} showed that while contributions
 from ${ \ell_{2} \!\gg\! \ell_{1} }$ are much smaller than that from
 $\ell_{1}$, i.e.\ ${ \bF_{\ell_{2}} \!\ll\! \bF_{\ell_{1}} }$, each
 contribution from a decade ${ \sum_{\ell = 0.1 \ell_{2}}^{10 \ell_{2}} \!\!
 \bF_{\ell} }$ is similar to any other decade ${ \sum_{\ell = 0.1 \ell_{1}}^{10
 \ell_{1}} \!\! \bF_{\ell} }$, in direct analogy with the \NR\ case.
 It seems
 that the $\ell$-convergence of  the kernel's width
 in ${ r/\rp }$ compensates the
 number of deflections set by the relevant patch of size ${ \pi/\ell }$, so that
 each decade of $\ell$ contributes roughly the same amount to the flux
 (cf Fig.~\ref{fig:Coulomb}). In all likelihood, ${ U_\ell (r, \rp) }$ inherited  this
 feature from the inverse-square law of the interaction, hence the same
 asymptotic ${ 1/\ell }$ contribution to the flux.  Clearly this rough argument
 warrants further detailed exploration.

At this point one might also interject and argue that the interactions between
quasi-stars ought not to be treated as bare Newtonian interactions at all, but
rather as dressed by collective effects, i.e.\ mediated via the dressed coupling
coefficients in Eq.~\eqref{def_mF}.  These collective effects will boost
the contribution from the top-left-panel configurations (low $\ell$) for certain
special pairs of small ${ \bn, \bnp }$, and this will have some impact on the low
$\ell$ flux.  While this is true, the argument of \S\ref{sec:RoleCollective
effects} suggests that in hot isotropic spheres it is only a modest effect.  A
posteriori, our calculations of ${ \p F / \p t }$ tell us
that the low $\ell$ flux
is insufficient to overwhelm the many contributions from configurations that
look like the bottom-right panel\footnote{For a disc, the
sum over $\ell$ disappears, while the strength of the wakes increases
significantly, hence the impact of low order resonances  can be more
significant. In contrast, for the sphere there is a $L$ volume factor in
Eq.~\eqref{def_DFred} reflecting the clusters' spherical symmetry.}.

Finally, we note that in
this picture, the effect of softening as investigated in~\S\ref{sec:soft} acts
as a minimum plane separation, effectively damping the divergent $1/\ell$ sum
corresponding to the Coulomb logarithm, as observed in
Fig.~\ref{fig:CoulombSoft}.
More quantitatively,  softening tends
to make the interaction kernel ${ U_{\ell} (r , \rp) }$ flatter,
so that when modulated by high-order resonances ${ (\bn , \bnp) }$
in $W_{\ell}^{\bn\bnp}(\bJ , \bJp)$ in Eq.~\eqref{def_WBare}, it gives a vanishing contribution.

Overall, a geometric understanding of the quasi-star orbit-orbit coupling
framework highlights  a posteriori why the \RR\ of spherical hot clusters can
indeed quantitatively match the predictions of \NR\ theory. A more quantitative
analysis would require investigating the relative dressed populations of quasi-stars at
each $\bn$, $\bnp$, their relative `tumbling' rates in the rotating frame, and
so on. This is left for future work.

\subsection{Relation to \citet{Hamilton+2018}}
\label{sec:H18}

\cite{Hamilton+2018} applied the \RR\ theory in the context of
spherical stellar systems (i.e.\ globular clusters). The main aim of their paper
was to evaluate the \BL\ prediction for ${ \p F/\p t }$ for the spherical
isochrone potential, and then compare this to the Landau prediction (to evaluate
the importance of collective effects for \RR\ processes)
and the Chandrasekhar prediction (to compare \RR\ with the canonical \NR\ theory).   
They performed this calculation for the same isotropic \DF\ that we employed here
(Eq.~\eqref{Diff_iso_E}), as well as for some anisotropic (but still
stable) isochrone \DFs\@. From the resulting maps of ${ \p F/\p t }$,
\cite{Hamilton+2018} claimed the following: (i) collective amplification,
particularly of ${ \ell \!=\! 1 }$ fluctuations, is strong in spherical clusters
so that the \BL\ prediction for the relaxation rate is much larger
than that of Landau;
and (ii) the typical \BL\ relaxation rate
is comparable in magnitude to, or even
greater than, the \NR\ prediction of Chandrasekhar,
while taking a very different form in $\bJ$-space.
Based on these results, \citet{Hamilton+2018}
concluded that large-scale self-gravitating collective motions provide a
crucial, and heretofore overlooked, contribution to the secular relaxation of
globular clusters. 

We would like to emphasise here that the formal results developed in the first
four sections of \citet{Hamilton+2018} are correct, and indeed we have based
much of the present paper on those foundations. However, the remainder of
\citet{Hamilton+2018}'s results should be revised. (i) In
\S\ref{sec:RoleCollective effects} we gave an analytical argument for why
collective amplification cannot be a dominant effect for the great majority of
stars in a cored spherical cluster, at least without a strong velocity anisotropy. We
justified this claim numerically in Fig.~\ref{fig:dFdtBL}. In other words, at
least in the great bulk of phase space there is no significant difference
between the \BL\ and Landau predictions of ${ \p F / \p t }$. (ii) The \RR\ and
\NR\ predictions are actually remarkably similar, once the contributions from
high-order harmonics and resonances are correctly accounted for.

Why do the conclusions of \citet{Hamilton+2018} differ so much from ours? The
simple answer is that their numerical computations of ${ \p F/\p t }$ were not
converged.  As in Fig.~\ref{fig:ROI} here, \citet{Hamilton+2018} verified their
computation of the response matrix ${ M_{pq}(\omega) }$ by recovering from it
the ${ \ell \!=\! 2 }$ radial orbit instability of~\cite{Saha1991}. Their
confidence in the accuracy of their code was strengthened by carrying out a
detailed convergence study around this instability, showing its recovery did not
depend on the code parameters beyond some threshold (see their Table 2).
However, when they then came to computing ${ \p F/\p t }$
for a stable ${ F (\bJ) }$ they
eased the heavy computational burden by making three parameter truncations, none
of which was truly justified. First, they considered only the largest scale
fluctuations, namely those with ${\ell \!=\! 0, 1, 2 }$ --- but as we
illustrated in Fig.~\ref{fig:Coulomb}, ${\ell_{\max} \!=\! 2 }$ is definitely
not sufficient to find any trace of the underlying small-scale logarithmic
divergence inherent in the \RR\ flux. Second, they truncated the sum over
resonance vectors in Eq.~\eqref{sum_bF} to ${|n_{1}|, |\np_{1}| \!\leq\! 2}$. As
illustrated in Fig.~\ref{fig:Coulombn1max}, this is not satisfactory, as the
contributions from high-order resonances are heavily under-estimated. Third,
they computed the coupling coefficients ${\Lambda_{\bn\bnp}^{\ell}}$ using a
finite basis expansion with ${ \ncut \!=\! 10 }$ radial basis elements. While
such a drastic truncation is sufficient to recover the large-scale ${ \ell \!=\!
2}$ mode (Fig.~\ref{fig:ROI}), it is not enough to resolve correctly the rest of
the $\Lambda_{\bn\bnp}^{\ell}$, as illustrated in
Fig.~\ref{fig:ErrorLambda}\footnote{There was also a small error in the code
used by~\cite{Hamilton+2018} to compute the \BL\ flux for isotropic \DFs\@,
${\Ftot \!=\! \Ftot (E) }$. The error was simply that the term
${\sin^{-1}(\sqrt{-E}) }$ in Eq.~\eqref{isoDF_iso} was coded as ${ \sin^{-1} (-E)}$.
This error biases the \DF\ by adding an extra population of weakly bound orbits,
which in turn leads to an erroneous boost in the self-gravitating amplification
of ${ \ell \!=\! 1 }$ fluctuations in the loosely-bound parts of phase space
(see the dashed contours in Fig.~\ref{fig:DF}). This error accounts for the
large dark blue triangle feature in the right panel of~\cite{Hamilton+2018}'s
Fig.~{12}. This error did not affect any non-isotropic calculations or the
recovery of the ${ \ell \!=\! 2 }$ instability.}.

\subsection{Relations to other  works}
\label{sec:literature}

In recent decades, many authors have compared the predictions of Chandrasekhar's
theory of two-body relaxation to direct $N$-body experiments of spherical
stellar clusters \citep[e.g.][]{HeggieHut2003,Vasiliev2015,Sellwood2015}. Often, but not
always, these studies have been geared towards describing the evolution of the
system using an orbit-averaged \FP\ equation in energy space (e.g.\@,
\citealt{Vasiliev2015} and references therein) or in energy-angular momentum
space (e.g.\@, \citealt{Takahashi1995, Drukier1999}).  There, the role of
Chandrasekhar's theory is to provide an estimate for the \NR\ diffusion
coefficients that enter the \FP\ equation.
The $N$-body experiments are
needed (i) to check that the orbit-averaged \FP\ theory is capable of
describing the evolution at least qualitatively, and (ii) to make the
description quantitative by calibrating the free parameter ${ \ln \Lambda }$. 
These
studies have found repeatedly that once ${ \ln \Lambda }$ is fixed,
\NR\ theory does a remarkably good job
of estimating the diffusion coefficients~\citep{Theuns1996}
and therefore of determining secular evolution~\citep{Kim2008,eSilva2017},
a finding corroborated by our Fig.~\ref{fig:dFdtNewtonian}. Moreover, the `free parameter' ${ \ln \Lambda }$ is
often well-constrained, such that any two reasonable values of it can
only produce \NR\ relaxation rates that differ
from one another in magnitude at the
level of ${ \!\sim\! 10\% }$, and often less~\citep{Giersz1994,eSilva2017}. 

Only occasionally has the agreement between \NR\ theory and $N$-body
simulation been called into question, except for systems that are rotating or
that have strongly anisotropic \DFs\
(see \S\ref{sec:Future}).
\cite{Theuns1996}
measured diffusion coefficients in energy space using $N$-body simulations
of King models, and compared these to the associated \NR\ theoretical prediction. He
found that while the agreement between the experimental diffusion coefficients
and their \NR\ theoretical counterparts is good, in King
models that have low central concentration, the \NR\ theory overestimates the
diffusion rate by a factor ${ \simeq 1.5\, \mbox{---} \, 2 }$. Noting that the
isochrone model has low central concentration also, this finding is in agreement
with what we found in Fig.~\ref{fig:dFdtNewtonian} in a different setting.
\citet{Theuns1996} attributed this additional relaxation to collective effects
(which should be accounted for by the \BL\ formalism), or to the scattering of
stars by global collective modes (which \BL\ does not cover --- see
\citealt{Hamilton+2020}).

While these studies may be enough to convince one that \NR\ theory is
a sufficiently good workhorse for most practical purposes, they do not really probe
in detail the underlying physics of relaxation.  That is because they confine
themselves to following the evolution of the cluster either in real space
(looking at, e.g.\@, the time-evolution of Lagrangian radii) or in the space of
energies $E$. Important though these quantities are, they are imperfect for
probing relaxation physics because changes in those quantities reflect adiabatic
changes in the mean field. A key novel feature of the present study is that we
calculated the secular evolution in the ${ (J_{r} , L) }$ action space, allowing us to
separate true relaxation from the adiabatic evolution of the mean field
potential ${ \psi (r) }$ --- whilst $E$ changes under slow changes in $\psi$, the
actions ${ (J_{r} , L ) }$ do not.  

Other than the present paper, to our knowledge, the only study in which the
action space evolution has been computed for a spherical stellar system from
$N$-body experiments is \cite{Lau+2019}.
They performed $10^{4}$ brief $N$-body
simulations of the isotropic isochrone cluster with ${ N \!=\! 10^{3} }$, and
stacked their results to build up good statistics. They drew their initial
conditions from a Poisson sampling of the underlying distribution ${ \Ftot (E)
}$. They showed very clearly that in the early stages of evolution, ${ \ell
\!=\! 1 }$ potential fluctuations are strongly amplified compared to the initial
bare Poisson noise (c.f. our Fig.~\ref{fig:NMat}). This amplification had not
yet saturated after ${\sim 3 \, \tcross }$ (where $\tcross$ is a typical
crossing time) which is when their simulations ended. 

\citet{Lau+2019} then compared their results to the \RR\ and \NR\ predictions
from~\cite{Hamilton+2018}. However, as they acknowledged, the fact that the
amplifying noise had not yet saturated meant that their simulations could not be
considered a fair test of the \BL\ theory, which assumes saturated noise from
the outset. Our simulations do not suffer from this shortcoming because the
larger $N$ value means that the dressing process has sufficient time to saturate
before the system relaxes significantly. Moreover, $\bF$ is a difficult quantity
to measure in simulations. Indeed, it seems that \cite{Lau+2019} may actually
have been measuring only the frictional contribution to $\bF$, i.e.\ the part
arising from the coefficient $\bD_1$, rather than the full $\bF$ (D.\ Heggie,
private communication). Indeed, the flux measurement reported by~\cite{Lau+2019}
matches qualitatively the \NR\ prediction for the frictional part of $\bF$, as
can be seen by comparing the middle panel of Fig.~{11} of \citet{Lau+2019} with
the upper panel of Fig.~{B1} of \citet{Hamilton+2018}.

In the present paper we chose to compute ${ \p F /\p t }$
rather than the flux $\bF$, because (i) one can measure it from $N$-body simulations in
an unambiguous way and (ii) its theoretical value is insensitive to the addition
of a $\bJ$-independent constant to the flux.  In so doing we arrived at the
conclusion that up to an overall scale factor the \NR\ prediction for
${ \p F / \p t }$ is remarkably similar to that measured in $N$-body
experiments.
While \RR\ differs from \NR\ at the largest scales, and while
collective amplification may play some minor role in the \RR\ prediction,
like most classical studies we have concluded that relaxation
does not differ fundamentally from the predictions of Chandrasekhar's
\NR\ theory, at least in an isotropic cored globular cluster.

\subsection{Future extensions}
\label{sec:Future}

Of course, the present work is only a first step
towards a complete description of the collective,
long-range and resonant relaxation of globular clusters.
Let us now list briefly several avenues that deserve further investigation.

First, for the sake of simplicity, we limited ourselves
to only considering isotropic non-rotating clusters,
i.e.\ clusters whose \DF\ follows ${ \Ftot \!=\! \Ftot (E) }$.
As recently highlighted in~\cite{Breen+2017},
clusters with (strong) tangential anisotropy
can undergo a much more efficient relaxation.
Accounting for anisotropic \DFs\@,
i.e.\ ${ \Ftot \!=\! \Ftot (E , L) }$,
would involve two main developments:
(i) in the \NR\ theory, as in Eq.~\eqref{local_DCs},
 a computation of Rosenbluth potentials 
involving ${3D}$ integrals is necessary;
(ii) in the \RR\ theory, e.g.\@, as highlighted in~\cite{Rozier+2019},
clusters can support an ever stronger self-gravitating amplification,
which may (or may not) lead to an efficient collective dressing
of the low-order harmonics.
All in all,
understanding the secular relaxation of rotating spheres would
be of genuine astrophysical interest:
the set of possible  resonances gets shifted  by rotation, 
and stars can extract free energy from the mean rotation  of the sphere.
This may impact 
the importance of collective effects, especially at low $\ell$.

Second, the present investigation was limited to the case
of an isochrone potential. It was picked for the convenience
of offering an explicit angular mapping,
as in Eq.~\eqref{nice_edge_iso},
making the orbital averages numerically much more sound.
Provided such explicit and well-behaved mappings can be designed,
the present work could then naturally be extended
to other cored potentials, as well as eventually cuspy ones.
In addition, we limited ourselves to only computing
the divergence of the diffusion flux at the initial time,
${ t \!=\! 0 }$.
It would be of interest to use the same kinetic theories
to integrate forward in time the dynamics of ${ F (\bJ , t) }$,
ideally up to the time of the cluster's core collapse.
Given the complexity of both the \NR\ and \RR\ formalisms,
this will be no easy task.

Third, when computing the \RR\ flux in Fig.~\ref{fig:dFdtLandauMultipole},
we emphasised that for $\ell$ large enough,
the maps of ${ - \p / \p \bJ \!\cdot\! \bFRR^{\ell} }$  resemble
those of ${ - \p / \p \bJ \!\cdot\! \bFNR }$,
up to an overall amplitude.
From the theoretical point of view, following \S\ref{sec:orbit-orbit}
it would therefore be interesting to understand in detail
how a global resonance condition between orbits,
${ \deltaD (\bn \!\cdot\! \bO \!-\! \bnp \!\cdot\! \bO^{\prime}) }$,
as captured by the \RR\ theory,
formally falls back on the orbit-averaged contributions
from local homogeneous deflections,
as captured by the \NR\ theory,
provided that one considers large enough harmonics $\ell$,
and large enough resonance numbers ${ (\bn , \bnp) }$.
Similarly, one should also better understand
the detailed origin of the scaling ${ \bFRR^{\ell} \!\propto\! 1/\ell }$,
observed in Fig.~\ref{fig:Coulomb}.

Fourth, while it is true that the \BL\ equation captures the amplification, and
that this amplification tends to be greatest when $\omega$ is close to the
pattern frequency of a weakly damped normal mode of the stellar system, it does
not account for the direct interaction between stars
and this continuously excited damped mode,
the subject of
\QL\ theory~\citep{Rogister+1968,Hamilton+2020}.
Weakly damped modes
are weakly damped precisely
because there are not many stars with which they resonate;
hence, it may be
expected that these \QL\ interactions do not contribute much to the global
evolution of $F$. However, their slow pattern speed means that they
will interact resonantly with stars that are on large, long-period orbits
and
therefore only weakly bound to the system;
hence possibly leading to excess evaporation
beyond the two-body prediction~\citep{Henon1960}.
Unfortunately, applying the \QL\ operator in practice
is no easy task
as it first requires a detailed characterisation of the damped modes of a given
cluster~\citep{Weinberg1994,Heggie+2020}
through the appropriate analytic continuation of linear response theory.

Finally, we emphasised here that the relaxation
of a star's `in-plane' actions, i.e.\ ${ \bJ \!=\! (J_{r} , L) }$,
up to a correction in the Coulomb logarithm,
is mainly driven by local, small-scale contributions.
Similarly, it would be of interest to determine
whether or not the relaxations of the `out-of-plane' actions,
i.e.\ $\hbL$ the instantaneous orientation of the orbital plane,
is also mainly driven by \NR\ effects, or \RR\ ones,
following the steps of~\cite{Meiron+2019,Fouvry+2019}.

\section{Conclusion}
\label{sec:Conclusion}

The study of the secular relaxation of globular clusters has a long history
dating back to~\cite{Chandrasekhar1943}. It might come as a surprise that almost
80 years later, this topic of research should remain so active. While it has
been claimed recently~\citep{Hamilton+2018,Lau+2019} that collective effects are
able to greatly amplify the efficiency of cluster relaxation, our present work
shows that for an isotropic isochrone sphere, Chandrasekhar's orbit-averaged
theory provides a good effective description, apart from an overall factor of
${ \sim\! 2 }$ in the relaxation rate (Fig.~\ref{fig:dFdtNewtonian}). However, the
physical basis of Chandrasekhar's theory should not be taken entirely literally,
since `collisions' on the scale of the cluster are certainly neither impulsive
nor local. Our implementations of both the \NR\ and \RR\ formalisms show that
the dominant contribution to the fluxes arises from the decades of high
$\ell$-harmonics. Interactions on these scales are barely affected by collective
amplification. The collective amplification on the largest scales (low
$\ell$-harmonics) are not totally negligible, but provide only a modest
correction to the overall relaxation of such a dynamically hot sphere. From our
softening analysis, we conclude that indeed the higher $\ell$-harmonics, i.e.\
small-scale perturbations, involving orbits captured in high-order resonances
contribute most of the flux. Finally we presented a mixed \NR\ and \RR\ approach
to effectively and simultaneously account for the joint effects of large scale,
resonant, dressed, non-local contributions and small scale, non-resonant,
bare, local contributions.

In future it will be of interest to extend our investigations to other cluster
models.  In particular, it is important to check how well our results hold for
colder, thinner, rotating or anisotropic systems --- in the rather extreme
case of old razor-thin discs, \cite{Fouvry+2015} have already shown that evolution
is dominated by the \RR\ processes driven by large-scale dressed fluctuations.
One should also aim to better understand and characterise from the analytical
point of view the deep connections between the \NR\ and \RR\ kinetic theories.

\section*{Acknowledgements}

This work is partially supported by the grant
Segal ANR-19-CE31-0017
of the French Agence Nationale de la Recherche,
and by the Idex Sorbonne Universit\'{e}.
We thank St\'ephane Rouberol for the smooth running of the
Horizon Cluster, where the simulations were performed.

\section*{Data availability}

The data and numerical codes
underlying this article were produced by the authors.
They will be shared on reasonable request
to the corresponding author.

\appendix

\section{Mean field dynamics}
\label{sec:MF}

In this Appendix
we spell out all our conventions
to describe the mean field dynamics
of a spherically symmetric ${3D}$ stellar system.

Following the notations from~\cite{TremaineWeinberg1984},
we define the ${3D}$ angle-action coordinates as
\begin{equation}
\obJ = (J_{r} , L , \Lz) ,
\label{def_obJ}
\end{equation}
with the associated angles ${ \obT \!=\! (\theta_{1} , \theta_{2} , \theta_{3})
}$, and orbital frequencies ${ \obO \!=\! (\Omega_{1} , \Omega_{2} , 0) }$. In
that expression, $J_{r}$ is the radial action, $L$ the norm of the angular
momentum vector, and $\Lz$ its projection
along a given $z$-direction. As a result of
spherical symmetry ${ \Omega_{3} \!=\! 0 }$, because mean field orbits remain
within their orbital plane. The other two frequencies are given by
\begin{align}
\frac{2 \pi}{\Omega_{1}} {} & = 2 \!\! \int_{\rperi}^{\rapo} \!\!\!\! \frac{\rd r}{\sqrt{2 (E - \psi (r)) - L^{2} / r^{2}}} ,
\nonumber
\\
\frac{\Omega_{2}}{\Omega_{1}} {} & = \frac{L}{\pi} \!\! \int_{\rperi}^{\rapo} \!\!\!\! \frac{\rd r}{r^{2} \sqrt{2 (E - \psi (r)) - L^{2} / r^{2}}} ,
\label{def_Omega}
\end{align}
where $\rperi$ (resp.\ $\rapo$)
is the orbit's pericentre (resp.\ apocentre).
Once the orbit has been characterised, the position of the star
is obtained through the angles
\begin{align}
\theta_{1} {} & = \!\! \int_{\mC} \!\! \rd r \, \frac{\Omega_{1}}{\sqrt{2 (E - \psi (r)) - L^{2} / r^{2}}} ,
\nonumber
\\
\theta_{2} - \vphi {} & = \!\! \int_{\mC} \!\! \rd r \, \frac{\Omega_{2} - L / r^{2}}{\sqrt{2 (E - \psi (r)) - L^{2} / r^{2}}} , 
\label{def_angles}
\end{align}
where $\mC$ is the contour going from the pericentre $\rperi$
up to the current position ${ r \!=\! r (\theta_{1}) }$,
along the radial oscillation.
The quantity $\vphi$ in Eq.~\eqref{def_angles} is the angle from
the ascending node to the current location of the particle
along the orbital motion --- see Fig.~{1} of~\cite{TremaineWeinberg1984}.

As mentioned in \S\ref{sec:Relaxation} one can take advantage of the spherical symmetry of the
problem and work exclusively with the in-plane angle-action coordinates.  Thus we
define
\begin{equation}
\bJ \equiv (J_{r} , L ) ;
\quad
\bT \equiv (\theta_{1} , \theta_{2}) ;
\quad
\bO \equiv (\Omega_{1} , \Omega_{2}) .
\label{def_inplane}
\end{equation}
Importantly, the shape of a mean field orbit
is characterised by just two quantities, the
actions $(J_r,L)$. It will sometimes be more convenient instead to label orbits
with the peri- and apocentre distances ${ (\rperi , \rapo) }$, which are related to
the energy $E$ and the angular momentum $L$ by
\begin{equation}
E = \frac{\rapo^{2} \, \psi (\rapo) - \rperi^{2} \, \psi (\rperi)}{\rapo^{2} - \rperi^{2}} ;
\quad
L = \sqrt{\frac{2 (\psi (\rapo) - \psi (\rperi))}{\rperi^{-2} - \rapo^{-2}}} .
\label{EL_rpra}
\end{equation}
One final way to label orbits, useful in numerical work, is via an effective
semi-major axis and eccentricity defined as
\begin{equation}
a = \frac{\rperi + \rapo}{2} ;
\quad
e = \frac{\rapo - \rperi}{\rapo + \rperi} .
\label{def_a_e}
\end{equation}
Such a rewriting proves particularly useful
in Eq.~\eqref{nice_edge_iso} to perform
numerically well-posed orbit-averages
in the isochrone potential.

\section{Linear response theory}
\label{sec:LinearResponseTheory}

\subsection{Basis method}
\label{sec:BasisMethod}

In order to characterise the linear stability
of a self-gravitating system,
we follow the basis method~\citep{Kalnajs1976}.
We introduce a set of potentials and densities
${ (\psi^{(\alpha)} , \rho^{(\alpha)}) }$
that satisfy the biorthogonality relation
\begin{align}
{} & \psi^{(\alpha)} (\br) = \!\! \int \!\! \rd \brp \, U (\br , \brp) \, \rho^{(\alpha)} (\brp) ,
\nonumber
\\
{} & \!\! \int \!\! \rd \br \, \psi^{(\alpha) *} (\br) \, \rho^{(\beta)} (\br) = - \delta_{\alpha\beta} ,
\label{def_basis}
\end{align}
with ${ U(\br , \brp) \!=\! - G / |\br - \brp| }$
the gravitational pairwise interaction.
In the case of a spherical system,
it is natural to write
\begin{align}
\psi^{(\alpha)} (\br) {} & = Y_{\ell}^{m} (\vtheta , \phi) \, U_{n}^{\ell} (r) ,
\nonumber
\\
\rho^{(\alpha)} (\br) {} & = Y_{\ell}^{m} (\vtheta , \phi) \, D_{n}^{\ell} (r) ,
\label{def_basis_sph}
\end{align}
with ${ (r , \vtheta , \phi) }$ the usual spherical coordinates and
$Y_\ell^m$ spherical harmonics normalised so that
${ \!\int\! \rd \vtheta \rd \phi \sin \vtheta | Y_\ell^m(\vtheta,\phi) |^{2} \!=\! 1}$. Equation~\eqref{def_basis_sph} also
involves the radial functions ${ (U_{n}^{\ell} , D_{n}^{\ell}) }$, which we take
to be real. As such, a given basis element is characterised by three integers:
the label $\alpha$ is a shorthand for the triplet $(\ell,m,n)$ where
${ \ell \!=\! 0,1,2,... }$ and ${ m \!=\! -\ell, -\ell+1, ..., \ell }$ describe the angular
dependence, and ${ n \!\geq\! 1 }$ gives the radial
dependence.

In practice, we use the radial basis elements
from~\cite{CluttonBrock1973}. 
With our present convention,
the radial functions of the basis elements read
\begin{align}
U_{n}^{\ell} (r) {} & = A_{n}^{\ell} \, \frac{(r/\Rb)^{\ell}}{(1 + (r / \Rb)^{2})^{\ell + 1/2}} \, C_{n-1}^{(\ell + 1)} (\rho) ,
\nonumber
\\
D_{n}^{\ell} (r) {} & = B_{n}^{\ell} \, \frac{(r / \Rb)^{\ell}}{(1 + (r/\Rb)^{2})^{\ell + 5/2}} \, C_{n-1}^{(\ell + 1)} (\rho) .
\label{def_CB73}
\end{align}
In that expression, $\Rb$ is a fixed scale radius,
and ${ -1 \!\leq\! \rho \!\leq\! 1 }$ is the rescaled variable
\begin{equation}
\rho = \frac{(r / \Rb)^{2} - 1}{(r / \Rb)^{2} + 1} .
\label{def_rho_CB73}
\end{equation}
Equation~\eqref{def_CB73} also involves the Gegenbauer
polynomials ${ C_{n}^{(\alpha)} (\rho) }$.
They can easily be computed through the upward stable
recurrence relation
\begin{equation}
(n \!+\! 1) C_{n+1}^{(\alpha)} \!(\rho) \!=\! 2 (n \!+\! \alpha) \rho \, C_{n}^{(\alpha)} \!(\rho) - (n \!+\! 2 \alpha \!-\! 1) C_{n-1}^{(\alpha)} \!(\rho) ,
\label{def_recc_Gegen}
\end{equation}
with the initial conditions
\begin{equation}
C_{0}^{(\alpha)} = 1 ;
\quad
C_{1}^{(\alpha)} = 2 \alpha \rho .
\label{init_Gegen}
\end{equation}
Finally, in Eq.~\eqref{def_CB73},
we introduced the normalisation coefficients
\begin{align}
A_{n}^{\ell} = {} & - \sqrt{G / \Rb} \, 2^{2 \ell + 3} \, \ell!
\label{Anl_CB73}
\\
\times {} & \bigg[\! \frac{(n \!-\! 1)! (n \!+\! \ell) }{(n \!+\! 2 \ell)! \big[ 4 (n \!-\! 1) (n \!+\! 2 \ell \!+\! 1) \!+\! (2 \ell \!+\! 1) (2 \ell \!+\! 3) \big]} \!\bigg]^{1/2} \!\! ,
\nonumber
\end{align}
as well as
\begin{align}
B_{n}^{\ell} = {} & \frac{1}{\sqrt{G} \, \Rb^{5/2}} \, \frac{2^{2 \ell + 3}}{4 \pi} \, \ell !
\label{Bnl_CB73}
\\
\times {} & \bigg[\! \frac{(n \!-\! 1)! (n \!+\! \ell) \big[ 4 (n \!-\! 1) (n \!+\! 2 \ell \!+\! 1) \!+\!  (2 \ell \!+\! 1) (2 \ell \!+\! 3) \big]}{(n \!+\! 2 \ell)!} \!\bigg]^{1/2} \!\! .
\nonumber
\end{align}

\subsection{Response matrix}
\label{sec:RepMat}

Having constructed basis elements,
they may now be used to represent the potential
fluctuations present in the system
so as to characterise its linear stability.
Following Eq.~{(37)} of~\cite{Hamilton+2018},
for a given harmonic $\ell$,
the linear stability of a stellar cluster is characterised by the response matrix,
${ \bM_{\ell} (\omega) }$, with coefficients
\begin{align}
M_{pq}^{\ell} (\omega) = \frac{2 (2 \pi)^{3}}{2 \ell + 1} \sum_{\mathclap{\substack{n_{1} \\ |n_{2}| \leq \ell \\ (\ell - n_{2}) \mathrm{ even}}}} |y_{\ell}^{n_{2}}|^{2} {} & \!\! \int \!\! \rd \bJ \, L \, \frac{\bn \!\cdot\! \p \Ftot / \p \bJ}{\omega - \bn \!\cdot\! \bO (\bJ)}
\nonumber
\\
{} & \times W_{\ell p}^{\bn} (\bJ) \, W_{\ell q}^{\bn} (\bJ).
\label{def_M}
\end{align}
Here,
${ y_{\ell}^{n} \!\equiv \! Y_{\ell}^{n} (\tfrac{\pi}{2} , 0) }$ are pure
numbers,
while~\cite[see Eq.~{(34)} of][]{Hamilton+2018}
\begin{equation}
W_{\ell n}^{\bn} (\bJ) = \!\! \int_{0}^{\pi} \!\! \frac{\rd \theta_{1}}{\pi} \, U_{n}^{\ell} (r) \, \cos \big( n_{1} \theta_{1} + n_{2} (\theta_{2} - \vphi) \big) ,
\label{def_W}
\end{equation}
whose computation relies on the angle mappings from Eq.~\eqref{def_angles}.
Having computed the response matrix,
we may finally define the susceptibility matrix as
\begin{equation}
\bN_{\ell} (\omega) = \big[ \bI - \bM_{\ell} (\omega) \big]^{-1} .
\label{def_N}
\end{equation}
This matrix characterises the amplitude of the self-gravitating dressing of
potential fluctuations, and is therefore involved in the dressed resonant
diffusion flux (see Eq.~\eqref{def_Lambda}).
In practice, for spherically symmetric systems
the susceptibility matrix satisfies the symmetry
\begin{equation}
\bN_{\ell} (- \omegaR) = \bN_{\ell}^{*} (\omegaR) ,
\label{sym_NMat}
\end{equation}
for ${ \omegaR \!\in\! \mathbb{R} }$,
so that we only need to pre-compute
the susceptibility matrix for ${ \omegaR \!\geq\! 0 }$.

\subsection{Numerical computation}
\label{sec:Comp}

The most demanding computation in Eq.~\eqref{def_M}
is the computation of the coupling coefficients,
${ W_{\ell n}^{\bn} (\bJ) }$, as defined in Eq.~\eqref{def_W}.
In order to accelerate their evaluation,
we follow an approach similar to the one of \S{B}
of~\cite{Rozier+2019}.

First, as already introduced in Eq.~\eqref{def_a_e},
we label the orbits using ${ (a , e) }$.
To compute any integral,
we follow the same trick as in~\cite{Henon1971},
and define an effective anomaly, ${ -1 \leq u \leq 1 }$,
through the explicit mapping
\begin{equation}
r (u) = a (1 + e f (u)) 
\quad \text{ with } \quad
f (u) = u \big( \tfrac{3}{2} - \half u^{2} \big) .
\label{def_u}
\end{equation}
Doing so, any integral over $\theta_{1}$ can be rewritten as
\begin{align}
\!\! \int_{0}^{\pi} \!\! \rd \theta_{1} \, F {} & = \!\! \int_{\rperi}^{\rapo} \!\!\!\! \rd r \, \frac{\rd \theta_{1}}{\rd r} \, F = \!\! \int_{- 1}^{1} \!\!\!\! \rd u \, \frac{\rd \theta_{1}}{\rd r} \, \frac{\rd r}{\rd u} \, F ,
\label{rewrite_av}
\end{align}
where the Jacobian, ${ \rd \theta_{1} / \rd r }$,
naturally follows from Eq.~\eqref{def_angles}.
Following such a change of variables,
integrands now have finite values
at the edge of the integration domain.
Furthermore, in order to increase the numerical stability of the scheme,
we use an exact and well-posed expression
for ${ \rd \theta_{1} / \rd u }$,
as presented in Eq.~\eqref{nice_edge_iso}
for the specific case of the isochrone potential.

Following this rewriting, one could still naively
interpret Eq.~\eqref{def_W} as involving nested integrals,
since one must also compute the values of ${ \theta_{1} [u] }$
and ${ (\theta_{2} \!-\! \vphi)[u] }$ following Eq.~\eqref{def_angles}.
Fortunately, we can use the same trick as in \S{B}
of~\cite{Rozier+2019} and interpret these joint integrals
simply as the forward integration of a single 3-vector.
This is the approach we pursued here.
In practice, we used the traditional RK4 scheme~\citep[see, e.g.,][]{Press2007},
using $K$ steps.
Owing to the analytical expression from Eq.~\eqref{nice_edge_iso},
the integrand is always numerically well-behaved,
which prevents any issues at the boundaries of the integration
where the radial velocity vanishes.

Having computed the coefficients ${ W_{\ell n}^{\bn} (\bJ) }$,
we now have at our disposal an efficient evaluation
of the integrand from Eq.~\eqref{def_M}.
In order to carry out the action integral present
in that expression,
we follow the same approach as in~\cite{Fouvry+2015}
up to three main improvements.
(i) The action space, $\bJ$, is remapped
to the dimensionless coordinates ${ (x , e) \!=\! (a/\bISO , e) }$,
with $\bISO$ the lengthscale of the considered isochrone model.
It is within these coordinates that the orbital domain
is tiled in small square regions of extension ${ \Delta x \!\times\! \Delta e }$.
(ii) In the expression of the approximated integrands,
derivatives, such as
${ \p W_{\ell n}^{\bn} / \p x }$ and ${ \p W_{\ell n}^{\bn} / \p e }$,
are not computed through finite differences
but rather through their analytical expressions
by computing explicitly the derivatives under the integral sign
in Eq.~\eqref{def_W}.
(iii) All angular integrals, including derivatives,
are computed efficiently using the effective anomaly from Eq.~\eqref{def_u}
and the associated integration trick.
Let us finally emphasise that, while Eq.~\eqref{def_M}
is a complicated function to compute,
once evaluated, one can store pre-computed
interpolation functions ${ (\ell , p , q , \omega) \!\to\! N_{pq}^{\ell} (\omega) }$,
which are then used to evaluate the dressed coupling coefficients
from Eq.~\eqref{def_Lambda}.

\begin{figure}
\centering
\includegraphics[width=0.45 \textwidth]{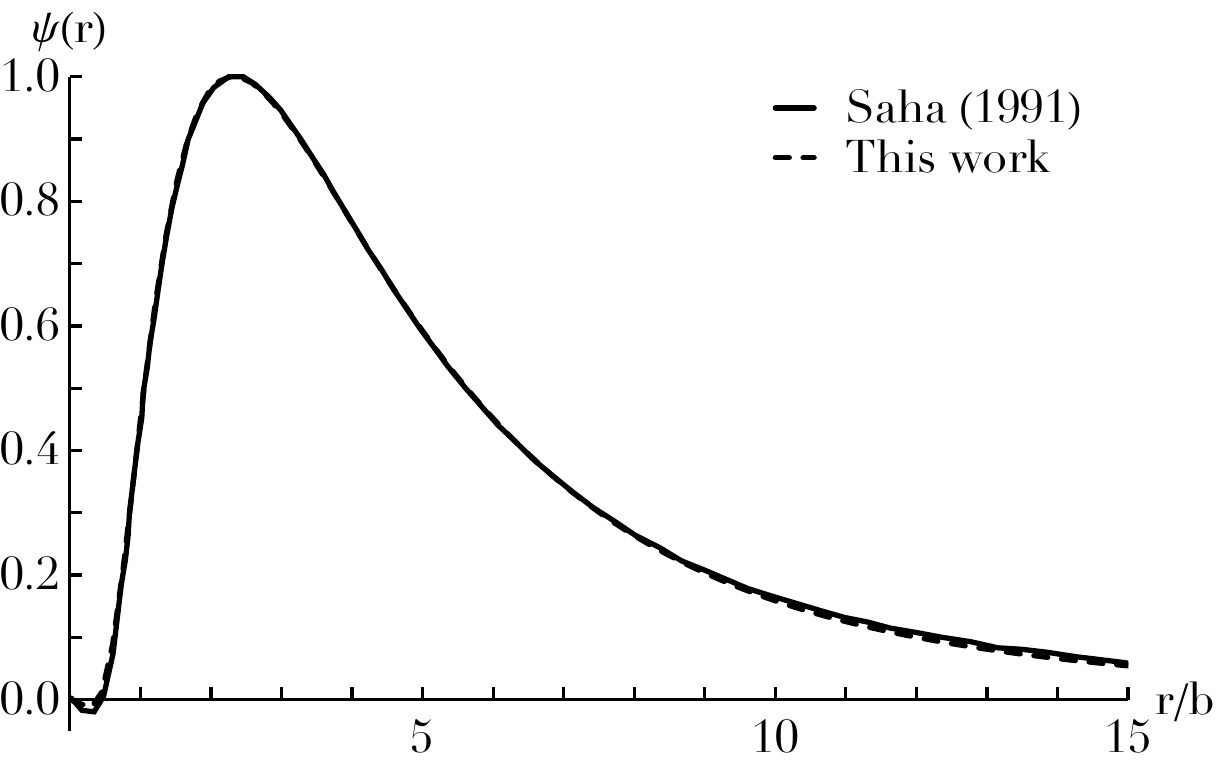}
   \caption{Illustration of the radial shape of the mode, ${ \psi (r) }$,
   as measured in Fig.~{4} of~\protect\cite{Saha1991}
   and compared with the present method,
   for the unstable ${ \ell \!=\! 2 }$ mode of the ${ \Ra \!=\! \bISO }$ model.
   The normalisation of the vertical axis is arbitrary.
   }
   \label{fig:ROI}
\end{figure}

In order to validate our implementation
of the response matrix,
we set out to reproduce the radial-orbit
instability of the isochrone potential
recovered in~\cite{Saha1991},
using the radially anisotropic \DF\ from Eq.~\eqref{anisoDF_iso}.
This is illustrated in Fig.~\ref{fig:ROI}.
For these calculations, 
following Eq.~\eqref{def_CB73},
we considered a total of ${ \nmax \!=\! 20 }$ basis elements
with ${ \Rb \!=\! 5 \, \bISO }$.
In Eq.~\eqref{def_M},
the orbital integral was performed
for ${ \ell \!=\! 2 }$
using a uniform grid in ${ (x,e)}$-space,
with ${ 0 \!\leq\! x \!\leq\! \xmax \!=\! 10 }$,
${ 0 \!\leq\! e \!\leq\! 1 }$,
with the step distances ${ \Delta x \!=\! 0.02 }$,
and ${ \Delta e \!=\! 0.002 }$.
In that same expression, the sum over resonances
was limited to ${ |n_{1}| \!\leq\! n_{1}^{\max} \!=\! 10 }$.
Finally, the orbital averages in Eq.~\eqref{def_W}
were performed using ${ K \!=\! 100 }$ steps.

In Fig.~\ref{fig:ROI}, we recover that the model ${ \Ra \!=\! \bISO }$
supports an unstable mode with growth rate
${ \eta \!\simeq\! 0.0245 \, \Omega_{0} }$
in good agreement with the value ${ 0.024 \, \Omega_{0} }$
measured in~\cite{Saha1991}.
The radial shape of the unstable mode
also offers a good agreement with~\cite{Saha1991}.
All in all, this shows the sanity of our present numerical implementation
of the response matrix.

We use this matrix method in \S\ref{sec:RoleCollective effects},
in particular to understand the inefficiency
of collective effects to enhance the \RR\ relaxation
in the core regions of isotropic clusters.
In the numerical calculation
presented in Fig.~\ref{fig:NMat},
we considered a total of ${ \nmax \!=\! 20 }$
basis elements, using the basis from Eq.~\eqref{def_CB73}
with the scale radius ${ \Rb \!=\! 10 \, \bISO }$.
The sum over resonances was limited
to ${ |n_{1}| \!\leq\! n_{1}^{\max} \!=\! 10 }$,
while the orbit-averages were performed with ${ K \!=\! 100 }$ steps.
Finally, the domain of orbital integration was limited
to ${ 0 \!\leq\! x \!\leq\! \xmax \!=\! 20 }$,
${ 0 \!\leq\! e \!\leq\! 1 }$,
with the uniform step distances ${ \Delta x \!=\! 0.02 }$
and ${ \Delta e \!=\! 0.002 }$.

\section{Fokker--Planck equation}
\label{sec:FP}

In this Appendix, we detail our implementation
of the orbit-averaged local diffusion coefficients
that appear in the \NR\ flux from Eq.~\eqref{def_bF_NR}.
Here, we follow in particular calculations from~\cite{BinneyTremaine2008}
for the local velocity diffusion coefficients,
and from~\cite{BarOrAlexander2016} for the
computation of the orbit-average.

Following Eq.~{(7.69)} of~\cite{BinneyTremaine2008},
we write the traditional Fokker--Planck equation in velocity space
with the convention
\begin{align}
\frac{\p P (\bv)}{\p t} = {} & - \sum_{i = 1}^{3} \frac{\p }{\p v_{i}} \big[ \big\langle \delta v_{i} \big\rangle \, P (\bv) \big]
\nonumber
\\
{} & + \frac{1}{2} \sum_{i , j = 1}^{3} \frac{\p^{2}}{\p v_{i} \p v_{j}} \big[ \big\langle \delta v_{i} \, \delta v_{j} \big\rangle \, P (\bv) \big] ,
\label{trad_FP}
\end{align}
where ${ P (\bv) }$ stands for an arbitrary \DF\@,
proportional to the number of particles within the volume ${ \rd \bv }$.

In a nutshell, the \NR\ flux is computed
through the following steps.
(i) At a given
phase space location ${ (\br , \bv) }$, one computes the local velocity
diffusion coefficients, ${ \langle \delta \bv \rangle (\br , \bv) }$ and ${
\langle (\delta \bv)^{2} \rangle (\br , \bv) }$,
a calculation made easier by our assumption
of an isotropic background \DF\@.
(ii) The local diffusion coefficients are translated
into local diffusion coefficients in integrals of motion, in practice the energy
and angular momentum, $E$ and $L$, e.g.\@, via ${ \langle \delta E \rangle (\br
, \bv) }$.
(iii) Contributions from all these kicks accumulate
as the star follows its mean field orbit,
leading to the associated orbit-averaged diffusion
coefficients, e.g.\@, ${ \langle \Delta E \rangle (\bJ) \!=\! \!\oint\!
\tfrac{\rd \theta_{1}}{2 \pi} \langle \delta E \rangle }$.
(iv) Finally, the diffusion coefficients in $\bJ$-space
are derived through the appropriate change of variables.
Let us now be more specific for each of these steps.

The first- and second-order diffusion coefficients
originate from local deflections.
Following Eq.~{(7.83a)} of~\cite{BinneyTremaine2008},
and paying a careful attention to our normalisation convention,
for a single-mass cluster, they read
\begin{align}
\big\langle \delta v_{i} \big\rangle {} & = 8 \pi G^{2} \, \mu \ln \Lambda \, \frac{\p h}{\p v_{i}} ,
\nonumber
\\
\big\langle \delta v_{i} \, \delta v_{j} \big\rangle {} & = 4 \pi G^{2} \mu \ln \Lambda \, \frac{\p^{2} g}{\p v_{i} \p v_{j}} ,
\label{local_DCs}
\end{align}
where we introduced ${ \ln \Lambda }$ as the Coulomb logarithm.
In practice, the Coulomb logarithm is fixed following
the prescription from Eq.~\eqref{exp_lnLambda_Classical}.
In that expression, the Rosenbluth potentials are given by
\begin{align}
h (\br , \bv) {} & = \!\! \int \!\! \rd \bvp \, \frac{\Ftot (\br , \bvp)}{|\bv  - \bvp|} ,
\nonumber
\\
g (\br , \bv) {} & = \!\! \int \!\! \rd \bvp \, \Ftot (\br , \bvp) \, |\bv - \bvp| .
\label{Rosenbluth}
\end{align}

For an isotropic \DF\@, ${ \Ftot (\br , \bv) \!=\! \Ftot (r , v) }$,
the diffusion coefficients are characterised
by only three quantities, namely
\begin{align}
\big\langle \delta v_{\parallel} \big\rangle {} & \!=\! - 2 \kappa \!\! \int_{0}^{v} \!\! \rd \vp \, \frac{v^{\prime 2}}{v^{2}} \, \Ftot (\vp) ,
\label{Diff_iso_v}
\\
\big\langle \big( \delta v_{\parallel} \big)^{2} \big\rangle {} & \!=\! \frac{2}{3} \kappa \bigg[ \!\! \int_{0}^{v} \!\! \rd \vp \, \frac{v^{\prime 4}}{v^{3}} \Ftot (\vp) + \!\! \int_{v}^{+ \infty} \!\!\!\!\!\!\!\! \rd \vp \vp \Ftot (\vp) \bigg] ,
\nonumber
\\
\big\langle \big( \delta v_{\perp} \big)^{2} \big\rangle {} & \!=\! \frac{2}{3} \kappa \bigg[ \!\! \int_{0}^{v} \!\!\!\! \rd \vp \bigg(\! \frac{3 v^{\prime 2}}{v} \!-\! \frac{v^{\prime 4}}{v^{3}} \!\bigg) \Ftot (\vp) \!+\! 2 \!\! \int_{v}^{+ \infty} \!\!\!\!\!\!\!\! \rd \vp \vp \Ftot (\vp) \bigg] .
\nonumber
\end{align}
where we introduced ${ \kappa \!=\! 16 \pi^{2} G^{2} \mu \ln \Lambda }$,
and did not write the dependence w.r.t.\ $r$
to shorten the expressions.

Fortunately, in the case of an isotropic \DF\@,
we can rewrite all these integrals as integrals over the energy.
Noting that ${ v \rd v \!=\! \rd E }$,
we can rewrite Eq.~\eqref{Diff_iso_v} as
\begin{align}
\big\langle \delta v_{\parallel} \big\rangle {} & = - 2 \kappa \frac{1}{v} \, \Iinf_{1} ,
\nonumber
\\
\big\langle \big( \delta v_{\parallel} \big)^{2} \big\rangle {} & = \frac{2}{3} \kappa \bigg[ \Iinf_{3} + \Isup_{0} \bigg] ,
\nonumber
\\
\big\langle \big( \delta v_{\perp} \big)^{2} \big\rangle {} &  = \frac{2}{3} \kappa \bigg[ 3 \Iinf_{1} - \Iinf_{3} + 2 \Isup_{0} \bigg] ,
\label{Diff_iso_E}
\end{align}
where we introduced the one-dimensional integrals
\begin{align}
\Iinf_{k} (r , v) {} & = \!\! \int_{\psi}^{E} \!\! \rd \Ep \, (\vp / v)^{k} \, \Ftot (\Ep) ,
\nonumber
\\
\Isup_{k} (r , v) {} & = \!\! \int_{E}^{0} \!\! \rd \Ep \, (\vp / v)^{k} \, \Ftot (\Ep) .
\label{def_Iinf_Isup}
\end{align}
In these expression,
the boundary of the integrals are given by
${ \psi \!=\! \psi (r) }$ and ${ E \!=\! \half v^{2} \!+\! \psi (r) }$.
We also note that all orbits are taken to be bound,
so that ${ E, \Ep < 0 }$.
In practice, these integrals are computed
using a midpoint rule with ${ K \!=\! 10^{3} }$ steps.

From these local diffusion coefficients in velocity,
we can now compute the local diffusion coefficients in ${ (E,L) }$.
To do so, we rely on the relations ${ E \!=\! \half v^{2} \!+\! \psi (r) }$,
and ${ L \!=\! |\br \!\times\! \bv| }$,
that are perturbed to first order.
Following Eqs.~{(85)--(89)} in~\cite{BarOrAlexander2016},
we obtain
\begin{align}
\big\langle \delta E \big\rangle & {} = v \, \big\langle \delta v_{\parallel} \big\rangle + \frac{1}{2} \big\langle \big( \delta v_{\parallel} \big)^{2} \big\rangle + \frac{1}{2} \big\langle \big( \delta v_{\perp} \big)^{2} \big\rangle , 
\nonumber
\\
\big\langle \delta L \big\rangle {} & = \frac{L}{v} \big\langle \delta v_{\parallel} \big\rangle + \frac{r^{2}}{4 L} \, \big\langle \big( \delta v_{\perp} \big)^{2} \big\rangle ,
\nonumber
\\
\big\langle \big( \delta E \big)^{2} \big\rangle {} & = v^{2} \big\langle \big( \delta v_{\parallel} \big)^{2} \big\rangle ,
\nonumber
\\
\big\langle \delta E \, \delta L \big\rangle {} & = L \, \big\langle \big( \delta v_{\parallel} \big)^{2} \big\rangle , 
\nonumber
\\
\big\langle \big( \delta L \big)^{2} \big\rangle {} & = \frac{L^{2}}{v^{2}} \big\langle \big( \delta v_{\parallel} \big)^{2}  \big\rangle+ \frac{1}{2} \bigg( r^{2} - \frac{L^{2}}{v^{2}} \bigg) \big\langle \big( \delta v_{\perp} \big)^{2} \big\rangle .
\label{Diff_iso_EL}
\end{align}

Having computed the local diffusion coefficients in ${ (E,L) }$,
we can now compute their orbit-average.
For an isotropic system,
it is given by the simple calculation
\begin{equation}
\big\langle \Delta E \big\rangle = \!\! \int_{0}^{\pi} \!\! \frac{\rd \theta_{1}}{\pi} \, \big\langle \delta E \big\rangle .
\label{orbit_average}
\end{equation}
In practice, to avoid any boundary issues,
we use the exact same technique as in Eq.~\eqref{rewrite_av},
and introduce an effective anomaly to perform the orbit-average.
These integrals are then computed using a midpoint rule
with ${ K \!=\! 10^{3} }$ steps.

At this stage, we have derived the orbit-averaged
diffusion coefficients in ${ (E,L)}$--space.
It now only remains to translate them in the ${ \bJ \!=\! (J_{r}, L) }$ coordinates.
Following Eqs.~{(122)} and~{(123)} of~\cite{BarOrAlexander2016},
under a coordinate transform of the form
${ \bxp \!=\! \bxp (\bx) }$,
the new diffusion coefficients are given by
\begin{align}
\big\langle \Delta \xp_{k} \big\rangle {} & = \frac{\p \xp_{k}}{\p x_{i}} \, \big\langle \Delta x_{i} \big\rangle + \frac{1}{2} \frac{\p^{2} \xp_{k}}{\p x_{i} \p x_{j}} \, \big\langle \Delta x_{i} \Delta x_{j} \big\rangle ,
\nonumber
\\
\big\langle \Delta \xp_{k} \Delta \xp_{l} \big\rangle {} & = \frac{\p \xp_{k}}{\p x_{i}} \, \frac{\p \xp_{l}}{\p x_{j}} \, \big\langle \Delta x_{i} \Delta x_{j} \big\rangle ,
\label{FP_transform}
\end{align}
where the sums over $i$ and $j$ are implied.
Fortunately, in the case of the isochrone potential,
we have at our disposal an explicit expression
for ${ J_{r} \!=\! J_{r} (E,L) }$,
as in Eq.~\eqref{Jr_iso},
which eases this change of coordinates.
Following all these manipulations,
we finally obtain the first-order diffusion coefficient,
${ \bD_{1} (\bJ) }$, and the second-order diffusion tensor,
${ \bD_{2} (\bJ) }$, as in Eq.~\eqref{def_bF_NR}.

\section{Deriving the Balescu--Lenard flux}
\label{sec:BLEqDerivation}

In this Appendix, we present the key steps
of the derivation of the \BL\ equation
from the Klimontovich equation~\citep{Chavanis2012},
focussing specifically on how, in spherically symmetric systems,
the resulting ${3D}$ kinetic equation
may rewritten as an effective ${2D}$ equation~\citep{Hamilton+2018}
involving a summation over $\ell$ harmonics.

\subsection{From Klimontovich to Balescu--Lenard}
\label{sec:KlimtoBL}

The state of the globular cluster can be fully described
by its discrete \DF\@,
\begin{equation}
\Fd (\bx , \bv , t) = \sum_{i = 1}^{N} \mu \, \deltaD (\bx - \bx_{i} (t)) \, \deltaD (\bv - \bv_{i} (t)) ,
\label{def_Ff}
\end{equation}
where ${ (\bx_{i} (t) , \bv_{i} (t)) }$ stands
for the location in phase space at time $t$ of particle $i$.
For a given realisation, the dynamics of that \DF\ is exactly described
by the Klimontovich equation that takes here the short form
\begin{equation}
\frac{\p \Fd}{\p t} + \big[ \Fd , \Hd \big] = 0 ,
\label{Klim_Eq}
\end{equation}
where the Poisson bracket is defined with the convention
\begin{equation}
\big[ f , h \big] = \frac{\p f}{\p \bx} \!\cdot\! \frac{\p h}{\p \bv} - \frac{\p f}{\p \bv} \!\cdot\! \frac{\p h}{\p \bx} .
\label{def_PoissonBracket}
\end{equation}
In Eq.~\eqref{Klim_Eq}, we introduced the (specific)
discrete Hamiltonian
\begin{equation}
\Hd (\bx , \bv , t) = \frac{1}{2} |\bv|^{2} + \Phid (\bx , t) ,
\label{def_Hd}
\end{equation}
where the instantaneous potential, ${ \Phid \!=\! \Phid [\Fd] }$,
self-consistently depends on the system's instantaneous \DF\@,
through the relation
\begin{equation}
\Phid (\bx , t) = \!\! \int \!\! \rd \bxp \rd \bvp \, \Fd (\bxp , \bvp , t) \, U (\bx , \bxp) ,
\label{def_Phid}
\end{equation}
with ${ U(\bx , \bxp) \!=\! - G / |\bx \!-\! \bxp| }$
the Newtonian pairwise interaction.

We now assume that the system's \DF\ and potential
can be decomposed into two components,
\begin{equation}
\Fd = \Ftot + \delta F ;
\quad
\Hd = \Htot + \delta \Phi ,
\label{QL_decomposition}
\end{equation}
such that ${ \delta F \!\ll\! \Ftot }$
and ${ \delta \psi \!\ll\! \Htot }$.
Perturbations are such that
${ \langle \delta F \rangle \!=\! 0 }$,
and ${ \langle \delta \Phi \rangle \!=\! 0 }$,
with ${\langle \,\cdot\, \rangle }$ the ensemble average over realisations.
Importantly, we note that the \DF\@'s and potential fluctuations
are self-consistent as, similarly to Eq.~\eqref{def_Phid}, one has
\begin{equation}
\delta \Phi (\bx , t) = \!\! \int \!\! \rd \bxp \rd \bvp \, \delta F (\bxp , \bvp , t) \, U (\bx , \bxp) .
\label{selfconsistent_psi}
\end{equation}
We assume that the mean system is in
an integrable mean field equilibrium,
so that ${ [ \Ftot , \Htot ] \!=\! 0 }$,
and there exist some ${3D}$ angle-action coordinates,
${ (\obT , \obJ) }$, 
as defined in Eq.~\eqref{def_obJ},
so that ${ \Ftot \!=\! \Ftot (\obJ , t) }$.
Similarly, the mean field Hamiltonian is such that
${ \Htot \!=\! \Htot (\obJ , t) }$,
which naturally defines the associated orbital frequencies
${ \obO \!=\! \p \Htot / \p \obJ }$.

Injecting the decomposition from Eq.~\eqref{QL_decomposition}
into Eq.~\eqref{Klim_Eq},
and assuming that the dynamics of perturbations
can be solved at linear order,
one obtains a set of two coupled evolution equations
reading respectively
\begin{align}
{} & \frac{\p \delta F}{\delta t} + \big[ \delta F , \Htot \big] + \big[ \Ftot , \delta \Phi \big] = 0 ,
\nonumber
\\
{} & \frac{\p \Ftot}{\p t} + \big\langle \big[ \delta F , \delta \Phi \big] \big\rangle = 0 .
\label{Klim_split}
\end{align}
The first equation is the linearised Klimontovich equation
that describes the combined effects of phase mixing
and collective amplification.
The second equation,
once the ensemble-averaged computed,
will give the long-term kinetic equation.

Introducing the Laplace-Fourier transform
with the convention
\begin{equation}
\delta \tF_{\obn} (\obJ , \omega) = \!\! \int_{0}^{+ \infty} \!\!\!\! \rd t \, \re^{\ri \omega t} \!\! \int \!\! \frac{\rd \obT}{(2 \pi)^{3}} \, \delta F (\obT , \obJ , t) \, \re^{- \ri \obn \cdot \obT} ,
\label{LaplaceFourier}
\end{equation}
with ${ \obn \!\in\! \mathbb{Z}^{3} }$,
one can rewrite Eq.~\eqref{Klim_split} as
\begin{equation}
\delta F_{\obn} (\obJ , \omega) = - \frac{\obn \!\cdot\! \p \Ftot / \p \obJ}{\omega - \obn \!\cdot\! \obO} \, \delta \tPhi_{\obn} (\obJ , \omega) - \frac{\delta F_{\obn} (\obJ , 0)}{\ri (\omega - \obn \!\cdot\! \obO)} ,
\label{LF_Klim}
\end{equation}
where ${ \delta F_{\obn} (\obJ , 0) }$
stands for the fluctuations of the \DF\ at the initial time.
Owing to self-consistency, ${ \delta \Phi \!=\! \delta \Phi [\delta F] }$,
one can rewrite Eq.~\eqref{LF_Klim} as
\begin{equation}
\delta \tPhi_{\obn} (\obJ , \omega) \!=\! - (2 \pi)^{3} \sum_{\obnp} \!\! \int \!\! \rd \obJp \, \frac{\delta F_{\obnp} (\obJp , 0)}{\ri (\omega \!-\! \obnp \!\cdot\! \obO (\obJp))} \, \psid_{\obn\obnp} (\obJ , \obJp , \omega) ,
\label{SC_deltaPhi}
\end{equation}
where the dressed susceptibility coefficients,
${ \psid_{\obn\obnp} (\obJ , \obJp , \omega) }$, read
\begin{equation}
\psid_{\obn\obnp} (\obJ , \obJp , \omega) = - \sum_{\alpha , \beta} \psi_{\obn}^{(\alpha)} (\obJ) \, \oN_{\alpha \beta} (\omega) \, \psi_{\obnp}^{(\beta) *} (\obJp) .
\label{def_psid}
\end{equation}
In that expression, the basis elements, ${ \psi^{(\alpha)} }$,
were introduced following the convention from Eq.~\eqref{def_basis},
and their Fourier transform defined with the convention
from Eq.~\eqref{LaplaceFourier}.
Equation~\eqref{def_psid} also involves the total susceptibility matrix,
${ \obN (\omega) \!=\! [\bI - \obM (\omega)]^{-1} }$,
with the response matrix
\begin{equation}
\oM_{\alpha \beta} (\omega) = (2 \pi)^{3} \sum_{\obn} \!\! \int \!\! \rd \obJ \, \frac{\obn \!\cdot\! \p \Ftot / \p \obJ}{\omega - \obn \!\cdot\! \obO (\obJ)} \, \psi^{(\alpha) *}_{\obn} (\obJ) \, \psi^{(\beta)}_{\obn} (\obJ) .
\label{def_obN}
\end{equation}

We may then inject the solutions from Eqs.~\eqref{LF_Klim}
and~\eqref{SC_deltaPhi} into the evolution equation
for ${ \p \Ftot / \p t }$ in Eq.~\eqref{Klim_split}.
Following this manipulation,
one gets an expression of the form
\begin{equation}
\frac{\p \Ftot (\obJ)}{\p t} = - \frac{\p }{\p \bJ} \cdot \bigg[ \obF_{1} (\obJ) + \obF_{2} (\obJ) \bigg] .
\label{injection_Klim}
\end{equation}
In that expression, we introduced the fluxes
${ \obF_{1} (\obJ) \!\propto\! \langle \delta \Phi \, \delta F (0) \rangle }$,
and 
${ \obF_{2} (\obJ) \!\propto\! \langle \delta \Phi \, \delta \Phi \rangle }$.
Here, on the one hand, ${ \obF_{1} (\obJ) }$ is the friction force
and is sourced by the correlations between one particular
fluctuation in the system's \DF\ and the associated
potential perturbations generated in the system.
As such, this captures the process of dynamical friction~\citep{TremaineWeinberg1984}.
On the other, the flux component, ${ \obF_{2} (\obJ) }$,
is sourced by the potential correlations between
the potential fluctuations.
This captures the process of resonant orbital diffusion~\citep{BinneyLacey1988}.

Following some lengthy and sometimes subtle manipulations,
and the use of the Poisson statistics
${ \langle \delta F (0) \delta F (0) \rangle \!\propto\! \mu \Ftot }$,
one can finally rewrite Eq.~\eqref{injection_Klim}
as the inhomogeneous \BL\ equation~\citep{Chavanis2012}
that generically reads
\begin{align}
{} & \frac{\p \Ftot (\obJ)}{\p t} \!=\! - \pi (2 \pi)^{3} \mu \frac{\p }{\p \obJ} \!\cdot\! \bigg[ \sum_{\obn , \obnp} \obn \!\! \int \!\! \rd \obJp \, \big| \psid_{\obn\obnp} (\obJ , \obJp , \obn \!\cdot\! \obO (\obJ)) \big|^{2} 
\nonumber
\\
\times {} & \deltaD (\obn \!\cdot\! \obO (\obJ) \!-\! \obnp \!\cdot\! \obO (\obJp)) \bigg(\! \obnp \!\cdot\! \frac{\p }{\p \obJp} \!-\! \obn \!\cdot\! \frac{\p }{\p \obJ} \!\bigg)\! \, \Ftot (\obJ) \Ftot (\obJp) \!\bigg] .
\label{def_BL}
\end{align}
Equation~\eqref{def_BL} is a key result,
as it is the master equation to describe
the long-term relaxation of self-gravitating systems
induced by the long-range, resonant, and dressed couplings
between Poisson fluctuations.

\subsection{From ${3D}$ to effectively ${2D}$  resonant relaxation}
\label{sec:3Dto2D}

While Eq.~\eqref{def_BL} is a very generic result,
it still requires to be tailored to ${3D}$ spherical systems
to benefit from these systems' additional symmetries,
i.e.\ orbits at fixed action remain coplanar
but gravitationally interact and resonate
with orbits in different orbital planes.
This is the calculation that was presented in \S{4}
of~\cite{Hamilton+2018},
and that we briefly reproduce here.

As emphasised in Eq.~\eqref{def_obJ},
in the absence of any perturbations,
stars are confined to their orbital planes.
This is the imprint of a dynamical degeneracy,
so that one has ${ \Omega_{3} (\obJ) \!=\! 0 }$,
i.e.\ ${ (\theta_{3} , L_{z})}$
(${ \propto\! \hbL }$ the orientation of the orbital plane)
are both conserved through the mean field dynamics.
Similarly, in spherically symmetric systems,
one has ${ \Ftot \!=\! \Ftot (\bJ) }$,
i.e.\ it depends only on the two in-plane actions
from Eq.~\eqref{bJ}.
We may then use these additional symmetries
to transform Eq.~\eqref{def_BL}
into an effectively ${2D}$ diffusion equation.

Let us highlight the key steps of~\cite{Hamilton+2018} relevant to the summation over orbital planes.
The first step is to transform the potential basis given by Eq.~\eqref{def_basis} into angle-action and Fourier transform w.r.t.\ to 
these angles to write for  ${ \alpha \!=\! (\ell , m , n) }$
\citep[see][]{TremaineWeinberg1984}  
\begin{equation}
{\psi}_{\obn}^{(\alpha)}(\obJ) \!=\!
\delta_{m}^{n_3} \, \ri^{m-n_2} y_{\ell}^{n_{2}} \,
R_{n_2 m}^{\ell}(\beta)\,W_{\ell n}^{\bn}(\bJ),
\nonumber
\end{equation}
with ${ y_{\ell}^{n} \!\equiv \! Y_{\ell}^{n} (\tfrac{\pi}{2} , 0) }$,
and where ${ R_{n_2 m}^{\ell}(\beta) }$
is the spin-$\ell$ Wigner rotation matrix, 
\begin{align}
R_{n m}^{\ell}(\beta) =&\sum_t  (-1)^t \frac{\sqrt{(\ell+n)!(\ell-n)!
(\ell+m)!(\ell-m)!}}{(\ell-m-t)!(\ell+n-t)!t!(t+m-n)!}
\nonumber\\
& \times [\cos(\beta/2)]^{2\ell+n-m-2t}[\sin(\beta/2)]^{2t+m-n}\,, \nonumber
\end{align}
and ${ W_{\ell n}^{\bn} (\bJ) }$ is defined by Eq.~\eqref{def_W}.

Since  $F_\mathrm{tot}$ is independent of $L_z$, the only $\beta$
dependence in the response matrix from Eq~\eqref{def_obN}
comes from the rotation matrices $R_{n m}^{\ell}$.
Given the orthogonality of these rotation matrices when
integrated over ${ \cos (\beta) }$~\citep[e.g.][]{Edmonds1996},
the ${3D}$ response matrix  
can be rewritten as Eq.~\eqref{def_M}
which depends explicitly on  $\ell$
and displays an extra $L$ volume element
in the integration over action space.
We can then proceed accordingly for the computation
of the dressed coupling coefficients
and write 
\begin{align}
&\big|  \psid_{\obn\obnp} (\obJ , \obJp ,\omega)\big|^2  =\delta_{n_3}^{\np_3} 
\nonumber
\\
&\hskip.75cm\times\sum_{ \substack{\ell^p}} \sum_{\substack{\ell^q}}  
\Lambda_{\bn\bnp}^{\ell^p} (\bJ , \bJp , \omega) 
 \,
 \Lambda_{\bn\bnp}^{\ell^q} (\bJ , \bJp , \omega) 
\nonumber
\\&\hskip.75cm \times 
R^{\ell^p}_{n_2 n_3} \!(\beta) \, 
R^{\ell^q}_{n_2n_3} \!(\beta) \, 
R^{\ell^p}_{\np_2 \np_3} \!(\betap) \, 
R^{\ell^q}_{\np_2 \np_3} \!(\betap) .
\label{exp_psid_3D}
\end{align} 
where  we now have two sums over the two plane orientations $\beta$ and $\betap$ while the ${2D}$ coupling coefficients
${ \Lambda_{\bn\bnp}^{\ell} (\bJ , \bJp , \omega) }$ are given by Eq.~\eqref{def_Lambda}.
Once again, since $\Ftot$ is independent of $L_z$ we can integrate Eq.~\eqref{exp_psid_3D} over ${ \cos (\beta) }$
and use the orthogonality conditions of $R$ to write
\begin{align}
\int & \rd \Jp_3\,\big| \psid_{\obn\obnp} (\obJ , \obJp , \obn \!\cdot\! \obO (\obJ))
\big|^2  =\delta_{n_3}^{\np_3} \, \Jp_2
\nonumber
\\
& \, \times \!\! \sum_{ \substack{\ell}} \frac{2}{2 \ell + 1}  
\big|\Lambda_{\bn\bnp}^{\ell} (\bJ , \bJp ,  \bn \!\cdot\! \bO (\bJ)) \big|^2 
\big| R^{\ell}_{n_2 n_3} \!(\beta) \big|^2 \!.
\label{integralz}
\end{align}
From Eq.~\eqref{def_BL} the only  
non-zero contributions to the flux are
proportional to ${ | R^{\ell}_{n_2 n_3} \!(\beta) |^2 }$.
For given $\bn$ and $\bnp$, we can sum over all values of
${ n_3 \!=\! \np_3 }$ and thanks to  the 
identities
\begin{equation}
\sum_{n_3} \big| R^{\ell}_{n_2n_3}(\beta) \big|^2 \!=\! 1;
\;
\sum_{n_3} n_3 \, \big| R^{\ell}_{n_2n_3}(\beta) \big|^2
\!=\! n_2 \cos (\beta) ,
\end{equation}
 integrate  Eq.~\eqref{def_BL} over ${ J_3 }$ so as to write 
\begin{align}
 \frac{\p F (\bJ)}{\p t} \!=\!-\!\! \frac{\p }{\p \bJ} \!\cdot\! \bigg[ {} & \sum_{\bn,\bnp} \! \bn \sum_\ell \frac{\pi (2 \pi)^{3} \mu}{2 \ell + 1} \! \!\! \int \!\! \rd \bJp  L  \Lp \nonumber
\\
\times {} & \big| \Lambda_{\bn\bnp}^{\ell} (\bJ , \bJp \!, \bn \!\cdot\! \bO (\bJ)) \big|^{2}  \deltaD (\bn \!\cdot\! \bO (\bJ) \!-\! \bnp \!\cdot\! \bO (\bJp))
\nonumber
\\
\times {} & \bigg(\! \bnp \!\cdot\! \frac{\p }{\p \bJp} \!-\! \bn \!\cdot\! \frac{\p }{\p \bJ} \!\bigg) \frac{F (\bJ)}{L} \frac{F (\bJp)}{\Lp} \bigg] ,
\label{2DBLappendix}
\end{align}
where we introduced the reduced \DF\
given by Eq.~\eqref{def_DFred}.
Equation~\eqref{2DBLappendix} is fully equivalent
to Eqs.~\eqref{sum_bF}--\eqref{def_mF}. 
This new (in-plane) \RR\ equation is formally very similar
to the generic Eq.~\eqref{def_BL},
 modulo a volume element $L$ in the integrand
 and an extra summation over $\ell$.
We can see from Eqs.~\eqref{exp_psid_3D}--\eqref{integralz} that this sum is sourced by the double integration over the ${ \beta, \betap }$  planes.
The dressed ${3D}$ coupling coefficient, $ \psid_{\obn\obnp} $, has been replaced by the ${2D}$ 
$\ell$-dependent coefficient, $\Lambda_{\bn\bnp}^{\ell} $,
given by Eq.~\eqref{def_Lambda}.

\section{Computing the Balescu--Lenard flux}
\label{sec:BLEq}

In this Appendix we detail our computation
of the inhomogeneous \BL\ flux for
spherical systems.

\subsection{Dressed coupling coefficients}
\label{sec:Lambda}

As already emphasised in Eq.~\eqref{def_mF},
the resonant diffusion flux
involves the dressed coupling coefficients,
${ \Lambda_{\bn\bnp}^{\ell} (\bJ , \bJp , \omega) }$.
Following Eq.~{(41)} of~\cite{Hamilton+2018},
they read
\begin{equation}
\Lambda_{\bn\bnp}^{\ell} (\bJ , \bJp , \omega) = y_{\ell}^{n_{2}} \, y_{\ell}^{\np_{2}} \, \sum_{\mathclap{p, q = 1}}^{+ \infty} W_{\ell p}^{\bn} (\bJ) \, N_{p q}^{\ell} (\omega) \, W_{\ell q}^{\bnp} (\bJp) ,
\label{def_Lambda}
\end{equation}
with ${ y_{\ell}^{n} \!\equiv \! Y_{\ell}^{n} (\tfrac{\pi}{2} , 0) }$.
This expression involves the in-plane coupling
coefficients, ${ W_{\ell n}^{\bn} (\bJ) }$,
introduced in Eq.~\eqref{def_W}.
These coefficients also
involve the susceptibility matrix,
${ \bN_{\ell} (\omega) }$,
already presented in Eq.~\eqref{def_N},
so that the pairwise coupling is said
to be dressed by collective effects.
Thanks to the prefactors
${ y_{\ell}^{n} \!=\! Y_{\ell}^{n} (\tfrac{\pi}{2} , 0) }$,
the  $\Lambda_{\bn\bnp}^{\ell}$ coefficients
are non-zero only
for ${ |n_{2}| , |\np_{2}| \leq \ell }$,
in conjunction with ${ (\ell - n_{2}) }$
and ${ (\ell - \np_{2}) }$ both even.

\subsection{Resonance condition}
\label{sec:ResCond}

In order to compute the resonant diffusion flux
from Eq.~\eqref{def_mF},
one must solve the resonance condition
${ \bn \!\cdot\! \bO \!=\! \bnp \!\cdot\! \bOp }$,
with the shortened notation ${ \bO \!=\! \bO (\bJ) }$
and ${ \bOp \!=\! \bO (\bJp) }$.
In order to ease that calculation,
we rewrite the integral from Eq.~\eqref{def_mF} as
\begin{align}
\mF_{\bn\bnp}^{\ell} {} & = \!\! \int \!\! \rd \bJp \, G (\bJp) \, \deltaD (\bn \!\cdot\! \bO - \bnp \!\cdot\! \bOp)
\label{shape_int}
\\
{} & = \!\! \int \!\! \rd \xp \rd \ep \, \frac{G}{\Omega_{1}} \, \bigg| \frac{\p (\Ep , \Lp)}{\p (\xp , \ep)} \bigg| \, \deltaD (\bn \!\cdot\! \bO - \bnp \!\cdot\! \bOp)
\nonumber
\\
{} & = \!\! \int_{\gamma} \!\! \rd \sigma (\xp,\ep) \, \frac{G}{\Omega_{1}} \, \bigg| \frac{\p (\Ep , \Lp)}{\p (\xp , \ep)} \bigg| \, \frac{1}{|\p (\bnp \!\cdot\! \bOp)/ \p (\xp , \ep) |} ,
\nonumber
\end{align}
where the function ${ G (\bJp) }$ directly follows from Eq.~\eqref{def_mF}.
In the second line, we used ${ (x,e) \!=\! (a/\bISO,e) }$
as our orbital coordinates, following Eq.~\eqref{def_a_e},
while the Jacobian of the transformation ${ (\Ep , \Lp) \!\to\! (\xp , \ep) }$
can be obtained from Eq.~\eqref{def_a_e}.
One interest of such a writing is that,
given that ${ x , e }$ are both dimensionless,
it is straightforward to integrate along the resonant line
in these coordinates.
This is highlighted in the third line of Eq.~\eqref{shape_int},
where we introduced the resonance line $\gamma$
as the ${1D}$ line in ${ (\xp , \ep) }$ space
along which the resonance condition
${ \bnp \!\cdot\! \bOp \!=\! \bn \!\cdot\! \bO \!=\! \omega }$
is satisfied,
with the associated measure ${ \rd \sigma }$.
In that expression, we also introduced the quantity
${ |\p (\omega)/ \p (\xp , \ep)| \!=\! \sqrt{(\p \omega / \p \xp)^{2} \!+\! (\p \omega / \p \ep)^{2}} }$.

In practice, in order to estimate the integral from Eq.~\eqref{shape_int},
we must then approximate the resonance line $\gamma$.
Examples of resonant lines are given in Fig.~\ref{fig:ResLines}.
This is done by determining a set ${ \{ \xp_{i} , \ep_{i} \}_{1 \leq i \leq \Kres} }$
of resonance locations along $\gamma$, with ${ \Kres \gg 1 }$.
We detail in \S\ref{sec:IsochronePotential}
how such a line can be efficiently constructed
in the case of the isochrone potential.
We then simply replace the integral from Eq.~\eqref{shape_int},
with ${ \Kres \!-\! 1 }$ straight lines connecting the points.
As such, we perform an estimation of the form
\begin{equation}
\!\! \int_{\gamma} \!\! \rd \sigmap \, g (\xp , \ep) \simeq \sum_{i = 1}^{\mathclap{\Kres - 1}} g (\oxp_{i} , \oep_{i}) \, \Delta \sigmap_{i} ,
\label{estimate_int}
\end{equation}
where we introduced the central location ${ (\oxp_{i} , \oep_{i}) }$
and length ${ \Delta \sigmap_{i} }$
\begin{align}
\big( \oxp_{i} , \oep_{i} \big) {} & = \big( \half (\xp_{i} \!+\! \xp_{i+1}) , \half (\ep_{i} \!+\! \ep_{i + 1}) \big) ,
\nonumber
\\
\Delta \sigmap_{i} {} & = \sqrt{(\xp_{i+1} - \xp_{i})^{2} + (\ep_{i+1} - \ep_{i})^{2}} .
\label{def_approx_int}
\end{align}

Given the numerical difficulty of these calculations,
it is important to limit as much as possible the number
of resonance pairs to consider.
Let us emphasise how these may be mitigated.

First,
we note that resonance pairs with ${ \bn \!=\! (0,0) }$
or ${ \bnp \!=\! (0,0) }$
do not contribute to the diffusion,
so that we limit our sums only to the pairs
such that ${ \bn \!\neq\! (0,0) }$ and ${ \bnp \!\neq\! (0,0) }$.
From Eq.~\eqref{def_mF},
we note that the resonance pairs ${ (\bn, \bnp) }$
and ${ (-\bn,-\bnp) }$ source the exact same flux.
As a consequence, we may account for only one of the two pairs,
and add an overall factor $2$ to the total flux.

Second, for a given resonance vector $\bn$,
the associated resonance frequency is equal to
${ \bn \!\cdot\! \bO \!=\! \Omega_{1} (n_{1} \!+\! \eta \, n_{2}) }$,
where, following the notation from Eq.~\eqref{def_O2_Iso}, we
introduced the ratio ${ \eta \!=\! \Omega_{2} / \Omega_{1} }$.
In the case of an outward decreasing cored density profile
such as the isochrone potential,
one generically has ${ \half \!\leq\! \eta \!\leq\! 1 }$.
Hence, by simply computing the two values
${ (n_{1} \!+\! \half n_{2}) }$ and ${ (n_{1} \!+\! n_{2}) }$,
one can determine whether the function
${ \bJ \!\mapsto\! \bn \!\cdot\! \bO (\bJ) }$
is always positive, always negative,
or changes sign.
Given this simple criterion,
we can finally keep only resonance pairs ${ (\bn , \bnp) }$
for which the resonance condition has a chance of being satisfied
given the two associated sign constraints.

\section{Computing the Landau flux}
\label{sec:LandauEq}

When collective effects are neglected,
the \BL\ flux becomes the Landau flux.
This allows us to accelerate greatly the
computation, as we now detail.

\subsection{Bare coupling coefficients}
\label{sec:LambdaBare}

Switching off collective effects at harmonic $\ell$ is equivalent to setting the response matrix
${ \bM_\ell (\omega) \!=\! 0 }$.
 Then the coefficients ${ \Lambda_{\bn\bnp}^{\ell} }$,
from Eq.~\eqref{def_Lambda} become independent
of the temporal frequency $\omega$ and read
\begin{equation}
\Lambda_{\bn\bnp}^{\ell} (\bJ , \bJp) = y_{\ell}^{n_{2}} y_{\ell}^{\np_{2}} \, \sum_{n} W_{\ell n}^{\bn} (\bJ) \, W_{\ell n}^{\bnp} (\bJp) ,
\label{def_Lambda_Landau}
\end{equation}
where we recall that $n$ runs over the basis elements.
These simplified coefficients can in practice
be computed without resorting to any
biorthogonal basis~\citep{Chavanis2013}.
Indeed, introducing the
basis elements generically amounts to assuming that the gravitational pairwise interaction,
${ U (\br , \brp) \!=\! - G / |\br - \brp| }$, can be decomposed under the separable
form
\begin{align}
U (\br , \brp) {} & = - \sum_{\alpha} \psi^{(\alpha)} (\br) \, \psi^{(\alpha) *} (\brp)
\nonumber
\\
{} & = - \sum_{\ell , m , n} Y_{\ell}^{m} (\hbr) \, Y_{\ell}^{m *} (\hbrp) \, U_{n}^{\ell} (r) \, U_{n}^{\ell} (\rp) ,
\label{sep_U}
\end{align}
with the usual notations ${ r \!=\! |\br| }$,
and ${ \hbr \!=\! \br / r }$.
Fortunately, using the Legendre expansion of the Newtonian
interaction kernel, as well as the addition theorem for spherical
harmonics, Eq.~\eqref{sep_U} can be rewritten as
\begin{equation}
U (\br , \brp) = - \sum_{\ell , m} Y_{\ell}^{m} (\hbr) \, Y_{\ell}^{m *} (\hbrp) \, U_{\ell} (r , \rp) ,
\label{U_Leg}
\end{equation}
where we introduced the function
\begin{equation}
U_{\ell} (r , \rp) = \frac{4 \pi G}{2 \ell + 1} \, \frac{\Min [r , \rp]^{\ell}}{\Max[r , \rp]^{\ell + 1}} .
\label{def_Ul}
\end{equation}

In the limit where collective effects can be neglected,
i.e.\ the limit ${ \bN_{\ell} (\omega) \to \bI }$,
the dressed coupling coefficients from Eq.~\eqref{def_Lambda}
then naturally become
\begin{equation}
\Lambda_{\bn\bnp}^{\ell} (\bJ , \bJp) = y_{\ell}^{n_{2}} y_{\ell}^{\np_{2}} \, W_{\ell}^{\bn\bnp} (\bJ , \bJp) ,
\label{def_LambdaBare}
\end{equation}
where we introduced the coefficients ${ W_{\ell}^{\bn\bnp} (\bJ , \bJp) }$
as
\begin{align}
W_{\ell}^{\bn\bnp} {} & (\bJ , \bJp) =\!\! \int_{0}^{\pi} \!\! \frac{\rd \theta_{1}}{\pi} \frac{\rd \thetap_{1}}{\pi} \, U_{\ell} (r , \rp)
\label{def_WBare}
\\
{} & \times \cos \big( n_{1} \theta_{1} \!+\! n_{2} (\theta_{2} \!-\! \vphi) \big) \, \cos \big( \np_{1} \thetap_{1} \!+\! \np_{2} (\thetap_{2} \!-\! \vphip) \big) ,
\nonumber
\end{align}
where $r$ and ${ (\theta_{2} \!-\! \vphi) }$
are both functions of $\theta_{1}$,
and similarly for the primed variables.
One of the drawbacks of such an expression
is that  ${ W_{\ell}^{\bn\bnp} (\bJ , \bJp) }$
is not explicitly  separable anymore,
compared to Eq.~\eqref{def_Lambda}
where both angular averages factor out.
Fortunately,
such coefficients can still be computed efficiently for the Newtonian interaction potential
using a traditional multipole approach,
owing to the (almost) separable form of the integrand
from Eq.~\eqref{def_WBare}.
We briefly detail this method in \S\ref{sec:Multipole}.

\subsection{Convergence of the basis function expansion}
\label{sec:BasisFunctionExpansion}

\begin{figure}
\centering
\includegraphics[width=0.45 \textwidth]{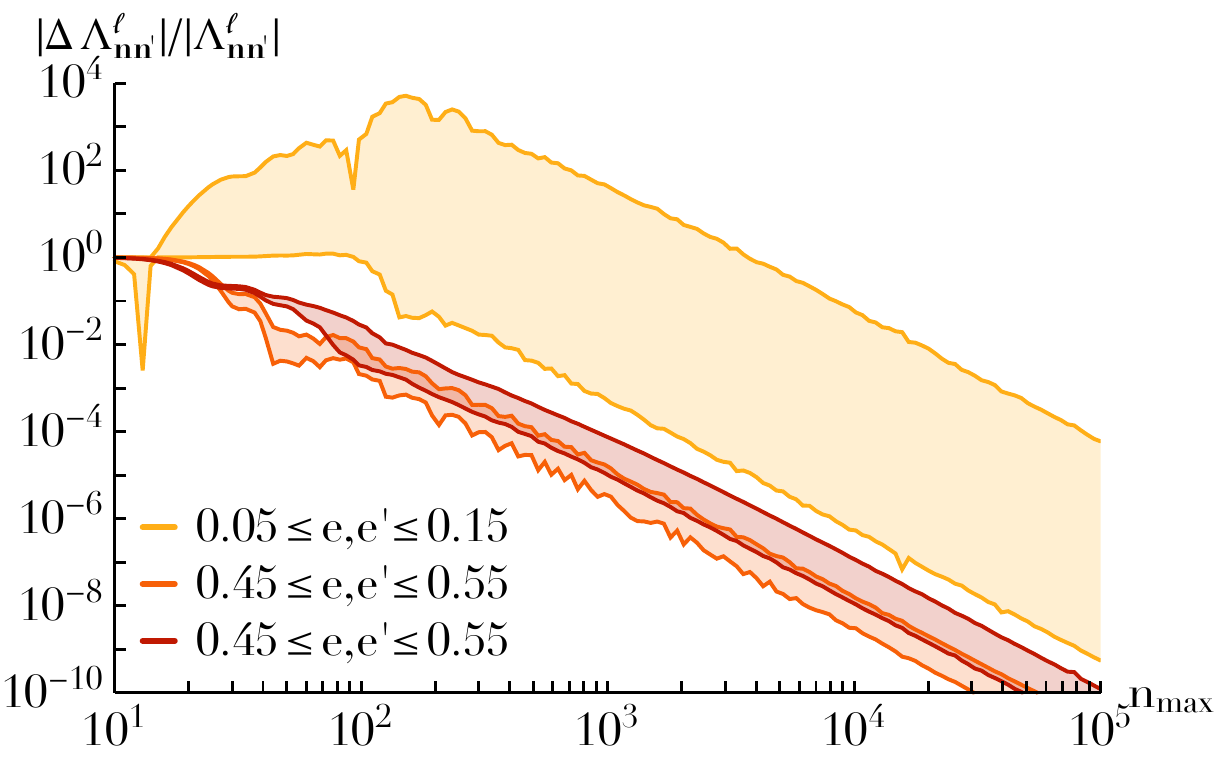}
   \caption{Illustration of the relative errors in
   the bare coupling coefficient, ${ \Lambda_{\bn\bnp}^{\ell} (\bJ , \bJp) }$,
   using the basis method as in Eq.~\eqref{def_Lambda_Landau}
   vs. the multipole expression from Eq.~\eqref{def_LambdaBare},
   as a function of the total number of basis elements, $n_{\max}$.
   Following Eq.~\eqref{rewrite_av}, the angular integrals
   were performed using ${ K \!=\! 10^{5} }$ steps,
   with ${ \ell \!=\! 1 }$, ${ \bn \!=\! (2,1) }$ and ${ \bnp \!=\! (3,1) }$,
   and a potential basis satisfying ${ \Rb \!=\! 10 \, \bISO }$.
   We considered $10^{4}$ pairs of orbits
   with ${ x \!=\! 1.0 }$ and ${ \xp \!=\! 1.2 }$,
   with the associated eccentricities, $e$ and $\ep$,
   taken uniformly within some finite range.
   Coloured regions correspond to the ${16\%}$ and ${84\%}$ levels
   among the pairs of orbits.
   We note in particular that the basis method converges
   significantly more slowly for quasi-circular orbits (in yellow).
   }
   \label{fig:ErrorLambda}
\end{figure}

In Fig.~\ref{fig:ErrorLambda}, we display
the errors in the bare coupling coefficients,
${ \Lambda_{\bn\bnp}^{\ell} (\bJ , \bJp) }$,
introduced by the finite truncation of the basis expansion.
In that figure, we note in particular that
for quasi-circular orbits,
the bare coupling coefficients
are affected by (very) significant errors
associated with the finite truncation of the number
of basis elements.

Let us now detail how one may mitigate these errors when computing the dressed
coupling coefficients as defined in Eq.~\eqref{def_Lambda}.  We assume that ${
\bN_{\ell} (\omega) \!\to\! \bI }$, for ${ p, q \!\geq\! \ncut }$, and compute the full
susceptibility matrix ${ \bN_{\ell} (\omega) }$ only for ${ 1 \!\leq\! p, q \!\leq\!
\ncut }$.  Then for the remaining $p,q  > \ncut$, we use the bare coefficients, so
that Eq.~\eqref{def_Lambda} becomes
\begin{align}
\Lambda_{\bn\bnp}^{\ell} (\bJ , \bJp , \omega) = y_{\ell}^{n_{2}} \, y_{\ell}^{\np_{2}} \bigg\{ {} & \sum_{\mathclap{p,q = 1}}^{\ncut} W_{\ell p}^{\bn} (\bJ) \, N_{pq}^{\ell} (\omega) \, W^{\bnp}_{\ell q} (\bJp)
\nonumber
\\
+ {} & \sum_{\mathclap{p = \ncut+1}}^{+\infty} W_{\ell p}^{\bn} (\bJ) \, W_{\ell p}^{\bnp} (\bJp) \bigg\} ,
\label{cut_Lambda}
\end{align}
Using Eq.~\eqref{def_Lambda_Landau}, this can be rewritten as
\begin{align}
\Lambda_{\bn\bnp}^{\ell} (\bJ , \bJp , \omega) = y_{\ell}^{n_{2}} \, y_{\ell}^{\np_{2}} \bigg\{ {} & \sum_{\mathclap{p,q = 1}}^{\ncut} W_{\ell p}^{\bn} (\bJ) \, \big[ N_{pq}^{\ell} (\omega) \!-\! \delta_{pq} \big] W_{\ell q}^{\bnp} (\bJp)
\nonumber
\\
+ {} & W_{\ell}^{\bn\bnp} (\bJ , \bJp) \bigg\} ,
\label{cut_Lambda_II}
\end{align}
where, importantly, the last term is obtained
through the multipole expression from Eq.~\eqref{def_WBare},
that does not require any basis elements.
We used such an expression to compute
the \BL\ fluxes in Fig.~\ref{fig:dFdtBL}.

\subsection{Convergence of the resonance truncation}

The \RR\ flux from Eq.~\eqref{sum_bF}
involves a sum over two resonance vectors
${ \bn , \bnp }$.
As highlighted in Eq.~\eqref{def_Lambda},
for a given harmonic number $\ell$,
the two resonance numbers $n_{2}$ and ${ \np_{2} }$
satisfy ${ |n_{2}| , |\np_{2}| \!\leq\! \ell }$.
Yet, there are no such constraints for other resonance
numbers $n_{1}$ and $\np_{1}$.
In practice, we mitigate this issue by limiting
ourselves to ${ |n_{1}| , |\np_{1}| \!\leq\! n_{1}^{\max} }$,
where $n_{1}^{\max}$ is a given threshold.

\begin{figure}
   \centering
  \includegraphics[width=0.45 \textwidth]{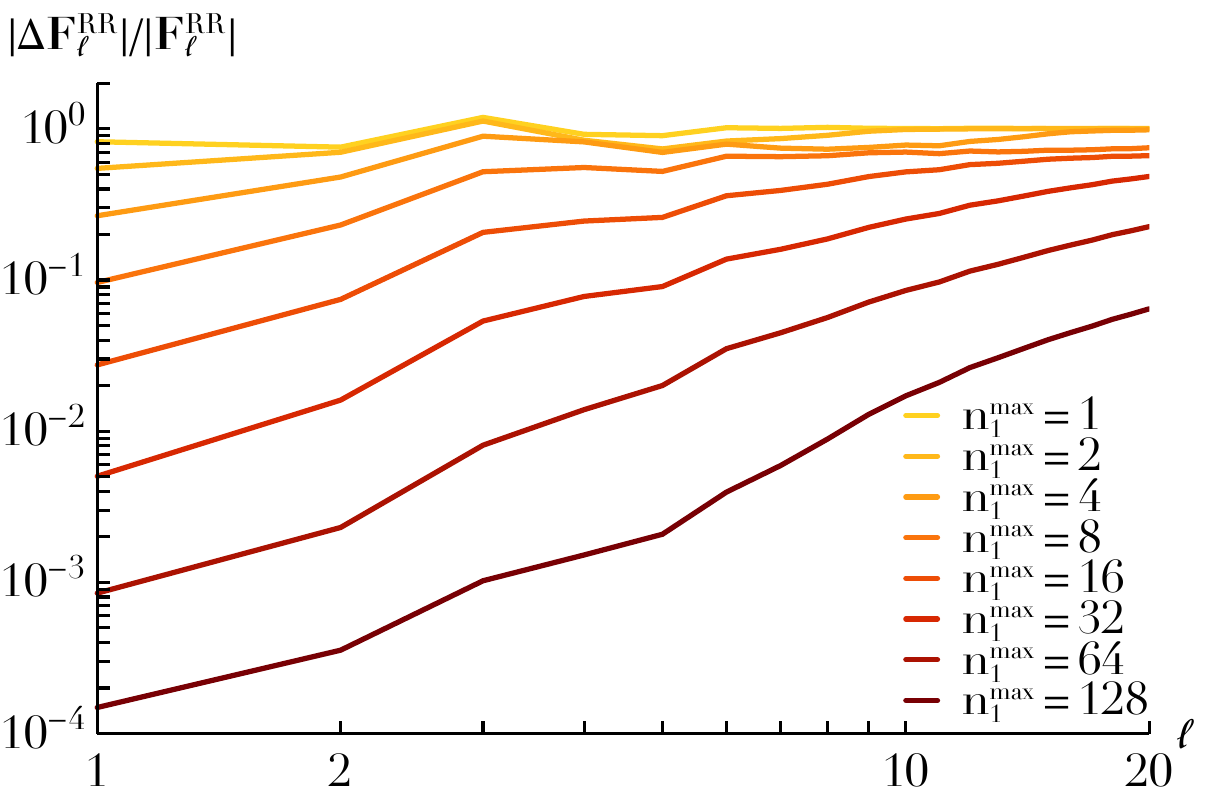}
  \caption{Illustration of the relative error in $\bFRR^{\ell}$,
  computed at the same orbital location as in Fig.~\ref{fig:Coulomb},
  as a function of the maximum resonance number, $n_{1}^{\max}$,
  and the considered harmonics $\ell$.
  Here, we used ${ K \!=\! 512 }$ nodes to compute
  the orbit-average and ${ \Kres \!=\! 512 }$ points
  to construct the resonance line.
  Relative errors are computed by comparison
  to a calculation with ${ n_{1}^{\max} \!=\! 256 }$.
  The larger $\ell$ is, the more high-order resonances contribute.
  }
  \label{fig:Coulombn1max}
\end{figure}

In Fig.~\ref{fig:Coulombn1max},
we explore the effect of this truncation
on the value of the bare \RR\ diffusion flux $\bFRR^{\ell}$.
First we computed the flux $\bFRR^{\ell}$ 
at the same orbital location as in Fig.~\ref{fig:Coulomb} and using ${ n_{1}^{\max} \!=\! 256 }$.
Different coloured lines in Fig.~\ref{fig:Coulombn1max} show the relative error w.r.t.\ this measurement 
that is induced by degrading the calculation of $\bFRR^{\ell}$
to smaller values of $n_{1}^{\max}$.
As expected, the larger $\ell$ is,
the larger $n_{1}^{\max}$ must be
for an accurate calculation of the diffusion flux.
In practice, it appears that picking ${n_{1}^{\max} \gtrsim 8 \!+\! 4 \ell}$
allows for a computation of the flux with a relative error
of the order of ${10 \%}$,
highlighting the role played by high-order resonances
in driving the \RR\ relaxation.
More important conceptually is the fact that,
at fixed $\ell$,
the infinite sum over $n_{1}$ and $\np_{1}$
in Eq.~\eqref{fig:Coulombn1max} converges,
in sharp constrast with the logarithmic divergence
w.r.t.\ the harmonic number $\ell$
highlighted in Fig.~\ref{fig:Coulomb}.

\subsection{Multipole expansion}
\label{sec:Multipole}

We now detail how the bare
coupling coefficients, ${ W_{\ell}^{\bn\bnp} (\bJ , \bJp) }$, from Eq.~\eqref{def_WBare}
may be efficiently computed using a multipole approach.
First, in order not to have to invert the implicit relation
${ \theta_{1} \!=\! \theta_{1} (r) }$ (see Eq.~\eqref{def_angles}),
and to avoid boundary divergences at the edge of the integration
domain where the radial velocity vanishes,
we rely on the same effective anomaly, ${ r \!=\! r (u) }$,
as in Eq.~\eqref{def_u}.
Equation~\eqref{def_WBare} then becomes
\begin{equation}
W_{\ell}^{\bn\bnp} (\bJ , \bJp) = \frac{4}{\pi^{2}} \!\! \int_{-1}^{1} \!\! \frac{\rd u}{2} \frac{\rd \up}{2} \, g (r) \, \gp (\rp) \, U_{\ell} (r , \rp) ,
\label{WBare_short}
\end{equation}
where we introduced the function
\begin{equation}
g (r) = \frac{\rd \theta_{1}}{\rd u} \, \cos \big( n_{1} \theta_{1} + n_{2} (\theta_{2} - \vphi) \big) ,
\label{def_g}
\end{equation}
and similarly for ${ \gp (\rp) }$.

To perform the two integrals from Eq.~\eqref{WBare_short},
we now sample uniformly each interval using $K$ nodes.
Specifically, for ${ 1 \!\leq\! k \!\leq\! K }$,
we sample the anomaly $u$ with
\begin{equation}
u_{k} = - 1 + \Delta u \, \big( k - \half \big)
\quad \text{ with } \quad
\Delta u = \frac{2}{K} .
\label{sampling_u}
\end{equation}
Following this discretisation, Eq.~\eqref{WBare_short} becomes
\begin{equation}
W_{\ell}^{\bn\bnp} \!\!=\! \frac{16 G }{\pi (2 \ell + 1)} \frac{1}{K^{2}} \sum_{i , j} g_{i} \, \gp_{j} \, \frac{\Min[r_{i} , \rp_{j}]^{\ell}}{\Max[r_{i} , \rp_{j}]^{\ell + 1}} ,
\label{WBare_discrete}
\end{equation}
where we used the shortened notation ${ g_{i} \!=\! g (r_{i}) }$.

Let us now use the particular structure of Eq.~\eqref{WBare_discrete}
to accelerate its evaluation.
To proceed forward, we order the set of radii ${ \{ r_{i} , \rp_{j}  \} }$
by increasing order.
We emphasise that this can be made in ${ \mO (K) }$ steps,
as the two sets ${ \{ r_{i} \} }$ and ${ \{ \rp_{j} \} }$
are already ordered, so that it only remains to merge the two lists.
Following this ordering, we construct the array $w_{j}$
which, for ${1 \!\leq\! j \!\leq\! K}$, is defined as
\begin{equation}
w_{j} = \Card \bigg\{ i \in \big\{ 1, ... , K \big\} \, \bigg| \, r_{i} \leq \rp_{j} \bigg\} ,
\label{def_w}
\end{equation}
with the boundary terms ${ w_{0} \!=\! 0 }$ and ${ w_{K + 1} \!=\! K }$.
We can now rewrite the double sum from Eq.~\eqref{WBare_discrete} as
\begin{equation}
W_{\ell}^{\bn\bnp} \!\!=\! \frac{16 G}{\pi (2 \ell + 1)} \frac{1}{K^{2}} \sum_{j = 1}^{K} \gp_{j} \big[ P_{j} + Q_{j} \big] ,
\label{rewrite_WBare_discrete}
\end{equation}
where we introduced
\begin{equation}
P_{j} = \sum_{i = 1}^{w_{j}} g_{i} \frac{r_{i}^{\ell}}{r_{j}^{\prime \ell + 1}} ;
\quad
Q_{j} = \sum_{\mathclap{i = w_{j}+1}}^{K} g_{i} \frac{r_{j}^{\prime \ell}}{r_{i}^{\ell + 1}} .
\label{def_P_Q}
\end{equation}
Here, it is essential to note that both ${ \{P_{j} \}_{1\leq j \leq K} }$
and ${ \{Q_{j}\}_{1\leq j \leq K} }$ series
can be computed with a complexity scaling linearly with $K$.
To highlight this point, we define for ${ 1 \!\leq\! j \!\leq\! K }$,
the partial sums
\begin{equation}
\delta P_{j} = \sum_{\mathclap{i = w_{j-1}+1}}^{w_{j}} g_{i} \, \frac{r_{i}^{\ell}}{r_{j}^{\prime \ell + 1}} ;
\quad
\delta Q_{j} = \sum_{\mathclap{i = w_{j} + 1 }}^{w_{j+1}} g_{i} \, \frac{r_{j}^{\prime \ell}}{r_{i}^{\ell + 1}} ,
\label{def_deltaP_deltaQ}
\end{equation}
which satisfy the recurrence relations
\begin{align}
P_{1} {} & = \delta P_{1} ;
\quad
P_{j+1} = \bigg[ \frac{\rp_{j}}{\rp_{j+1}} \bigg]^{\ell + 1} \, P_{j} + \delta P_{j+1} ,
\nonumber
\\
Q_{K} {} & = \delta Q_{K} ;
\quad
Q_{j - 1} = \bigg[ \frac{\rp_{j-1}}{\rp_{j}} \bigg]^{\ell} \, Q_{j} + \delta Q_{j - 1} .
\label{recc_P_Q}
\end{align}
Owing to these explicit recurrence relations,
we are in a position to compute
the bare coupling coefficients,
${ W_{\ell}^{\bn\bnp} (\bJ , \bJp) }$,
with a complexity in ${ \mO (K) }$.

As a closing remark,
let us detail a bit more the preparation of Eq.~\eqref{WBare_discrete}.
In that equation,
one must compute ${ \theta_{1} [u_{k}] }$
and ${ (\theta_{2} \!-\! \vphi) [u_{k}] }$.
This is done using the same method
as in Eq.~\eqref{rewrite_av},
i.e.\ using a RK4 integration of each expression.
In practice, the initial value of the various
integrals are obtained through a first `warm-up'
starting from ${ u \!=\! -1 }$ with one RK4 step
of length ${ \Delta u / 2 }$.

In practice, for the maps presented in Fig.~\ref{fig:dFdtLandauMultipole},
we considered the resonances up to ${ |n_{1}|, |\np_{1}| \!\leq\! 40 }$
contained within the domain ${ 0 \!\leq\! x \!\leq\! \xmax \!=\! 10 }$.
The orbit-average was performed with ${ K \!=\! 200 }$ points,
and the resonance lines constructed with ${ \Kres \!=\! 200 }$ points.

\subsection{Softened bare coupling coefficients}
\label{sec:LambdaBareSoftened}

It is also possible to obtain 
the bare coupling coefficients associated
with a softened pairwise interaction
of the form
\begin{equation}
U (\br , \brp , \veps) = - \frac{G}{\sqrt{|\br - \brp|^{2} + \veps^{2}}} ,
\label{Usoft}
\end{equation}
with $\veps$ the considered softening length.
Following \S{B} of~\cite{Weinberg1986}
(see also~\cite{Wachlin2006}),
in the case of a softened interaction,
Eq.~\eqref{def_Ul} becomes
\begin{equation}
U_{\ell} (r , \rp , \veps) = \frac{4 \pi G}{2 \ell + 1} \frac{r_{\beta}^{\ell}}{r_{\alpha}^{\ell + 1}} ,
\label{def_Ul_soft}
\end{equation}
where we introduced the notations
\begin{align}
\ralpha {} & = \!\bigg[\! \half \bigg(\! r^{2} \!+\! r^{\prime 2} \!+\! \veps^{2} \!+\! \sqrt{ \!\big( (r \!+\! \rp)^{2} \!+\! \veps^{2} \big)  \big( (r \!-\! \rp)^{2} \!+\! \veps^{2} \big)} \bigg) \!\bigg]^{1/2} ,
\nonumber
\\
\rbeta {} & = \frac{r \, \rp}{\ralpha} .
\label{def_ralpha_rbeta}
\end{align}
From Eq.~\eqref{def_ralpha_rbeta},
one can immediately recover the unsoftened limit
presented in Eq.~\eqref{def_Ul}.
Unfortunately, the expression from Eq.~\eqref{def_Ul_soft}
is far from being (almost) separable,
so that it cannot benefit from the fast evaluation
permitted by the multipole approach
from \S\ref{sec:Multipole}.

We finally note that,
even for the softened interaction kernel
from Eq.~\eqref{def_Ul_soft},
one can perform the same asymptotic expansion as in Eq.~\eqref{shape_U_Leg},
as already detailed in Eq.~{(B8)} of~\cite{Weinberg1986}.
As such, let us assume that ${ r \!=\! \bISO }$, ${ \rp \!=\! \bISO (1 \!-\! \alpha) }$,
with ${ \alpha \!>\! 0 }$.
Assuming that ${ \mO (\alpha) \!\simeq\! \mO (\veps) }$,
one can write the following expansions
\begin{align}
\ralpha {} & \simeq \bISO + \mO (\veps) ,
\nonumber
\\
\frac{\rbeta}{\ralpha} {} & \simeq 1 - \sqrt{\alpha^{2} + (\veps/\bISO)^{2}} + \mO (\veps^{2}) .
\label{expansion_ratio_softened}
\end{align} 
As a consequence, in the limit ${ \alpha , \veps \!\ll\! 1 }$ and ${ \ell \!\geq\! 1 }$,
one can expand Eq.~\eqref{def_Ul_soft} as
\begin{align}
U_{\ell} (\alpha) {} & \propto \frac{1}{\bISO} \, \Big( 1 - \sqrt{\alpha^{2} + (\veps / \bISO)^{2}} \Big)^{\ell}
\nonumber
\\
{} & \simeq \frac{1}{\bISO} \, \re^{ - \ell \sqrt{\alpha^{2} + (\veps/\bISO)^{2}}} .
\label{asymp_softened}
\end{align}
This is the direct equivalent of Eq.~\eqref{asymp_U_Leg}
in the case of a softened interaction.
In particular, we note that for interparticle separations, $\alpha$,
smaller than the softening length, $\veps$,
the pairwise coupling tends to a constant value.
This prevents the system
from sustaining any relaxation
on scales smaller than the softening scale.

\section{Isochrone potential}
\label{sec:IsochronePotential}

In this section, we follow~\cite{Henon1959},
and recall some of the key analytical expressions
of the isochrone potential
used throughout the paper.
It is defined as
\begin{equation}
\psi (r) = - \frac{G M}{\bISO + \sqrt{\bISO^{2} + r^{2}}} ,
\label{def_psi_iso}
\end{equation}
with $M$ the system's total active mass,
and $\bISO$ its lengthscale.
In practice, for all the numerical applications,
we pick units so that ${ G \!=\! M \!=\! \bISO \!=\! 1 }$.
The isochrone Hamiltonian can be explicitly written
as a function of the action coordinates.
It reads
\begin{equation}
H (\bJ) = - \frac{(G M)^{2}}{2 {[J_{r} + \half (L + \sqrt{L^{2} + 4 G M \bISO})]}^{2}} .
\label{H_iso}
\end{equation}
Fortunately, that same expression
also provides us with an explicit inversion
of the expression of the radial action,
so that
\begin{equation}
J_{r} = \frac{G M}{\sqrt{- 2 E}} - \frac{1}{2} \bigg( L + \sqrt{L^{2} + 4 G M \bISO} \bigg) .
\label{Jr_iso}
\end{equation}
The radial frequency is given by
\begin{equation}
\Omega_{1} = \omega (a , e) \, \Omega_{0},
\label{def_O1_Iso}
\end{equation}
with the frequency scale ${ \Omega_{0} \!=\! \sqrt{G M / \bISO^{3}} }$.
In Eq.~\eqref{def_O1_Iso}, we introduced the dimensionless function
\begin{equation}
\omega(a , e) = \bigg( \frac{E}{\Emin} \bigg)^{3/2}
\!\! = \bigg( \frac{2}{\speri \!+\! \sapo} \bigg)^{3/2} ,
\label{def_omega_Iso}
\end{equation}
where we introduced
${ \speri \!=\! \sqrt{1 \!+\! \xperi^{2}} }$ (similarly for $\sapo$)
with the dimensionless pericentre, ${ \xperi \!=\! \rperi/\bISO }$,
as well as the minimum energy ${ \Emin \!=\! - G M / (2 \bISO) }$.
The azimuthal frequency is given by
\begin{equation}
\Omega_{2} = \omega (a , e) \, \eta (a , e) \, \Omega_{0} ,
\label{def_O2_Iso}
\end{equation}
where we introduced the frequency ratio ${ \eta \!=\! \Omega_{2} / \Omega_{1} }$.
In the isochrone case, it follows the explicit form
\begin{align}
\eta (a , e) {} & = \frac{1}{2} \bigg( 1+ \frac{L}{\sqrt{L^{2} + 4 G M \bISO}} \bigg)
\label{def_eta_Iso}
\\
{} & = \frac{1}{2} \bigg( 1 \!+\! \frac{\xperi \xapo}{\big(1 + \speri \big) \big( 1 + \sapo \big)} \bigg) .
\nonumber
\end{align}
We note that along circular (resp.\ radial) orbits,
i.e.\ for ${ e \!\to\! 0 }$ (resp.\ ${ e \!\to\! 1 }$),
the isochrone frequencies take the simple forms
\begin{align}
\begin{cases}
\displaystyle \omegacirc (x) = \bigg(\! \frac{1}{\sqrt{1 \!+\! x^{2}}} \!\bigg)^{3/2} ,
\\[2.0ex]
\displaystyle \etacirc (x) = \frac{\sqrt{1 \!+\! x^{2}}}{1 \!+\! \sqrt{1 \!+\! x^{2}}} ,
\end{cases}
\!\!\!
\begin{cases}
\displaystyle \omegarad (x) = \bigg(\! \frac{2}{1 \!+\! \sqrt{1 \!+\! 4 x^{2}}} \!\bigg)^{3/2} \!\! ,
\\[2.0ex]
\displaystyle \etarad (x) = \frac{1}{2} . \label{circ_rad_limits}
\end{cases}
\end{align}

In the specific case of the isochrone potential,
one can also get numerically well-posed expressions
for ${ E \!=\! E (\rperi, \rapo) }$ and ${ L \!=\! L (\rperi,\rapo) }$
from Eq.~\eqref{EL_rpra}. They read
\begin{equation}
E = \frac{E_{0}}{\speri \!+\! \sapo} ;
\quad
L = \sqrt{2} \, L_{0} \, \frac{\xperi \xapo}{\sqrt{(1 \!+\! \speri) (1 \!+\! \sapo) (\speri \!+\! \sapo)}} ,
\label{EL_rpra_ISO}
\end{equation}
with the energy scale ${ E_{0} \!=\! - G M / \bISO }$,
and the action scale ${ L_{0} \!=\! \sqrt{G M \bISO} }$.
These explicit expressions finally allow us to obtain
exact expressions for the Jacobian ${ \rd \theta_{1} / \rd u }$
appearing in Eq.~\eqref{rewrite_av}
\begin{equation}
\frac{\rd \theta_{1}}{\rd u} = \frac{3}{\sqrt{2}} \frac{\Omega_{1}}{\Omega_{0}} \, \frac{x_{r}}{\sqrt{4 \!-\! u^{2}}} \, \frac{\sqrt{(s_{r} \!+\! \speri) (s_{r} \!+\! \sapo) (\speri \!+\! \sapo)}}{\sqrt{(x_{r} \!+\! \xperi) (x_{r} \!+\! \xapo)}} ,
\label{nice_edge_iso}
\end{equation}
where we introduced ${ x_{r} \!=\! r/b }$,
and ${ s_{r} \!=\! \sqrt{1 \!+\! x_{r}^{2}} }$.
Importantly, we note that this expression
is numerically well-behaved for any ${ -1 \!\leq\! u \!\leq\! 1 }$.

Following Eq.~{(4.54)} of~\cite{BinneyTremaine2008},
the isotropic \DF\ of the isochrone potential reads
\begin{align}
\Ftot (E) = {} & \, \frac{M}{(G M \bISO)^{3/2}} \, \frac{1}{128 \sqrt{2} \, \pi^{3}} \, \frac{\sqrt{\mE}}{(1 \!-\! \mE)^{4}}
\nonumber
\\
\times {} & \, \bigg[ 27 \!-\! 66 \mE \!+\! 320 \mE^{2} \!-\! 240 \mE^{3} + 64 \mE^{4}
\nonumber
\\
{} & + \frac{3 \sin^{-1} (\sqrt{\mE})}{\sqrt{\mE (1 \!-\! \mE)}} (- 9 \!+\! 28 \mE \!+\! 16 \mE^{2}) \bigg] . 
\label{isoDF_iso}
\end{align}
where we introduced the rescaled energy ${ \mE \!=\! E / E_{0} }$.
Owing to Eq.~\eqref{isoDF_iso},
one can compute all the gradients ${ \p F / \p \bJ }$
that appear both in the response matrix from Eq.~\eqref{def_M}
and in the \RR\ diffusion flux from Eq.~\eqref{def_mF}.

In \S\ref{sec:Comp},
we validate our implementation of the response matrix
by recovering the radial orbit instability in a radially
anisotropic isochrone cluster following~\cite{Saha1991}.
In that case, we consider an anisotropic \DF\ defined as
\begin{align}
{} & \Ftot (Q) = \frac{M}{(G M \bISO)^{3/2}} \frac{1}{128 \sqrt{2} \, \pi^{3}} \frac{\sqrt{Q}}{(1 \!-\! Q)^{4}} 
\label{anisoDF_iso}
\\
\times {} & \bigg\{ 27 \!+\! 77 \gamma \!-\! (66 \!+\! 286 \gamma) Q \!+\! (320 \!+\! 136 \gamma) Q^{2}
\nonumber
\\
{} & \!-\! (240 \!+\! 32 \gamma) Q^{3} \!+\! 64 Q^{4}
\nonumber
\\
+ {} & \frac{3 \sin^{-1} \big( \sqrt{Q} \big)}{\sqrt{Q (1 \!-\! Q)}} \big[ (- 9 \!+\! 17 \gamma) \!+\! (28 \!-\! 44 \gamma)Q \!+\! (16 \!-\! 8 \gamma) Q^{2} \big] \bigg\} ,
\nonumber
\end{align}
where we introduced
\begin{equation}
Q = \frac{1}{E_{0}} \bigg( E + \frac{L^{2}}{2 \Ra^{2}} \bigg) ;
\quad
\gamma = \bigg( \frac{\bISO}{\Ra} \bigg)^{2} .
\label{def_Q_gamma}
\end{equation}
Here $\Ra$ is the so-called \textit{anisotropy radius}.  Stars orbiting at radii
much smaller than $\Ra$ tend to have isotropically distributed velocities while
stars at radii much larger than $\Ra$ are nearly all on highly radial orbits.
In the limit ${ \Ra \!\to\! + \infty }$,
one has ${ Q \!\to\! \mE }$ and ${ \gamma \!\to\! 0 }$,
so that Eq.~\eqref{anisoDF_iso} reduces to Eq.~\eqref{isoDF_iso}.

In order to compute the resonant flux from Eq.~\eqref{def_mF},
one has to compute a resonance condition of the form
${ \deltaD (\bn \!\cdot\! \bO (\bJ) \!-\! \bnp \!\cdot\! \bO (\bJp)) }$.
As already defined in Eq.~\eqref{shape_int},
this amounts to finding all the resonant locations ${ (\xp , \ep) }$
such that the resonance condition ${ \bnp \!\cdot\! \bO (\xp , \ep) \!=\! \bn \!\cdot\! \bO (\bJ) }$ is satisfied. In conjunction with the computation of the response matrix,
this is one of the most cumbersome tasks in the estimation
of the \RR\ diffusion flux.
Fortunately, such a search can be eased in the case
of the isochrone potential,
owing to the explicit expressions of the
associated orbital frequencies
obtained in Eqs.~\eqref{def_omega_Iso} and~\eqref{def_eta_Iso}.
Let us now briefly detail our scheme to construct
the system's resonance lines.

Following Eq.~\eqref{def_mF},
a resonance is characterised by a resonance vector
${ \bnp \!=\! (\np_{1} , \np_{2}) }$.
Following Eqs.~\eqref{def_O1_Iso} and~\eqref{def_O2_Iso},
the associated resonance condition reads
\begin{equation}
\omegares (\xp , \ep) = \varpi ,
\label{res_cond}
\end{equation}
where we introduced the
resonance frequency ${ \omegares \!=\! \omega (\np_{1} \!+\! \np_{2} \eta) }$,
as well as the rescaled frequency
${ \varpi \!=\! \bn \!\cdot\! \bO (\bJ) / \Omega_{0} }$.

First, we compute the quantity
\begin{equation}
\nu \equiv \np_{1} + \half \np_{2} .
\label{def_nu}
\end{equation}
Owing to the simple expression of $\etarad$
from Eq.~\eqref{circ_rad_limits},
we have the inequality ${ |\omegares (x , e \!=\! 1)| \leq |\nu| }$,
and the function ${ \xp \!\mapsto\! \omega(\xp, e \!=\! 1) }$
is a monotonic function.
As a consequence, dealing appropriately with the case
${ \nu \!=\! 0 }$, we may conclude that the resonance line
goes up to the radial orbit if one has
\begin{equation}
 0 \leq \frac{\varpi}{\nu} \leq 1 .
\label{ineq_nu}
\end{equation}
If this constraint is satisfied,
the resonance line reaches radial orbits
for ${ \xp \!=\! \xprad }$
so that ${ \omegares (\xprad , \ep) \!=\! \varpi }$
with the explicit expression
\begin{equation}
\xprad = \frac{\sqrt{1 - (\varpi / \nu)^{2/3}}}{(\varpi/\nu)^{2/3}} .
\label{exp_xprad}
\end{equation}

Having determined whether or not the resonance line
reaches the radial orbits,
we must now consider how it reaches the circular orbit.
Along circular orbits,
the resonance frequency takes the form
\begin{equation}
\omegares (\xp , \ep \!=\! 0) = \frac{1}{q^{3/2}} \bigg[ \np_{1} + \np_{2} \, \frac{q}{1 + q} \bigg] \equiv h(q) ,
\label{omegares_circ}
\end{equation}
where we introduced ${ q \!=\! \sqrt{1 + x^{2}} }$.
We note therefore that the boundary terms
are given by
${ \omegares (\xp \!=\! 0 , \ep \!=\! 0) \!=\! \nu }$
and
${ \omegares (\xp \!=\! + \infty , \ep \!=\! 0) \!=\! 0 }$.
Since these are the same bounds
as in Eq.~\eqref{ineq_nu},
we conclude that any resonance line that reaches
the radial orbits necessarily reaches the circular orbits.

For some resonance vector, ${ (\np_{1}, \np_{2}) }$,
the function ${ q \!\mapsto\! h (q) }$ might not be monotonic.
Yet, we have the systematic bound
${ | h(q)| \!\leq\! (|\np_{1} \!+\! \np_{2}|)/q^{3/2} }$,
so that introducing
\begin{equation}
\qb = \bigg( \frac{2 (|\np_{1}| \!+\! |\np_{2}|)}{|\varpi|} \bigg)^{2/3} ,
\label{def_qmax}
\end{equation}
one gets that ${ q \!\leq\! \qb }$ implies ${ |h (q)| \!\leq\! |\varpi| / 2 }$.
In the case where Eq.~\eqref{ineq_nu} is satisfied,
such a bound provides us with an explicit
interval within which to perform a bisection search
in order to obtain the location at which the resonance
line intersects the circular orbits, $\xpcirc$.
Once ${ (\xpcirc , \xprad) }$ have been determined,
we sample uniformly the range ${ 0 \!\leq\! \ep \!\leq\! 1 }$
with $\Kres$ points,
and we use bisection searches to identify precisely
the resonance locations.
In practice, we also ensure that the resonant search
is constrained to the domain ${ 0 \!\leq\! \xp \!\leq\! \xmax }$.

Even if the condition from Eq.~\eqref{ineq_nu}
is not met, the system can still support
a resonance line that would not reach radial orbits,
but would rather connect two circular orbits.
To find such lines, we follow Eq.~\eqref{omegares_circ}
and write
\begin{equation}
\frac{\rd h}{\rd q} = \frac{P (q)}{2 q^{5/2} (1 + q)^{2}} ,
\label{calc_dhdq}
\end{equation}
where we introduced the second-order polynomial
\begin{equation}
P (q) = - 3 (\np_{1} \!+\! \np_{2}) q^{2} - (6 \np_{1} + \np_{2}) q - 3 \np_{1} .
\label{def_Pq}
\end{equation}
We note that if ${ \np_{2} \!=\! - \np_{1} }$,
${ P (q) }$ becomes linear in $q$.
In the range ${ 1 \!\leq\! q \!\leq\! + \infty }$
this function is of constant sign,
i.e.\ the function ${ h (q) }$ is monotonic.
As a consequence, for such resonances,
we may use the exact same criteria as in Eq.~\eqref{ineq_nu},
and all the resonance lines connect the circular orbits
to the radial ones.

Let us then assume ${ \np_{2} \!\neq\! \np_{1} }$.
The polynomial ${ P(q) }$ from Eq.~\eqref{def_Pq}
is then a true second-order polynomial,
and its discriminant reads
\begin{equation}
\Delta = \np_{2} (\np_{2} - 24 \np_{1}) .
\label{def_Delta_P}
\end{equation}

If ${ \Delta \!\leq\! 0 }$, ${ P(q) }$ does not change sign.
As a consequence, ${ h(q) }$ is monotonic.
The criteria from Eq.~\eqref{ineq_nu} applies again,
and all the resonance lines connect the circular orbits
to the radial ones.

If ${ \Delta \!>\! 0 }$,
${ P (q) }$ has two roots, i.e.\ it changes of sign.
It only remains to determine whether or not this change
of sign occurs within the domain ${ 1 \!\leq\! q \!<\! + \infty }$.
The two roots of ${ P(q) }$ are given by
\begin{equation}
q_{\pm} = \frac{-(6 \np_{1} + \np_{2}) \pm \sqrt{\Delta}}{6(\np_{1} + \np_{2})} ,
\label{def_qpm}
\end{equation}
which are subsequently ordered as
\begin{equation}
\qmin = \Min [q_{-} , q_{+}] ;
\quad
\qmax = \Max [q_{-} , q_{+}] .
\label{def_qmin_qmax}
\end{equation}
It is only if ${ 1 \!<\! \qmax }$
that the system can support resonance line
joining two circular orbits.
In that case, we use a bisection search
to identify these two radii $\xpmin$ and $\xpmax$
where the resonance condition is met.
We then sample uniformly the range
${ \xpmin \!\leq\! \xp \!\leq\! \xpmax }$
with $\Kres$ points
and we use bisection searches
to identify precisely the resonance locations.
In practice, we also enforce the additional
constraint that the resonance line
is limited to the domain ${ 0 \!\leq\! \xp \!\leq\! \xmax }$.

\begin{figure}
    \centering
   \includegraphics[width=0.45 \textwidth]{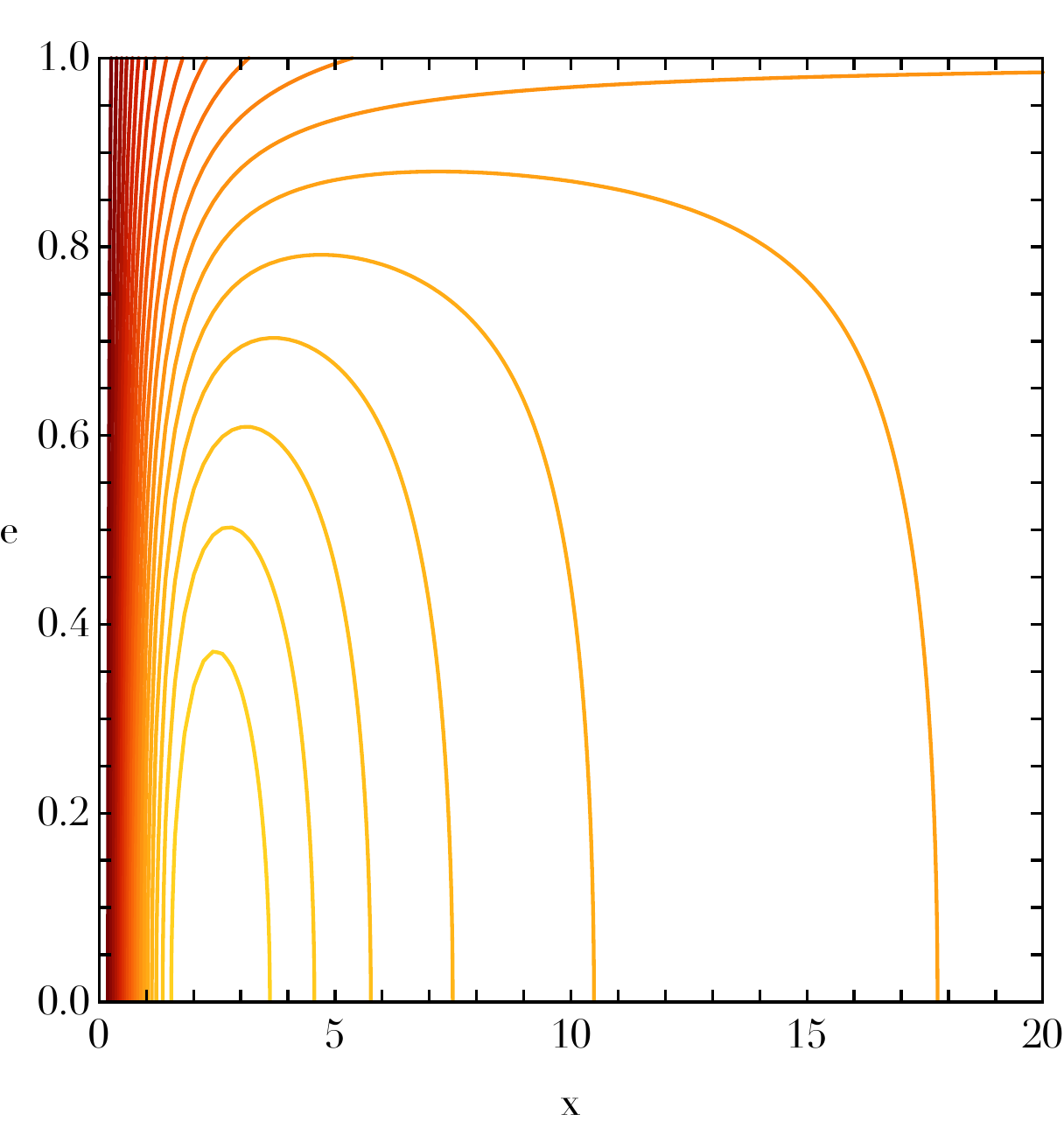}
   \caption{Illustration of the level lines of the resonance frequency
   ${ \omegares \!=\! \bn \!\cdot\! \bO (x , e) }$
   for an isochrone cluster,
   considering the resonance vector ${ \bn \!=\! (4,-7) }$.
   In particular, one can note that there exist two types
   of resonance lines, depending on whether or not
   they reach radial orbits (${ e \!=\! 1 }$). 
   }
   \label{fig:ResLines}
\end{figure}

To conclude this Appendix,
we briefly illustrate in Fig.~\ref{fig:ResLines}
an example of resonance lines,
where one can note, as previously discussed,
the presence of two types of resonance lines
depending on whether or not they reach radial orbits.

\section{Numerical simulations}
\label{sec:NumericalSimulations}

In this Appendix, we briefly detail the properties
of our numerical simulations.

In order to ease the description of each of these setups,
for this Appendix, units are displayed
in H\'enon units ($\HU$)~\citep{Henon1971}.
For the isochrone potential from Eq.~\eqref{def_psi_iso},
the virial theorem gives the system's velocity dispersion
as~\citep[see, e.g., Eq.~{(B36)} in][]{Hamilton+2018}
\begin{equation}
\sigma^{2} = \frac{G M}{\bISO} \bigg( \frac{\pi}{4} - \frac{2}{3} \bigg) .
\label{sigma_iso}
\end{equation}
As a result, the lengthscale
of the Henon units is given by
\begin{equation}
\rH = \frac{6}{3 \pi - 8} \, \bISO \simeq 4.21 \, \bISO .
\label{def_rh}
\end{equation}

In order to sample the cluster's initial condition,
we adapted the publicly available code \texttt{PlummerPlus}\footnote{\url{https://github.com/pgbreen/PlummerPlus}}
to the case of an isotropic isochrone cluster.
All simulations were performed with a total of
${ N \!=\! 10^{5} }$ particles of equal mass.

\subsection{Collisional simulations}
\label{sec:NBODY6SIM}

In order to simulate the dynamics of the globular cluster
as driven by the (unsoftened) Newtonian interaction,
we used the direct code \texttt{NBODY6++GPU}~\citep{Wang+2015}, with the control parameters
\begin{align}
{} & \texttt{NNBOPT} \!=\! 400;
\;
\texttt{ETAI} \!=\! 0.02;
\;
\texttt{ETAR} \!=\! 0.01;
\nonumber
\\
{} & \texttt{RS0} \!=\! 0.10;
\;
\texttt{DTADJ} \!=\! 2.0;
\;
\texttt{QE} \!=\! 2 \!\times\! 10^{-4} ;
\nonumber
\\
{} & \texttt{DTMIN} \!=\! 10^{-5};
\;
\texttt{RMIN} \!=\! 2 \!\times\! 10^{-4};
\;
\texttt{ETAU} \!=\! 0.1 ;
\label{params_NBODY}
\\
{} & \texttt{ECLOSE} \!=\! 1.0;
\;
\texttt{GMIN} \!=\! 10^{-6};
\;
\texttt{GMAX} \!=\! 0.01 ;
\;
\texttt{SMAX} \!=\! 0.125.
\nonumber
\end{align}
Each individual realisation was run on a node
with a single GPU and a 40-core CPU.
Integrating one realisation up to ${ \tmax \!=\! 10^{3} \, \HU }$
required about $26$h of computation.
We performed a total of ${ \Nreal \!=\! 10^{2} }$
different realisations.

\subsection{Collisionless simulations}
\label{sec:gyrfalconSIM}

In order to simulate the dynamics of the globular cluster
as driven by the softened Plummer interaction kernel
from Eq.~\eqref{Usoft},
we used the collisionless code
\texttt{gyrfalcON}~\citep{Dehnen2000}.
Simulations were performed using the Plummer softening kernel $P_{0}$,
which matches exactly with Eq.~\eqref{Usoft}.
Integration parameters were chosen so that the integration time step is ${ \Delta t \!=\! 2^{-8} \, \HU }$,
while the minimum tolerance parameter
was fixed to ${ \theta \!=\! 0.6 }$.

To fix the softening length, we followed the same approach
as in~\cite{Theuns1996} (see Eq.~{(14)} therein).
In the very core of the cluster, the typical interparticle distance
is given by
\begin{equation}
d = \big[ \mu / \rho (0) \big]^{1/3} .
\label{def_interdist}
\end{equation}
For the isochrone potential~\citep[see Eq.~{(2.49)} of][]{BinneyTremaine2008},
this simply becomes
\begin{align}
d {} & = \big[ 16 \pi / (3 N) \big]^{1/3} \, \bISO 
\nonumber
\\
{} & \simeq 0.013 \, \HU \;\;\;\;\;\;\; \text{for} \;\; N = 10^{5} .
\label{d_iso}
\end{align}
It is then appropriate to consider a softening length
comparable with this scale.
In practice, for the fiducial runs,
we used ${ \veps \!=\! d/2 }$.
Each individual realisation was run on a single CPU-core.
Integrating one realisation up to ${ \tmax \!=\! 10^{3} \, \HU }$
required about $17$h of computation.
We performed a total of ${ \Nreal \!=\! 168 }$
different realisations with ${ \veps \!=\! 0.0065 \, \HU }$.

\subsection{Measuring the diffusion rate}
\label{sec:MeasureFlux}

As recently investigated in~\cite{Heggie+2020}
(see also references therein),
the density centre of the cluster undergoes
a correlated random walk throughout
its relaxation.
It is therefore of prime importance
to correctly centre the coordinate system
before attempting any measurement of the particles' actions.

In the collisional runs, to estimate the position
of the density centre, we followed the algorithm
from~\cite{CasertanoHut1985}
using the ${ j \!=\! 6 }$ nearest neighbours to estimate
the local densities.
In the collisionless runs,
we used the same algorithm with ${ j \!=\! 32 }$.

Once the origin of the coordinate system has been determined,
we checked that the cluster's mean potential
had scarcely changed
from the initial mean isochrone potential.
We could then use the particles' position and velocities ${ (\br , \bv) }$
to estimate their energy and angular momentum
${ (E,L) }$, and finally the associated actions ${ (J_{r} , L) }$,
following Eq.~\eqref{Jr_iso}.
Having determined the particle's actions at the initial time ${ t \!=\! 0 }$,
and at the late time ${ t \!=\! 10^{3} \, \HU }$,
we binned particles into ${ 20 \!\times\! 20 }$ bins
within the domain
${ 0 \!\leq\! J_{r} \!\leq\! 0.4 \, \HU }$
and ${ 0 \!\leq\! L \!\leq\! 0.7 \, \HU }$.
This allowed us to compute the variation
in the number of particles
within each action bin through a simple difference
between both times.
Once averaged over available realisations,
this led to a direct estimation of ${ \p F (\bJ) / \p t }$,
as presented in Figs.~\ref{fig:dFdtNewtonian} and~\ref{fig:dFdtSoftened}.

\end{document}